\documentclass[prb,twocolumn,notitlepage,superscriptaddress]{revtex4-2}
\usepackage{amsmath,amsfonts,amssymb,amsthm,epsfig,array}
\usepackage{dsfont}
\usepackage{slashed}
\usepackage{graphics}
\usepackage{float}
\usepackage{verbatim}
\usepackage{color}
\usepackage{bm}
\usepackage[dvipsnames]{xcolor}
\usepackage[hidelinks,colorlinks,linkcolor=blue,
citecolor=blue,urlcolor=blue]{hyperref}
\usepackage[titletoc,title]{appendix}
\usepackage{multirow}
\usepackage{bibentry}
\usepackage{tabularx}
\usepackage[mathscr]{euscript}
\usepackage{mathtools}
\usepackage{hhline}
\usepackage[normalem]{ulem}
\usepackage{chemformula}

\newcommand{\bsl}[1]{\boldsymbol{#1}}

\renewcommand{\mod}{\,\mathrm{mod}\,}

\newcommand{\ii}{\mathrm{i}}

\newcommand{\Tr}{\mathop{\mathrm{Tr}}}

\newcommand{\eqnref}[1]{Eq.\,\eqref{#1}}
\newcommand{\figref}[1]{Fig.\,\ref{#1}}

\newcommand{\appref}[1]{Appendix.\,\ref{#1}}
\newcommand{\refcite}[1]{Ref.\,\cite{#1}}

\newcommand{\mat}[1]{\left(\begin{matrix}#1\end{matrix}\right)}

\newcommand{\eq}[1]{\begin{equation} #1 \end{equation}}

\newcommand{\eqa}[1]{\begin{align}\begin{split} #1 \end{split}\end{align}}

\usepackage{environ}
\NewEnviron{eqs}{%
\begin{equation}\begin{split}
    \BODY
\end{split}\end{equation}
}

\let\oldAA\AA
\renewcommand{\AA}{\text{\normalfont\oldAA}}

\newcommand{\ie}{{\emph{i.e.}}}
\newcommand{\eg}{{\emph{e.g.}}}

\newcommand{\eV}{\text{eV}}

\newcommand{\be}{\begin{equation}}
\newcommand{\ee}{\end{equation}}
\newcommand{\bea}{\begin{equation} \begin{aligned}}
\newcommand{\eea}{\end{aligned} \end{equation} }
\newcommand{\bi}{\begin{itemize}}
\newcommand{\ei}{\end{itemize}}

\renewcommand{\be}{\beta}
\newcommand{\al}{\alpha}
\newcommand{\bpm}{\begin{pmatrix}}
\newcommand{\epm}{\end{pmatrix}}

\newcommand{\lp}{\left(}
\newcommand{\rp}{\right)}

\newcommand{\del}{\partial}

\newcommand{\mbf}[1]{\mathbf{#1}}
\renewcommand{\d}{\downarrow}

\usepackage{color}
\usepackage{graphicx}
\usepackage{verbatim}
\usepackage{amsmath}
\usepackage{amssymb}
\usepackage{wasysym}
\usepackage{mathrsfs}
\usepackage[caption=false]{subfig}
\usepackage{url}
\usepackage{bbold}
\usepackage{slashed}
\usepackage{epstopdf}
\usepackage{braket}
\usepackage{float}
\usepackage[percent]{overpic}

\DeclareRobustCommand{\Sec}[1]{Sec.~\ref{#1}}

\DeclareRobustCommand{\App}[1]{App.~\ref{#1}}

\DeclareRobustCommand{\Fig}[1]{Fig.~\ref{#1}}

\DeclareRobustCommand{\Eq}[1]{Eq.~(\ref{#1})}

\DeclareMathAlphabet\mathbfcal{OMS}{cmsy}{b}{n}

\RequirePackage[normalem]{ulem} 
\RequirePackage{color}\definecolor{RED}{rgb}{1,0,0}\definecolor{BLUE}{rgb}{0,0,1} 

\begin{document}

\title{Pb$_{9}$Cu(PO$_4$)$_6$(OH)$_2$: Phonon bands, Localized Flat Band Magnetism, Models, and Chemical Analysis}

\author{Yi Jiang}
\thanks{These authors contributed equally.}
\affiliation{Donostia International Physics Center (DIPC), P. Manuel de Lardizabal 4, 20018 Donostia-San Sebastian, Spain}

\author{Scott B. Lee}
\thanks{These authors contributed equally.}
\thanks{SBL grew crystals and solved the SCXD solutions reported in this paper.}
\affiliation{Department of Chemistry, Princeton University, Princeton, New Jersey 08544, USA}

\author{Jonah Herzog-Arbeitman}
\thanks{These authors contributed equally.}
\affiliation{Department of Physics, Princeton University, Princeton, New Jersey 08544, USA}

\author{Jiabin Yu}
\thanks{These authors contributed equally.}
\affiliation{Department of Physics, Princeton University, Princeton, New Jersey 08544, USA}

\author{Xiaolong Feng}
\thanks{These authors contributed equally.}
\affiliation{Max Planck Institute for Chemical Physics of Solids, 01187 Dresden, Germany}

\author{Haoyu Hu}
\thanks{These authors contributed equally.}
\affiliation{Donostia International Physics Center (DIPC), P. Manuel de Lardizabal 4, 20018 Donostia-San Sebastian, Spain}

\author{Dumitru C\u{a}lug\u{a}ru}
\thanks{These authors contributed equally.}
\affiliation{Department of Physics, Princeton University, Princeton, New Jersey 08544, USA}

\author{Parker S. Brodale}
\thanks{These authors contributed equally.}
\affiliation{Department of Chemistry and Biochemistry, University of Oregon, Eugene, OR, 97403}

\author{Eoghan L. Gormley}
\thanks{These authors contributed equally.}
\affiliation{Department of Chemistry and Biochemistry, University of Oregon, Eugene, OR, 97403}

\author{Maia G. Vergniory}
\affiliation{Donostia International Physics Center (DIPC), P. Manuel de Lardizabal 4, 20018 Donostia-San Sebastian, Spain}
\affiliation{Max Planck Institute for Chemical Physics of Solids, 01187 Dresden, Germany}

\author{Claudia Felser}
\affiliation{Max Planck Institute for Chemical Physics of Solids, 01187 Dresden, Germany}

\author{S. Blanco-Canosa}
\affiliation{Donostia International Physics Center (DIPC), P. Manuel de Lardizabal 4, 20018 Donostia-San Sebastian, Spain}
\affiliation{IKERBASQUE, Basque Foundation for Science, 48013 Bilbao, Spain}

\author{Christopher H. Hendon}
\affiliation{Department of Chemistry and Biochemistry, University of Oregon, Eugene, OR, 97403}

\author{Leslie M. Schoop}
\thanks{These authors contributed equally.}
\affiliation{Department of Chemistry, Princeton University, Princeton, New Jersey 08544, USA}

\author{B. Andrei  Bernevig}
\thanks{These authors contributed equally.}
\affiliation{Department of Physics, Princeton University, Princeton, New Jersey 08544, USA}
\affiliation{Donostia International Physics Center (DIPC), P. Manuel de Lardizabal 4, 20018 Donostia-San Sebastian, Spain}
\affiliation{IKERBASQUE, Basque Foundation for Science, 48013 Bilbao, Spain}

\date{\today}

\begin{abstract}
In a series of recent reports, doped lead apatite (LK-99) has been proposed as a candidate ambient temperature and pressure superconductor. However, from both an experimental and theoretical perspective, these claims are largely unsubstantiated. To this end, our synthesis and subsequent analysis of an LK-99 sample reveals a multiphase material that does not exhibit high-temperature superconductivity. We study the structure of this phase with single-crystal X-ray diffraction (SXRD) and find a structure consistent with doped \ch{Pb10(PO4)_6(OH)2}. However, the material is transparent which rules out a superconducting nature. From \emph{ab initio} defect formation energy calculations, we find that the material likely hosts \ch{OH-} anions, rather than divalent $O^{2-}$ anions, within the hexagonal channels and that Cu substitution is highly thermodynamically disfavored. Phonon spectra on the equilibrium structures reveal numerous unstable phonon modes. Together, these calculations suggest it is doubtful that Cu enters the structure in meaningful concentrations, despite initial attempts to model LK-99 in this way. However for the sake of completeness, we perform \emph{ab initio} calculations of the topology, quantum geometry, and Wannier function localization in the Cu-dominated flat bands of four separate doped structures. In all cases, we find they are atomically localized by irreps, Wilson loops, and the Fubini-Study metric. It is unlikely that such bands can support strong superfluidity, and instead are susceptible to ferromagnetism (or out-of-plane antiferromagnetism) at low temperatures, which we find in \emph{ab initio} studies. In sum,  Pb$_{9}$Cu(PO$_4$)$_6$(OH)$_2$ could more likely
be a magnet, rather than an ambient temperature and pressure superconductor.
\end{abstract}

\maketitle

\section{Introduction}

A wave of scientific and social interest has followed a recent claim that LK-99\cite{LEE23a, LEE23b}, with the proposed composition of Pb$_{10-x}$Cu$_x$(PO$_4$)$_6$O, exhibits ambient temperature and pressure superconductivity~\cite{ABR23, BAS23, BRU95, CAB23, GRI23, GUO23, HIR23, HOU23, KUM23, KUM23a, KUR23, LAI23, LIU23a, OH23, SI23, SUN23, TAO23, TAV23, WU23}. Although the data presented in the original reports is insufficient to support such a remarkable claim, a large body of immediate work has followed. Experimentally, it is unclear what the structure and composition of the material really are; most likely it is a multi-phase sample. Experiments have suggested diamagnetic behavior, and levitation experiments have discussed that it may arise from either diamagnetism or small ferromagnetic impurities \cite{GUO23}. As the sample is likely to contain multiple phases, it is possible that two different compounds contribute to each property, \textit{i.e.} one part is diamagnetic, and the other metallic. Thus, clarification of the composition of the material is necessary \emph{before} trusting the models based on postulated structures. Assuming some of the numerous possible compositions, a number of \textit{ab initio} band structures have been produced. Claims of flat bands -- and their positive influence on superconductivity -- have been made. 

However, given the stakes, computation and prediction of physical properties requires an elevated level of accuracy. While flat bands provide a theoretical platform for high-temperature superconductivity, non-trivial quantum geometry is a compulsory for superfluidity, i.e. the Meissner effect\cite{2015NatCo...6.8944P,2022NatRP...4..528T,2021NatRP...3..462Z,2018PhRvB..98m4513T,2016PhRvB..94x5149T,2017PhRvB..95b4515L,PhysRevB.106.014518,PhysRevLett.123.237002,PhysRevLett.131.016002,2023PNAS..12017816M,2023PhRvL.130v6001H}. This is because the mass of the condensing Cooper pair is inversely proportional to the minimal Fubini-Study metric \cite{2018PhRvB..98v0511T,2022arXiv220900007H}, a rigorous measure of quantum geometry. A variety of nonzero lower bounds on the minimal Fubini-Study metric exist for non-atomic bands \cite{2015NatCo...6.8944P,PhysRevB.105.104515,PhysRevB.107.L201106,2020PhRvL.124p7002X,2022PhRvL.128h7002H}. However, flat bands are in fact detrimental to superfluidity if they are atomically localized, i.e. if their narrow dispersion comes from Wannier localization as opposed to destructive interference \cite{2022NatPh..18..185C,PhysRevB.104.L081104,2021PhRvB.104h5144H}. The inescapability of this conclusion is seen in the limiting case of a tight-binding model with all hoppings vanishing. The perfectly flat band that results cannot support transport of any kind, much less superconductivity, even in the presence of attractive Hubbard interactions. The key physics of flat band superconductivity thus lies in deviations from this limit, measured by quantum geometry \cite{2011EPJB...79..121R}. Furthermore, if Coulomb repulsion overwhelms the would-be attractive interaction, flat band ferromagnetism, rather than superconductivity, would be favored \cite{PhysRevLett.62.1201,1993CMaPh.158..341M}. Thus a comprehensive understanding of the active bands at the Fermi level is required for predictions of the many-body state.

\begin{figure}[htbp]
    \centering
    \includegraphics[width=\columnwidth]{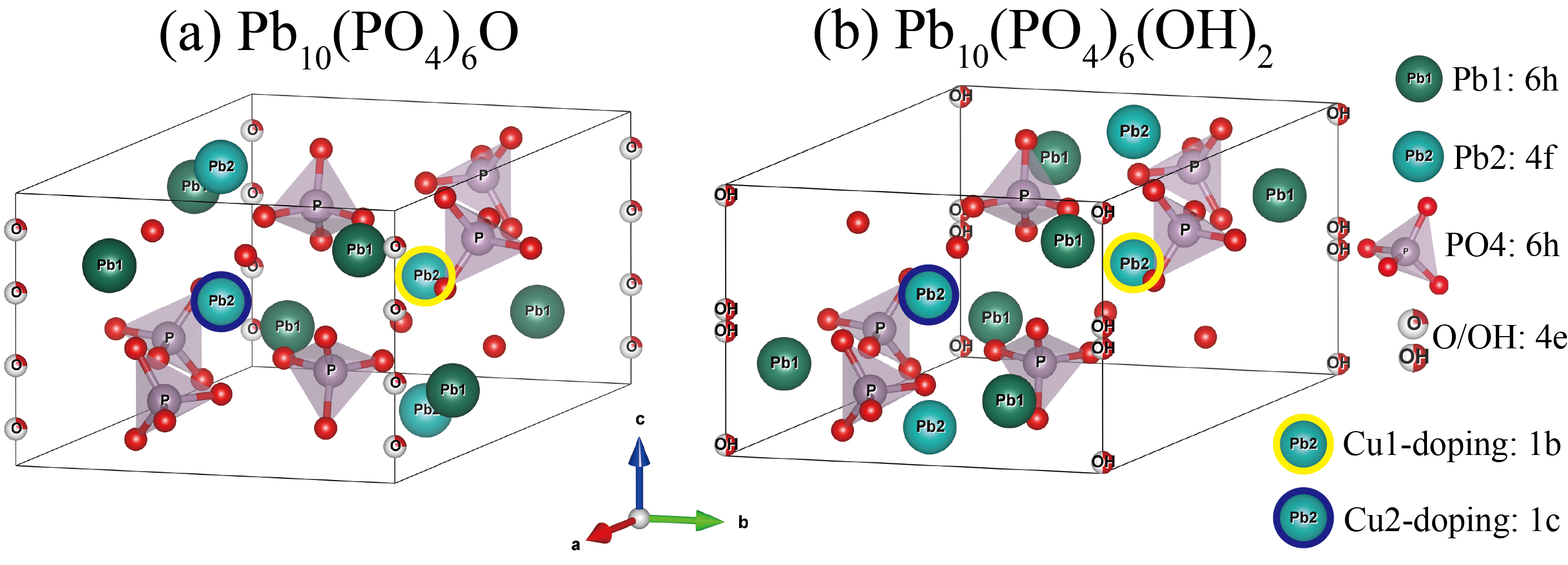}
    \caption{The crystal structure for (a) \ch{Pb10(PO4)6O}\cite{krivovichev2003crystal} and (b) \ch{Pb10(PO4)6(OH)2}\cite{bruckner1995crystal}. The Wyckoff positions of atoms in space group 176 $P6_3/m$ are labelled, with their coordinates given in Table. \ref{table:crystal_structure}. The O atoms surrounding P atoms that form PO$_4$ are at $6h$ or generic positions and are not labeled for simplicity. We also mark two possible Cu doping positions on Pb$_2$ with yellow and blue circles, which are $1b$ and $1c$ Wyckoff positions in space group 143 $P3$ and will be called Cu$_1$- and Cu$_2$-doping, respectively. 
    The H atoms are not shown in (b) for simplicity, which are close to the trigonal O atoms at $4e$ Wyckoff position. 
    }
    \label{fig:crystal_structure}
\end{figure}

Due to new advances in understanding and classification of band structures \cite{bradlyn2017topological, elcoro2021magnetic}, we can analyze their symmetry, localization, and topology with extreme accuracy that only depends on the accuracy of the DFT calculation. In this work, we study two lead apatites (see \Fig{fig:crystal_structure}) available in materials databases, \ch{Pb10(PO4)6O}\cite{krivovichev2003crystal} and \ch{Pb10(PO4)6(OH)2}\cite{bruckner1995crystal} -- which we emphasize may not be the ultimate material structure. Making assumptions about the location of the fractionally occupied O anion and the Cu dopant, we obtain several microscopic models for the electron and phonon bands in this system. Our results can be summarized as follows.

Firstly in the \ch{Pb10(PO4)6O} initial structure (\Fig{fig:crystal_structure}(a)), we consider Cu replacing Pb at two possible positions, either the 1b or 1c Wyckoff positions, referred to as the \ch{Cu1}- and \ch{Cu2}-dopings respectively.  
Both possibilities result in a set of two bands at the Fermi level dominated by Cu (see Sec. \ref{sec:electronic_structure}). 
They are narrow with a $\sim100$-meV bandwidth, and form an elementary band representation \cite{bradlyn2017topological} of the Cu $d$-orbitals. We compute their Fubini-Study metric and non-abelian Wilson loops, which show strong localization, although the the Cu$_2$ structure does have a significantly reduced gap to the O bands below.  For both possibilities, we obtain 4-band, short-ranged, symmetric tight-binding models which demonstrate that the weak dispersion of the Cu orbitals arises primarily from hybridization with nearby O bands. Secondly we consider the \ch{Pb10(PO4)6(OH)2} structure (\Fig{fig:crystal_structure}(b)). For both locations of the Cu dopant, we again find a set of Cu bands in an elementary band representation with $\sim100$-meV bandwidth. However, the gap to the nearby O bands is much larger, and we provide a 2-band model built entirely from Cu Wannier functions. Again, the Fubini-Study metric indicates atomic localization. See \Sec{sec:abinitio} and \Sec{sec:TBmodels} for more details.

In all cases, the two bands at the Fermi level lack strong quantum geometry. Due to their flat, localized nature, ferromagnetism seems to be the preferred configuration of these states in ab initio studies (see Sec. \ref{sec:electronic_structure}). The absence of extended states in these bands does not support a theory of high-temperature superconductivity based on the flat bands we obtain in these structures. However, preliminary calculations of the phonon spectrum show that more careful relaxation of the doped compound is required to fully stabilize their structures, which may result in changes to the band geometry. A phonon-driven mechanism for superconductivity must also compete with the strong Hubbard repulsive interaction, which we also construct using \textit{ab initio} Hubbard-Kanamori parameters. 

\begin{figure}[H]
    \centering
    \includegraphics[width=\columnwidth]{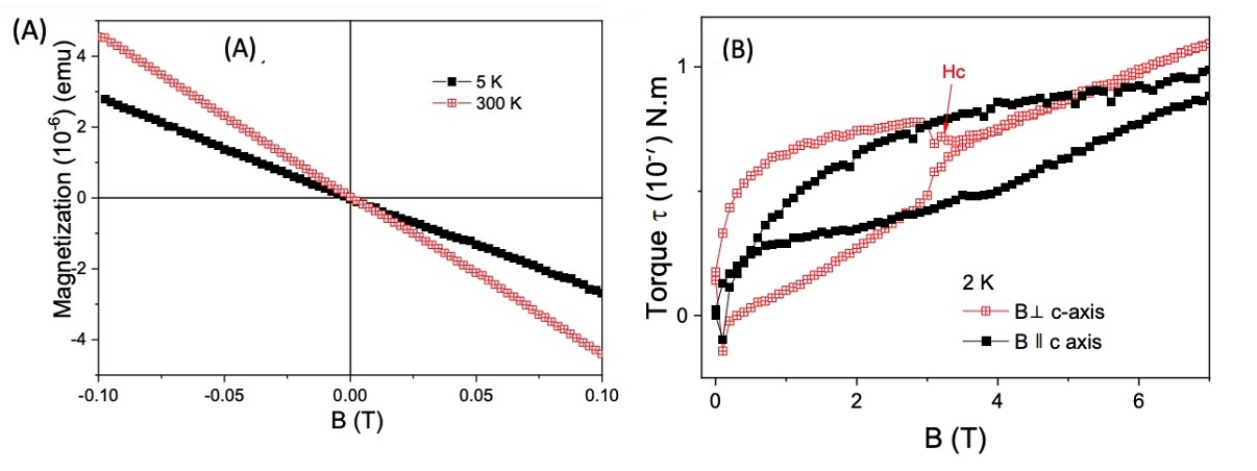}
    \caption{(A) Magnetization \textit{vs} field for (HOPG) Highly Oriented Pyrolityc Graphite, showing a diamagnetic response. (B) Hystereses in the curves of the torque ($\tau$) \textit{vs} H for NbSe$_2$.}
    \label{fig:Torque}
\end{figure}

In Refs.\,\cite{LEE23a,LEE23b}, LK-99 shows a sharp drop of the resistivity around 400 K, towards a state claimed to be superconducting. Nevertheless, the reported value of resistivity is 2-3 orders of magnitude higher than that of good metals; for instance, Cu presents a resistivity value of 10$^{-6}$ $\Omega$cm. Moreover, the analysis of the reported specific heat shows no transition up to 400 K, in principle, at odds with a jump expected due to the release of entropy of a second order phase transition. Indeed, the specific heat seems to drop with temperature, which adds more controversy to the claim of room-temperature superconductivity. Ref. \cite{2023arXiv230804353Z} has recently found evidence for the \ch{Cu2S} present in the multi-phase compound sourcing the resistivity transition. Moreover, the presence of a diamagnetic signal does not necessarily imply SC. In \figref{fig:Torque}(A), we show the magnetic response of the diamagnetic compound HOPG (Highly Oriented Pyrolityc Graphite). As we can see, the diamagnetism of HOPG is clearly different from any superconductor \cite{tinkham2004introduction}. 

On the other hand, besides the typical hysteric behavior of the magnetization expected in a SC, in \figref{fig:Torque}{B}, we show the magnetic torque response $\pmb{\tau}=\mbf{M} \times \mbf{B}$ of the type II SC NbSe$_2$. The hysteric behavior between the sweep-up and -down curves defines the different field regions of a type II SC (vortex solid, vortex liquid, critical field, etc) \cite{OngPNAS}.

\section{Chemical Structure}
Pb-apatite is structurally related to the parent compound, Ca-apatite, and much can be learned through their comparison. This crystal type features 1-dimensional channels filled with requisite charge balancing anions. For example, the calcium-oxo analogue has been shown to only exist as \ch{Ca10(PO4)6O} at temperatures exceeding 1000$^oC$,\cite{LIAO19991807} below of which it rapidly includes water into the lattice, forming the neutral hydrate \ch{Ca10(PO4)6(OH)2}. While the Pb-based system may have different dehydration temperatures to interconvert between Pb$_{10}$(PO$_4$)$_6$O\cite{krivovichev2003crystal} and \ch{Pb10(PO4)6(OH)2}\cite{bruckner1995crystal} (see \Fig{fig:crystal_structure}), historical data would indicate that the hydrated material is thermodynamically preferred. To this end, we performed a number of  DFT calculations to assess the formation enthalpy of the hydroxy and oxo Pb apatites considered here. Our calculations reveal that the inclusion of \ch{H2O}, forming \ch{Pb10(PO4)6(OH)2}, is exothermic with its inclusion favored by 38.5 kcal/mol. These data suggest while Pb$_{10}$(PO$_4$)$_6$O may be a metastable form, direct syntheses of that material will tend to form the hydrate if sufficient protons are available. And even if Pb$_{10}$(PO$_4$)$_6$O is formed it will interconvert to the hydrate upon exposure to air.

Structurally, the oxo and hydroxy apatites are better distinguished by their lattice parameters, and less so by direct crystallographic measurements. Ca-apatite exhibits a dramatic lattice contraction in both the \textit{a} and \textit{c} parameters\cite{negas1968high}, proportional to the extent of hydration, \textit{i.e.} the conversion from O$^{2-}$ to OH$^-$ contracts the lattice.\cite{cell_contraction}. Such contraction has been attributed to Cu-inclusion within the Pb$_{10}$(PO$_4$)$_6$O lattice\cite{LEE23b,GRI23}, but a similar effect is predicted through simple hydration.

While theoretical studies thus far have focused on the pure Pb$_{10-x}$Cu$_x$(PO$_4$)$_6$O or Pb$_{10-x}$Cu$_x$(PO$_4$)$_6$(OH)$_2$ phase, the reported synthesis methods cannot result in single phase samples of either of them. In the final step of the reaction, Lanarkite (Pb$_2$(SO$_4$)O) is mixed in a 1:1 molar ratio with Cu$_3$P, but this 2:1 Pb/P ratio is stoichiometrically inconsistent with the desired product. Even loosening the ratio of the reactants would create a significant amount of Cu impurities: the final product has 6 P, so there will be 18 Cu per formula unit of Pb$_{9}$Cu$_1$(PO$_4$)$_6$O or Pb$_{9}$Cu$_1$(PO$_4$)$_6$(OH)$_2$. The powder X-ray diffraction (PXRD) pattern presented in the original reports\cite{LEE23a, LEE23b} suggests that the majority of the sample exists in a structure related to \ch{Pb10(PO4)6(OH)2}, but also features many crystalline impurities. Due to the severe off-stoichiometry of the reaction, additional amorphous phases are also to be expected. Only few Rietveld analyses have thus far been performed\cite{KUM23a}, and it is difficult to separate the structural effects of Cu inclusion versus hydration without high-quality single crystal data. To this end, is unclear if \textit{any} Cu atoms were actually incorporated into the structure. 

\begin{figure}[H]
     \centering
     \includegraphics[width=0.8\columnwidth]{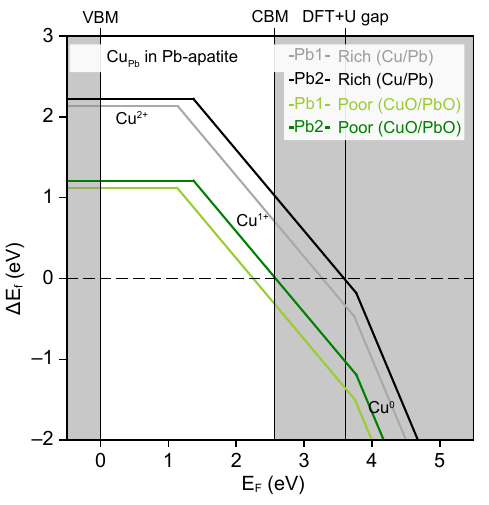}
     \caption{The defect formation enthalpies for Cu substitution in Pb lattice sites at both rich and poor potentials show that Cu$_\mathrm{Pb}$ is unfavorably incorporated into either Pb site within the apatite lattice.}
     \label{fig:defect}
 \end{figure}

To address this, we are able to compute the formation enthalpies of Cu substitutions at Pb lattice sites. Following the standard procedure\cite{RevModPhys.86.253}, the Cu defects were computed using a 2$\times$2$\times$2 super cell containing 1831 electrons for the neutral substitution. Computations were referenced to bulk Cu/Pb (rich potentials) and CuO/PbO (poor potentials) and the formation enthalpy was computed using a 2$\times$2$\times$2 k-mesh, with the FNV correctional scheme employed for charged defects and averaged diagonalized dielectric tensors\cite{walsh_correcting_2021,PhysRevLett.102.016402}, and Hubbard model (U = 4.0 eV for Cu). The formation enthalpies for the defects are depicted in \ref{fig:defect}. Notably, the experimentally observed Cu$^{2+}$ (\textit{i.e.} charge-neutral) substitution is predicted to form in \textit{p}-type conditions, at minimum occurring with a 1.2 eV penalty depending on the reference potentials and amounting to many orders of magnitude sub-stoichiometric Cu concentration, essentially forbidding its inclusion within the material. However, the charge transition from inclusion of Cu$^{2+}$ to Cu$^{1+}$ occurs very near the Fermi level for the undoped Pb-apatite material, and Cu$^{1+}$ inclusions becomes more favored in \textit{n}-type conditions. The experimental absence of appreciable Cu$^{1+}$ suggests that if Cu$^{2+}$ is incorporating, it is being aided by other correlated defects not considered here.  Additionally, other mechanisms of Cu inclusion are possible (\textit{e.g.} interstitial formation, as observed in Pb$^{2+}$-containing in lead halide perovskites\cite{bian2022unveiling}). Pb-apatite may host other dopants \textit{e.g.} S, which is also present during the reaction\cite{Cerny1994}. These may be studied in future work.

To address whether Cu and OH$^-$ are included, and to generally deduce the structure, we performed a modified synthesis of the putative LK-99 compound described in  Ref.\cite{LEE23a, LEE23b}. Sample purity of all intermediate reagents was confirmed via powder X-ray diffraction using a STOE Stadi P powder X-ray diffractometer equipped with a Mo K$\alpha$ ($\lambda$=0.71073 \AA) sealed-tube x-ray source and graphite monochromator at room temperature in either Debye-Scherrer or transmission geometry \Fig{fig:Pb2SO5}, \Fig{fig:Cu3P}. In the final step, Cu$_3$P and Pb$_2$(SO$_4$)O in a 1:1 molar ratio were ground into a fine powder in mortar and pestle. The powders were loaded into an alumina crucible, placed in a quartz tube, and sealed under dynamic vacuum without any argon back-filling. The tube was then loaded in a furnace and heated to 950 $^{\circ}$C over 4 hours, kept at this temperature for 6 hours, and then shut off to cool quickly. (Upon pulling the sample out of the furnace we observed that our sample, due to its loading in an alumina crucible, did not attack our quartz tube in contrast to 
Refs. \cite{LEE23a, LEE23b}. This suggests the possibility of further chemical reactions in the protocol of Refs. \cite{LEE23a, LEE23b}.) The resulting product inside the crucible contained mostly white and orange colored powder in addition to metallic gray lumps as seen in \Fig{fig:Multiphase}. A PXRD pattern of this sample can be seen in \Fig{fig:ExOver}, showing that our sample has more impurity phases than the original LK-99 sample, but that the characteristic peaks are present and thus it seems possible to isolate the main phase reported in the original sample from our sample as well. Energy-dispersive X-ray spectroscopy (EDS) was also utilized to examine these impurity phases. In  \Fig{fig:redandwhite}, we show that the transparent-orange phase likely gets its color from pockets of metallic Cu. Likewise, in \Fig{fig:metallic} we observe islands of Cu$_2$S embedded in a matrix of the transparent-white phase. As noted previously\cite{zhu2023order, jain2023phase}, these Cu$_2$S islands are likely the cause of the tenfold drop in resistivity observed at 104.8 $^{\circ}$C in the original preprint. 
 
\begin{figure}
    \centering
    \includegraphics[width=0.8\columnwidth]{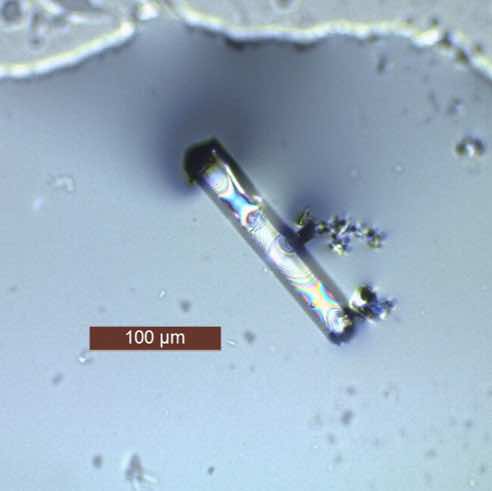}
    \caption{A single crystal similar in transparency, shape, and size to what we picked for SXRD.}
    \label{fig:SingleXtal}
\end{figure}

Under magnification, the white powder was observed to be transparent single crystals with a clear hexagonal rod habit (\Fig{fig:SingleXtal}). Knowing that Pb$_{10}$(PO$_4$)$_6$O was reported to crystallize in a hexagonal space group, a transparent single crystal of dimensions (0.55 $\times$ 0.092 $\times$ 0.176 mm) was picked for single crystal X-ray diffraction (SXRD) analysis using an APEX2 CCD diffractometer equipped with a Mo K$\alpha$ ($\lambda$=0.71073 \AA) sealed-tube X-ray source and graphite monochromator at room temperature. Initial unit cell refinement obtained a unit cell in a hexagonal setting with lattice parameters \textbf{a}=9.8508(1) \AA and \textbf{c}=7.4395(2) \AA, agreeing with other previously reported structures for lead apatite \cite{bruckner1995crystal, krivovichev2003crystal}. Indexation and integration proceeded smoothly for a full hemisphere collection out to a resolution of 0.5 \AA.  Run list generation and frame data processing were done in APEX 2 \cite{Bruker2012}. An analytical absorption correction was used to scale the data before importing the peak list into JANA2020 \cite{Petricek2023}.  

\begin{figure}
    \centering
    \includegraphics[width=\columnwidth]{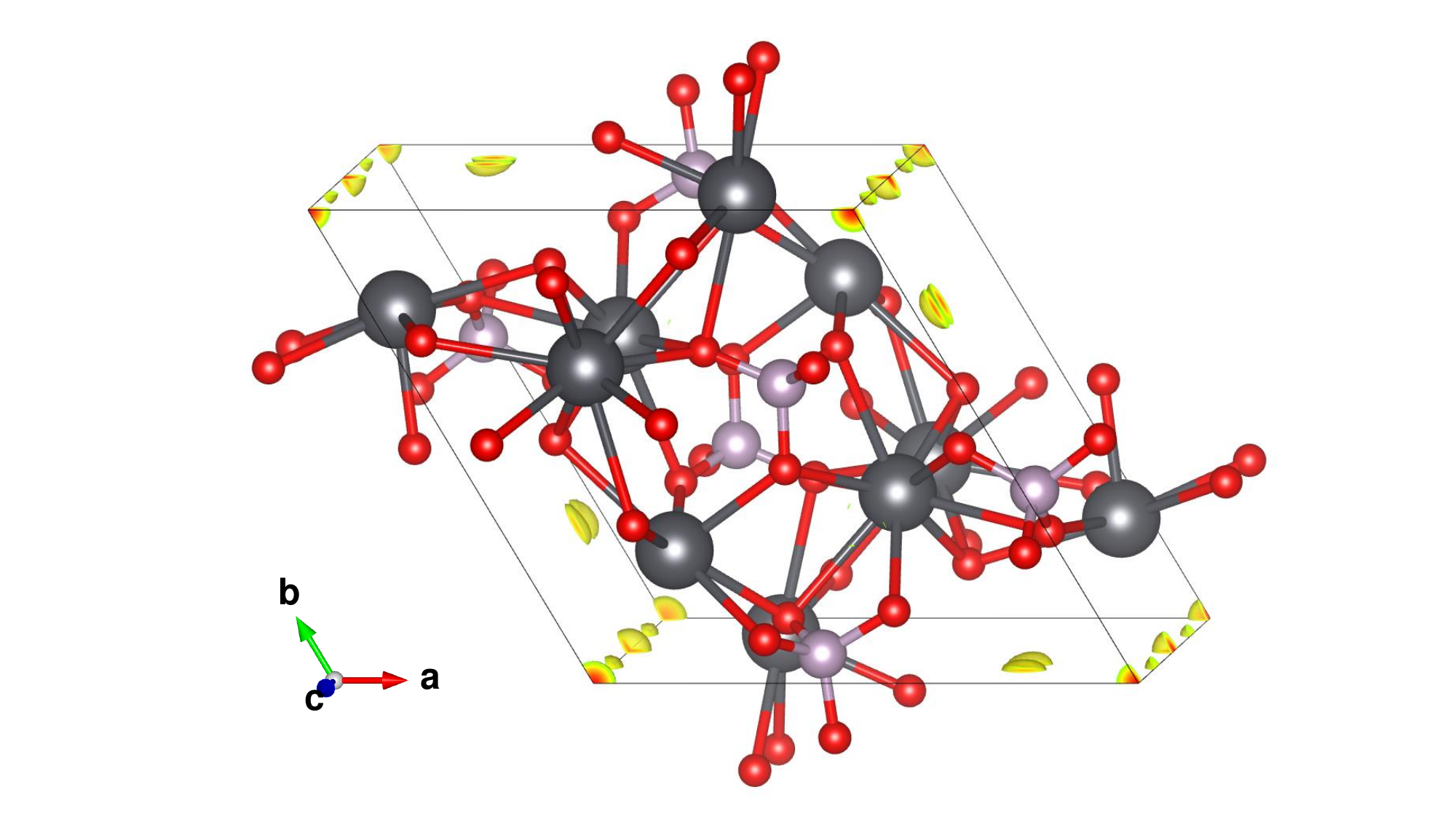}
    \caption{Residual electron density located at the 2a and 2b positions along the c axis. Positive isosurfaces are drawn at 7e \AA$^{-3}$.}
    \label{fig:ReseDen}
\end{figure}

The initial structure solution was obtained in space group P6$_3$/$m$, consistent with observed systematic absences \Fig{fig:Prec}, from the charge-flipping algorithm as implemented in SUPERFLIP \cite{Palatinus2007}. An initial refinement, done on $F^{2}$, was needed to obtain a valid initial structure with 6 distinct sites (2 Pb, 1 P, and 3 O). Notably, during this initial refinement Oxygen atoms along the c-axis were removed. Running a Fourier transform of residual electron density, visualized with the Vesta software package \cite{Momma2008}, reveals 2 unique crystallographic pockets of significant electron density around points [0 0 0] (Wyckoff position 2$b$) and [0 0 $1/4$] (2$a$) in P6$_3$/$m$, shown in \Fig{fig:ReseDen}, suggestive of a Pb$_{10}$(PO$_4$)$_6$(OH)$_2$ structural solution. After Oxygen atoms were added to these points, freely refining their occupancies results in a site occupancy factor of the 2$b$ Oxygen centered at [0 0 0] of 1.163, an unphysical value for an OH$^-$ molecular unit. Clearly, an atom with more electron density is needed. Replacing the oxygen instead with sulfur, a chemically similar element that is also present in the reaction, seems like a likely candidate, although further elemental analysis methods will be needed to distinguish between an SH$^-$, PH$_2^-$ or other possible dopants at this site. 

Freely refining occupancies for both an O at [0 0 $1/4$] and S located at [0 0 0] results in site occupancy factors of 0.54 and 0.56, respectively. The freely refined occupancies having a summation very close to 1 seems suggestive that this could be a physically real interpretation of the structural solution. After refining the isotropic parameters anisotropically, a final refinement converges with a goodness of fit (GoF) parameter of 1.22 and R value of 3.94 compared to all reflections. Here, our site occupancies for O (1.08) and S (1.12) sum to above 1. If we decide to restrict the occupancies of the 2$b$ S and 2$a$ O to sum to a site occupancy factor of 1, we end refinement with a final composition Pb$_{10}$(PO$_4$)$_6$(OH)$_{0.94}$(SH)$_{1.06}$. This refinement has nominally the same refinement statistics (GoF(all) $= 1.22$, R(all) $=3.95$). 

We also investigated two ways of possible Cu doping within our refinement: one in which the Cu atom substitutes on the Pb lattice site, and another which the Cu atom is inserted along the chain.  Attempts to dope both Pb sites with Cu results in a refined composition of Pb$_{9.55}$Cu$_{0.45}$(PO$_4$)$_6$(OH)$_{0.94}$(SH)$_{1.06}$ with similar statistics (GoF(all)$= 1.21$, R(all) $= 3.93$). We found the Cu has to be added in as a split site after Pb position has been refined completely. The last two steps of occupancy and anisotropic parameter refinement had to be done by restraining the Pb/Cu split site position with automatic refining keys switched off. 

\begin{figure*}
    \centering
    \includegraphics[width=2\columnwidth]{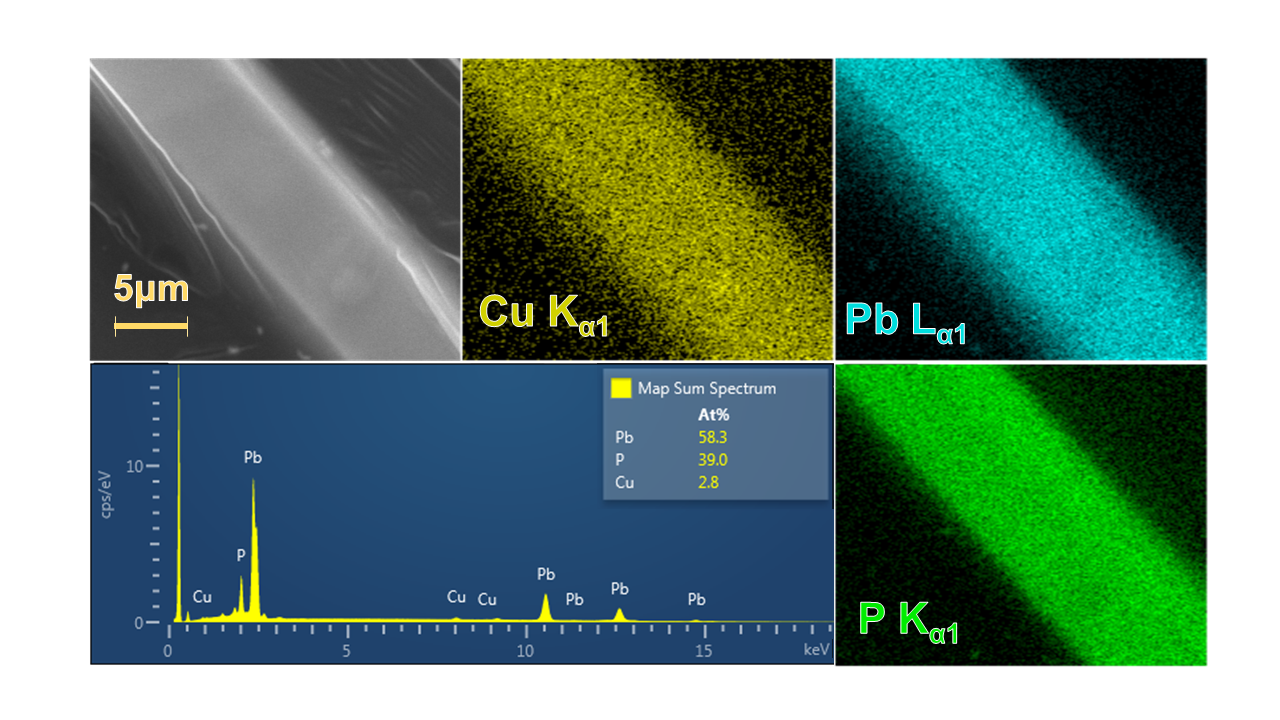}
    \caption{Energy dispersive X-ray Spectroscopy map of a translucent crystal coated in carbon.}
    \label{fig:EDX}
\end{figure*}

The compositional refinement of this structure certainly warrants skepticism. Unfortunately for us, this system has complications for EDS, namely, the Pb L$_{\alpha1}$ edge lies at the same energy as the S K$_{\alpha1}$ edge. This, along with neutron diffraction being needed to confirm H incorporation means future studies are needed. However, in search of confirming the refinements Cu incorporation, we carbon coated a small transparent needle and ran EDS on a Quanta environmental scanning electron microscope. Using the elemental mapping technique, we were surprised to find that we were able to detect Cu, and it is homogeneously distributed within the crystal. Furthermore, in \Fig{fig:EDX} the Pb:Cu ratio we detected over a roughly 15 minute period was 58.3:2.8, which roughly translates to a ratio of 9.55:0.46, an indicator that the Cu incorporation in our refined composition of Pb$_{9.55}$Cu$_{0.45}$(PO$_4$)$_6$(OH)$_{0.94}$(SH)$_{1.06}$ could be correct. However, we would like to acknowledge the fact that 2.8 atomic percent is extremely low for EDS characterization and future studies utilizing synchrotron radiation are needed to confirm incorporation. In addition, as there is a lot of Cu in the reaction it cannot be ruled out that Cu is at the surface of the crystal.

We can also obtain a reasonable refinement to the data if we place Cu into the channel with composition Pb$_{10}$(PO$_4$)$_6$(OH) $_{1.11}$ Cu$_{0.49}$  (GoF $= 1.22$, R $= 3.93$). Still, as the crystals are transparent, charge balance needs to be maintained and thus this last solution is not chemically reasonable. Full collection, integration, and refinement statistics can be found in \App{app:ExpDet}. We would to stress here again that the transparent nature of the crystal should rule out superconductivity as a property, it rather indicates a wide band gap.

Next, we compare simulated powder patterns of our structural solution, the ICSD reported structures, and the relaxed DFT structures of Cu doped variants with the published PXRD pattern from \cite{LEE23a, LEE23b}. To do this, structural position files were loaded into VESTA, and patterns simulated for using a Cu K$\alpha$ ($\lambda=1.5406 \AA$) wavelength. Simulated patterns were then overlaid atop the experimental pattern, extracted using an in-house Mathematica code. For ease of visualization, we adjust the simulated patterns via a zero point shift to match the peaks expected around 18 degrees in the experimental data. This zero-point correction is not uncommon for Rietveld refinements, and is needed if the diffractometer used in the experimental pattern is misaligned. Most zero-point corrections were minimal, i.e. $0.2^\circ$. We find a good agreement (see \figref{fig:ExpRefinements} and \Fig{fig:CompRefinements}) of our SXRD solution with the reported data. Other structures fit the data less well and a discussion is given in the SI. This analysis is preliminary as the cropped data from the original preprint are not of high enough quality to perform a Rietveld analysis. It does however show that our structure obtained from SRXD, measured in a \emph{transparent} crystal, agrees with the powder pattern published in the original LK-99 paper. Further analysis of our samples will follow. 

Due to the numerous uncertainties regarding the ultimate structural composition of the LK-99 material, we will investigate a variety of scenarios using ab initio density functional theory. In our ab initio studies, we still focus on structures on which Cu substitutes for Pb, as we come to different conclusions as previous theoretical works assuming the same substitutions.

\begin{figure}[htbp]
    \centering
    \includegraphics[width=\columnwidth]{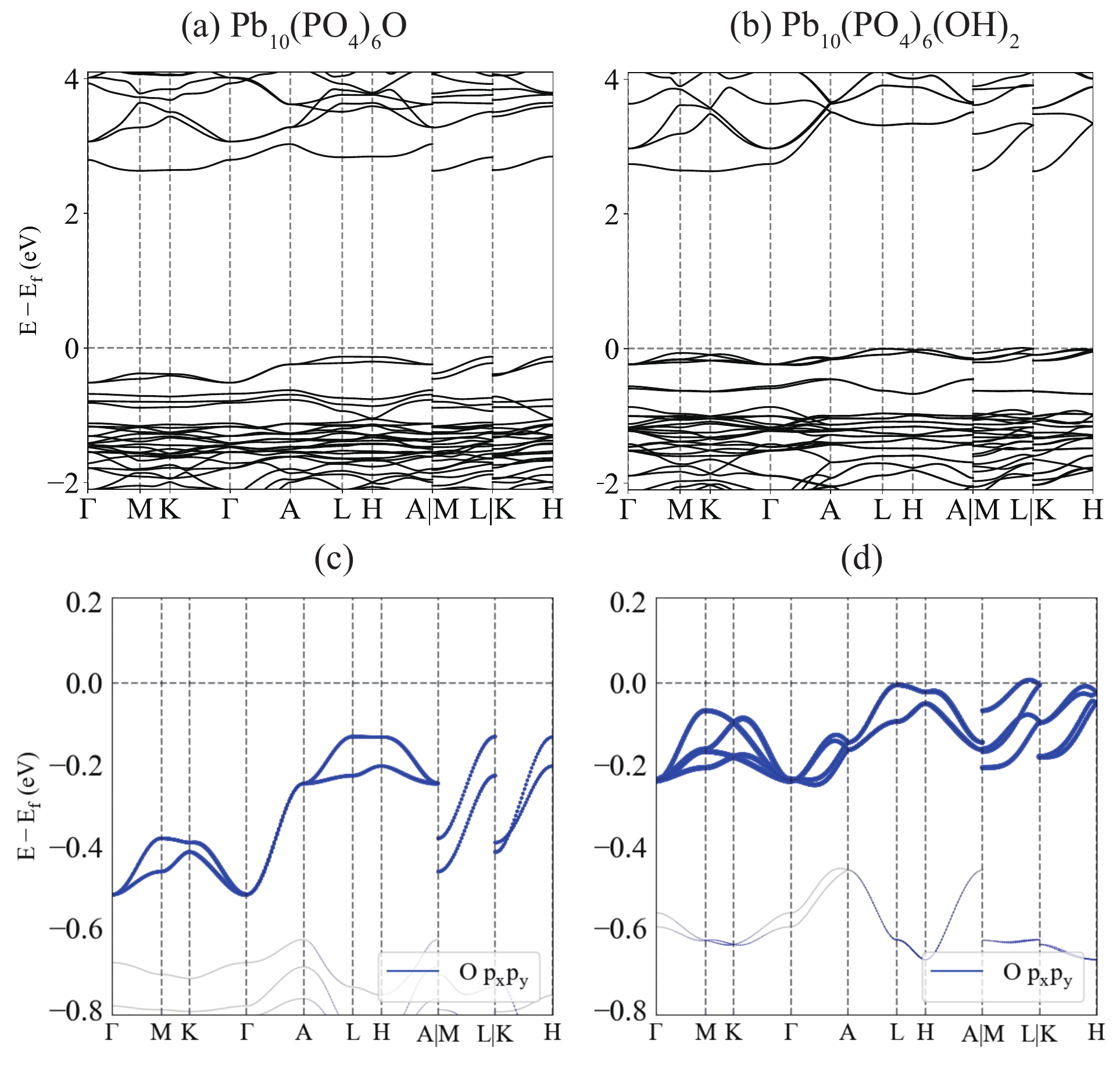}
    \caption{Electronic band structure for (a) \ch{Pb10(PO4)6O} and (b) \ch{Pb10(PO4)6(OH)2}, and zoom in plots in (c) for \ch{Pb10(PO4)6O} and (d) for \ch{Pb10(PO4)6(OH)2}. The quasi-flat bands near $E_f$ are mainly formed by the $(p_x,p_y)$ orbitals of the trigonal-O atoms, which are shown with blue weights. 
    }
    \label{fig:undoped_bands}
\end{figure}
\section{ab initio Results}
\label{sec:abinitio}

\begin{table}[htbp]
\begin{tabular}{c|c|c}
\hline\hline
Compound & Atom   & Wyckoff position  \\ \hline
\multirow{4}{*}{\ch{Pb10(PO4)6O}}
& Pb$_1$ & $6h$ \\ \cline{2-3} 
& Pb$_2$ & $4f$, $z=0.004$  \\ \cline{2-3} 
& P      & $6h$  \\ \cline{2-3} 
& tri-O  & $4e$, $z=0.134$, $\frac{1}{4}$-occu \\ \hline
\multirow{4}{*}{\ch{Pb10(PO4)6(OH)2}} 
& Pb$_1$ & $6h$ \\ \cline{2-3} 
& Pb$_2$ & $4f$, $z=0.994$   \\ \cline{2-3} 
& P      & $6h$  \\ \cline{2-3} 
& tri-(OH)$_2$ & $4e$, $z=0.040$, $\frac{1}{2}$-occu \\ \hline\hline
\end{tabular}
\caption{The atomic positions in \ch{Pb10(PO4)6O}\cite{krivovichev2003crystal} and \ch{Pb10(PO4)6(OH)2}\cite{bruckner1995crystal}. The lattice constants for \ch{Pb10(PO4)6O} are $a=9.865, c=7.431$ \AA, and for \ch{Pb10(PO4)6(OH)2} are $a=9.866, c=7.426$ \AA. 
The Wyckoff positions in SG 176 have the following coordinates: $4f=(\frac{1}{3},\frac{2}{3},z),(\frac{2}{3},\frac{1}{3},z+1/2),(\frac{2}{3},\frac{1}{3},-z),(\frac{1}{3},\frac{2}{3},-z+1/2)$, $4e=(0,0,\pm 1),(0,0,\pm 1+\frac{1}{2})$, and $6h=(x,y,\frac{1}{4}),(-y,x-y,\frac{1}{4}),(-x+y,-x,\frac{1}{4}),(-x,-y,\frac{3}{4}), (y,-x+y,\frac{3}{4}),(x-y,x,\frac{3}{4})$. 
For both compounds, the Pb$_2$ atoms at $4f$ approximately lie on the honeycomb lattices on $z=0,\frac{1}{2}$ planes. The `tri' in the table stands for trigonal lattice. The tri-O and tri-OH in the two compounds have fractional occupancies, i.e., $\frac{1}{4}$ for tri-O and $\frac{1}{2}$ for tri-(OH)$_2$.
}
\label{table:crystal_structure}
\end{table}

We consider two different experimental structures of lead apatite for \textit{ab-initio} calculations in this work, i.e., \ch{Pb10(PO4)6O}\cite{Villars2023:sm_isp_sd_1616772, krivovichev2003crystal} and \ch{Pb10(PO4)6(OH)2}\cite{Villars2023:sm_isp_sd_1003884, bruckner1995crystal}, both have space group (SG) 176 $P6_3/m$ symmetry.  Their crystal structures are shown in Fig. \ref{fig:crystal_structure}. The O atoms at $4e$ Wyckoff position in \ch{Pb10(PO4)6O} have $\frac{1}{4}$ occupancy, while \ch{(OH)2} atoms at $4e$ in \ch{Pb10(PO4)6(OH)2} have $\frac{1}{2}$ occupancy. The atomic positions are summarized in Table. \ref{table:crystal_structure}. 
We remark that lead apatite has many experimentally reported structures, with the position of trigonal-O or (OH)$_2$ being slightly different. For example, the structures in Ref.\cite{Villars2023:sm_isp_sd_1714473, zhu2007hydrothermal, Villars2023:sm_isp_sd_1631620, zhu2010crystallographic} has trigonal-(OH)$_2$ located at Wyckoff position $2a=(0,0,\frac{1}{4}),(0,0,\frac{3}{4})$ without fractional occupancy. 
The LK-99\cite{LEE23a, LEE23b} Pb$_{10-x}$Cu$_x$\ch{(PO4)6O} ($0.9<x<1.1$) is hypothesized to be synthesized by doping Pb atoms with Cu at $4f$ Wyckoff positions.

\begin{figure}[htbp]
    \centering
    \includegraphics[width=\columnwidth]{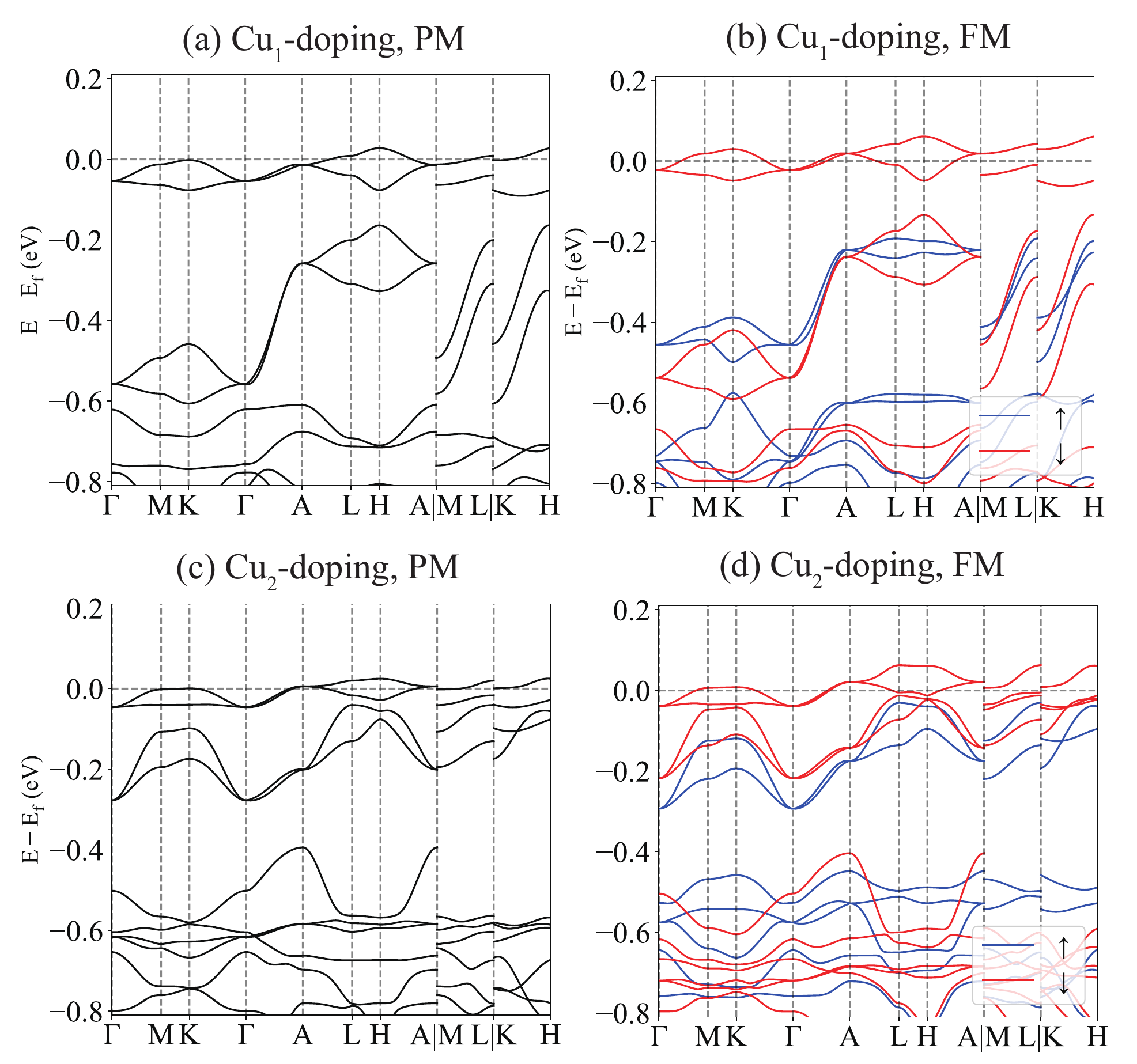}
    \caption{Electronic band structure for \ch{Pb9Cu1(PO4)6O}, where (a) is the PM phase and (b) is the FM phase of Cu$_1$-doping. (c) and (d) are similar for Cu$_2$-doping. The atomic positions of two Cu-doping can be found in Table. \ref{table:relaxed_atom_positions}.
    }
    \label{fig:1O_doped_bands}
\end{figure}

\subsection{Electronic Structure}\label{sec:electronic_structure}

In this section, we discuss the electronic structures for \ch{Pb10(PO4)6O} and \ch{Pb10(PO4)6(OH)2} in both undoped and Cu-doped phases.

We first consider the undoped phase. As reported in literature\cite{krivovichev2003crystal, bruckner1995crystal}, the trigonal-O in \ch{Pb10(PO4)6O} has $\frac{1}{4}$ occupancy, while the trigonal-(OH)$_2$ in \ch{Pb10(PO4)6(OH)2} has $\frac{1}{2}$ occupancy. Such fractional occupancy is difficult to treat in DFT. For simplicity, we fix their positions to remove the fractional occupancy, i.e., fix O at $(0,0,0.634)$, and fix O in OH$_2$ at $(0,0,0.04),(0,0,0.54)$ and H at $(0,0,-0.10),(0,0,0.40)$, s.t the total number of electrons in the unit cell is the same as the fractionally occupied structure. 
Mind that the positions of H atoms are not given in the original experimental structure\cite{bruckner1995crystal} we use and are manually added using the $O$-$H$ bond length in H$_2$O molecular, i.e., about 1\AA, which also agrees with the $O$-$H$ length reported in another experiment structure\cite{kim1997neutron}.
Notice that after fixing the trigonal-O and (OH)$_2$ positions, the original SG 176 $P6_3/m$ symmetry is lowered to SG 143 $P3$ and SG 173 $P6_3$ for two structures, respectively. 
Remark that fixing trigonal-O at any of the four $4e$ positions are equivalent, as they are related by $\{C_{6z|00\frac{1}{2}}\}$ and $\{M_z|000\}$ in SG 176. For trigonal (OH)$_2$ the scenario is similar. Thus we will focus on the aforementioned positions of trigonal-O and (OH)$_2$ in the following. 

We relax the structure without fractional occupancy and obtain the relaxed lattice constants and atomic positions summarized in Appendix. \ref{Sec:appendix_relaxed_crystal_struct} 
(Table. \ref{table:relaxed_latt_const}, \ref{table:relaxed_atom_positions}). The symmetry is maintained during the relaxation, i.e., SG 143 for \ch{Pb10(PO4)6O} and SG 173 for \ch{Pb10(PO4)6(OH)2}. 
The band structures for two relaxed structures are shown in Fig. \ref{fig:undoped_bands}, which host a large band gap of 2.761 and 2.635 eV, respectively. The highest occupied bands are quasi-flat and mainly come from the $(p_x,p_y)$ orbitals of the trigonal-O atoms.

We then consider the Cu-doped phases. As hypothesized in Ref.\cite{LEE23a, LEE23b}, the Cu-doped LK-99 Pb$_{10-x}$Cu$_x$\ch{(PO4)6O} ($0.9<x<1.1$) has Cu doping the Pb atoms at $4f$ position. By fixing $x=1$ for simplicity, there exist four possible Cu-doping structures by placing the Cu atom at one of the $4f$ positions. The four $4f$ positions are equivalent in SG 176. However, after fixing (OH)$_2$ in \ch{Pb10(PO4)6(OH)2}, the SG is lowered to SG 173 $P6_3$, which gives two inequivalent Cu doping positions, which we call the one at $(\frac{1}{3}, \frac{2}{3}, z)$ Cu$_1$-doping and $(\frac{2}{3}, \frac{1}{3}, -z)$ Cu$_2$-doping, as marked using yellow and blue circles in Fig. \ref{fig:crystal_structure}.  
For \ch{Pb10(PO4)6O}, after fixing the position of trigonal-O, the SG is lower to 143 $P3$ which makes all 4 positions inequivalent. For simplicity, we only consider the Cu$_1$-doping and Cu$_2$-doping, as the other two possible dopings have similar band structures as reported in Ref.\cite{LAI23}.

\begin{figure}[htbp]
    \centering
    \includegraphics[width=\columnwidth]{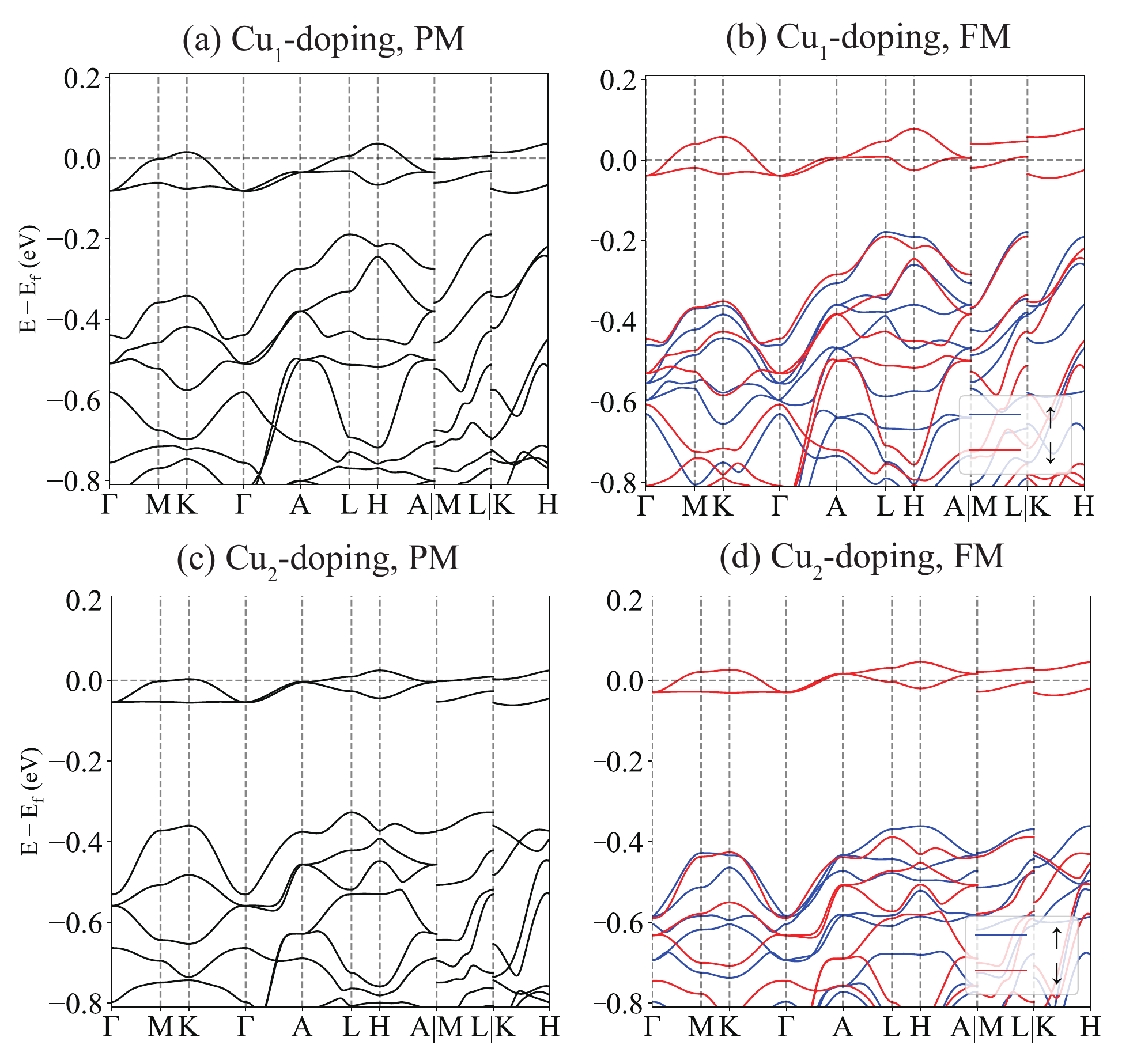}
    \caption{Electronic band structure for \ch{Pb9Cu1(PO4)6(OH)2}, where (a) is the PM phase and (b) is the FM phase of Cu$_1$-doping. (c) and (d) are similar for Cu$_2$-doping. The atomic positions of two Cu-doping can be found in Table. \ref{table:relaxed_atom_positions}.
    }
    \label{fig:2OH_doped_bands}
\end{figure}

We relax the structure of Cu$_1$- and Cu$_2$-doping for \ch{Pb10(PO4)6O} and \ch{Pb10(PO4)6(OH)2}, both having SG 143 $P3$ symmetry, with lattice constants and atomic positions summarized in Appendix. \ref{Sec:appendix_relaxed_crystal_struct} 
(Table. \ref{table:relaxed_latt_const}, \ref{table:relaxed_atom_positions}). 
The four relaxed structures are used to perform DFT calculations. We consider the paramagnetic (PM) phase and ferromagnetic (FM) for each structure, with their band structures shown in Fig. \ref{fig:1O_doped_bands}, \ref{fig:2OH_doped_bands}, and the orbital projections given in Appendix. \ref{Sec:appendix_orb_proj} 
(Fig. \ref{fig:1O_orbital_proj}, \ref{fig:2OH_orbital_proj}).

In the PM phase, for all four structures, there exist two quasi-flat bands with $\frac{3}{4}$ filling at the Fermi level $E_f$, contributed mainly by the $(d_{xz},d_{yz})$ orbitals of Cu (with Cu $(d_{xz},d_{yz})$ weight about $50\%$, $(d_{xy},d_{x^2-y^2})$ about $20\%$, and $p$ orbital of O atoms that close to Cu weight about $30\%$), forming elementary band representation (EBR) $^1E^2E@1b$ in SG P3 (notice in the PM phase the time-reversal symmetry exists and enforces $^1E$ and $^2E$ to be degenerate) for Cu$_1$ doping and $^1E^2E@1c$ for Cu$_2$ doping. 
For two Cu dopings of \ch{Pb10(PO4)6O}, there are two bands close the to Cu bands with a small band gap, mainly formed by the bands of $(p_x,p_y)$ of the trigonal-O and form EBR $^1E^2E@1a$, as shown in Fig. \ref{fig:1O_doped_bands}, \ref{fig:1O_orbital_proj}. Although here these four bands formed by $(d_{xz},d_{yz})$ of Cu and $(p_x,p_y)$ of the trigonal-O do not have band inversions and are topologically trivial, we can adjust the hoppings and make them topological, which we show in Sec. \ref{sec:TBmodels} using minimal tight-binding models. 
For two Cu dopings of \ch{Pb10(PO4)6(OH)2}, however, the bands below the Cu bands have a larger band gap and weak hybridization with Cu bands.

The FM phases of the four structures all have a lower total energy compared with the PM phase, as shown in Table. \ref{table:total_energy_compare}, suggesting that the FM phase is energetically more favored than the PM phase. Evidence for soft ferromagnetism, consistent with the well-localized bands and strong Coulomb repulsion we find, has recently appeared in experiments \cite{GUO23}. In the spin-polarized orbital projections shown in Fig. \ref{fig:1O_orbital_proj}, \ref{fig:2OH_orbital_proj}, the quasi-flat bands of Cu near $E_f$ have a large spin-splitting of about $0.6$ eV, while the bands of trigonal-O have negligible spin-splitting. The calculated magnetic moment is about 0.6 $\mu_B$ for Cu and small moments for O atoms surrounding Cu which sum to about 0.3 $\mu_B$ in total, in agreement with the fact that the two flat bands near $E_f$ in the PM phase also have about $30\%$ weight from O. 
We also calculate the total energy of the A-type antiferromagnetic phase (AFM), i.e., FM inplane and AFM out of the plane, as shown in Table. \ref{table:total_energy_compare}. The FM and AFM phases have almost the same energy per unit cell, both being lower than the PM phase. We leave for future studies to investigate the magnetic ground state.

\begin{table}[htbp]
\begin{tabular}{c|c|c|c|c}
\hline\hline
Compound &  Phase      & PM     & FM  & AFM   \\ \hline
\multirow{2}{*}{\ch{Pb10(PO4)6O}} & Cu$_1$-doping  & 0      & -0.129  & -0.130 \\ \cline{2-5}
& Cu$_2$-doping  & 0.195  & 0.089  & 0.092 \\ \hline
\multirow{2}{*}{\ch{Pb10(PO4)6(OH)2}} & Cu$_1$-doping  & 0  & -0.136 & -0.136 \\ \cline{2-5}
& Cu$_2$-doping  & -0.177 & -0.332 & -0.332 \\ \hline\hline
\end{tabular}
\caption{Comparison between total energy of PM, FM, and A-type AFM phases per unit cell (in eV) calculated in DFT. The total energies in the Cu$_1$-doping phase in \ch{Pb10(PO4)6O} and \ch{Pb10(PO4)6(OH)2} are used as zero. It can be seen that in all four cases, the FM and AFM phases have very close total energy, which are both lower than the PM phase. In \ch{Pb10(PO4)6O} the Cu$_1$ doping has lower energy while \ch{Pb10(PO4)6)(OH)2} the Cu$_2$ phase is lower.}
\label{table:total_energy_compare}
\end{table}

\subsection{Phonon Spectrum \label{ssec:pho}}

\begin{figure*}[ht]
\centering
\subfloat[Pb-(OH)$_2$ at low-T.]{
\label{fig:2OHLowT}
\includegraphics[width=0.32\textwidth]{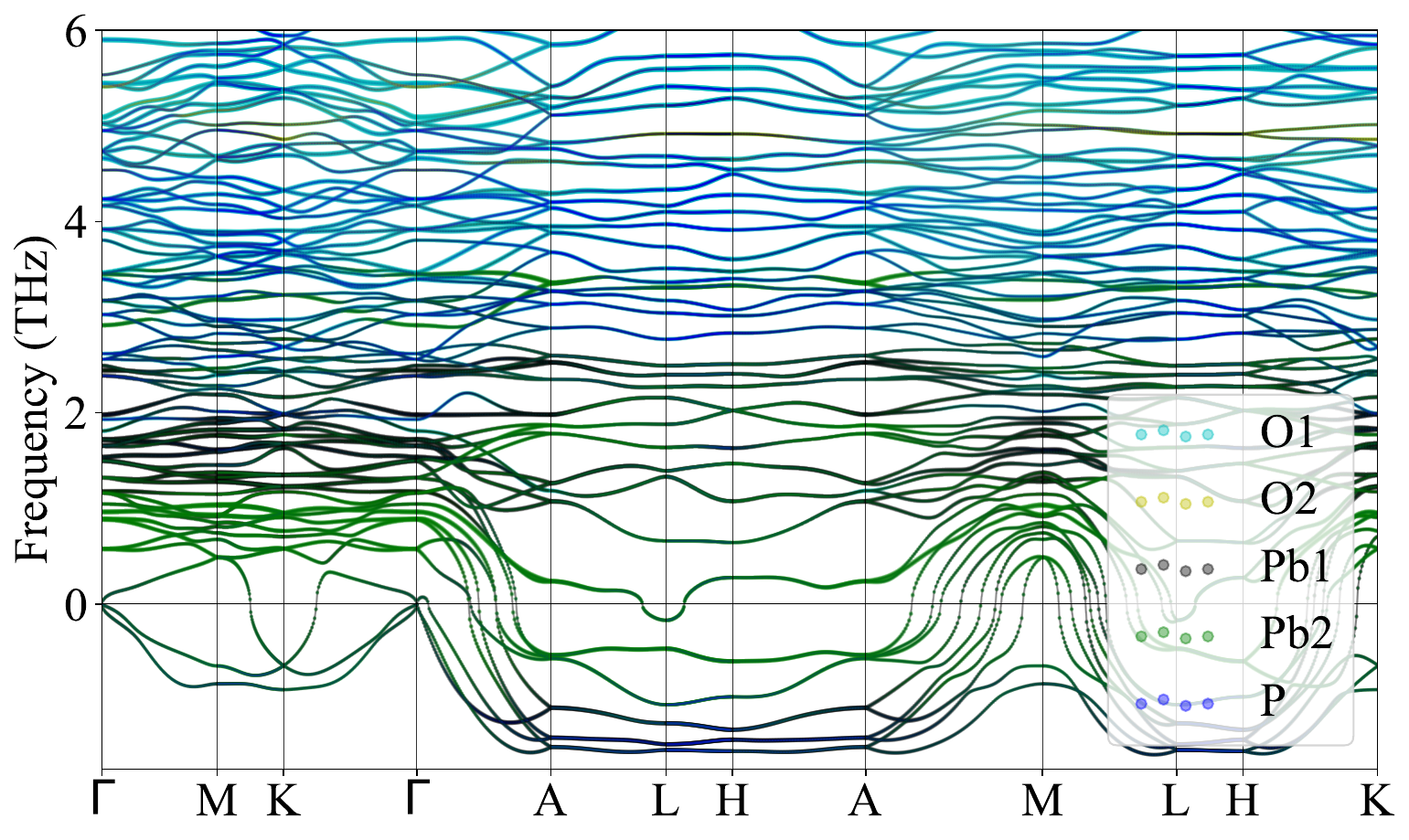}
}
\subfloat[Pb-(OH)$_2$ at high-T.]{
\label{fig:2OHhighT}
\includegraphics[width=0.32\textwidth]{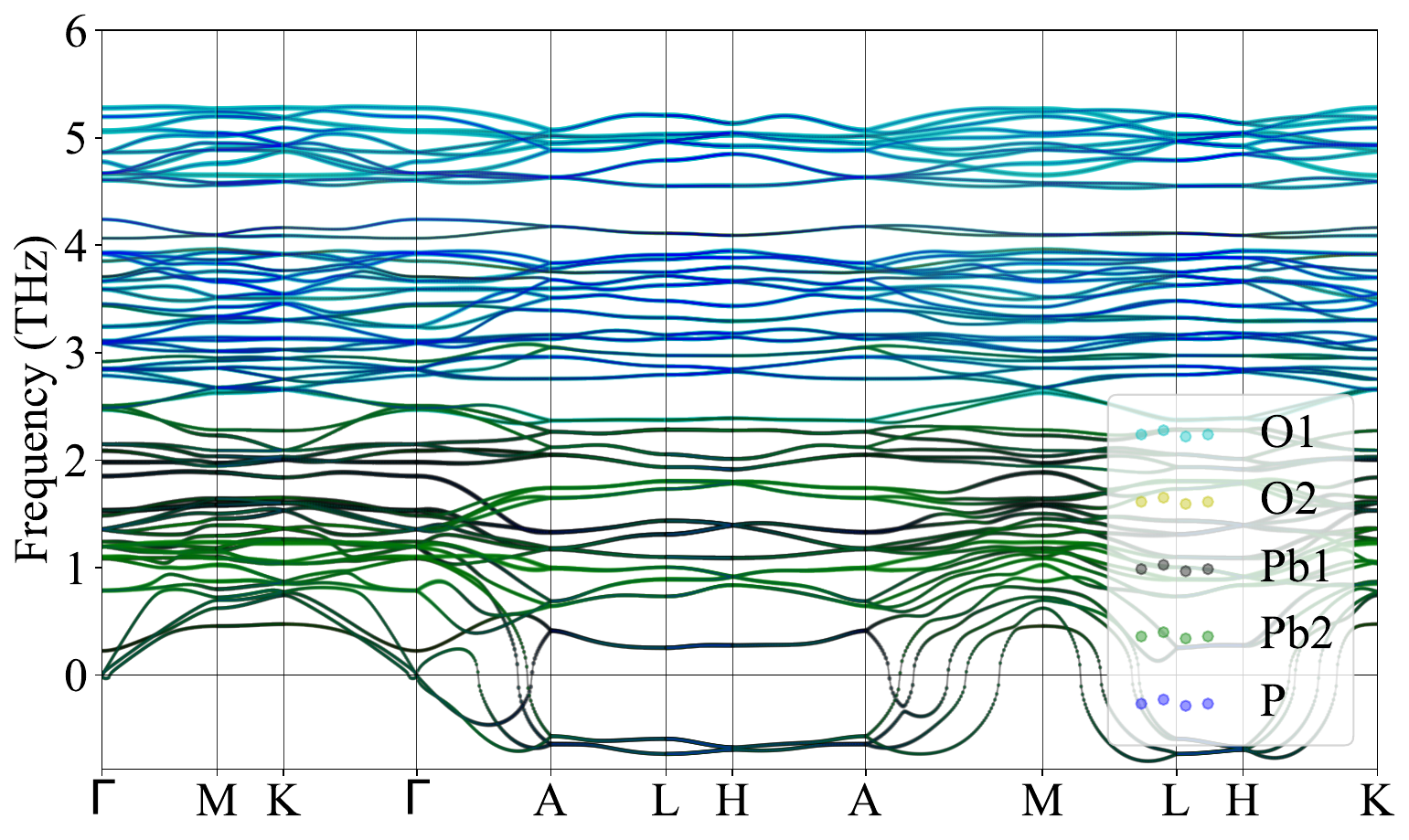}
}
\subfloat[Pb-(OH)$_2$ at high-T in 112 supercell.]{
\label{fig:2OHhighT112main}
\includegraphics[width=0.32\textwidth]{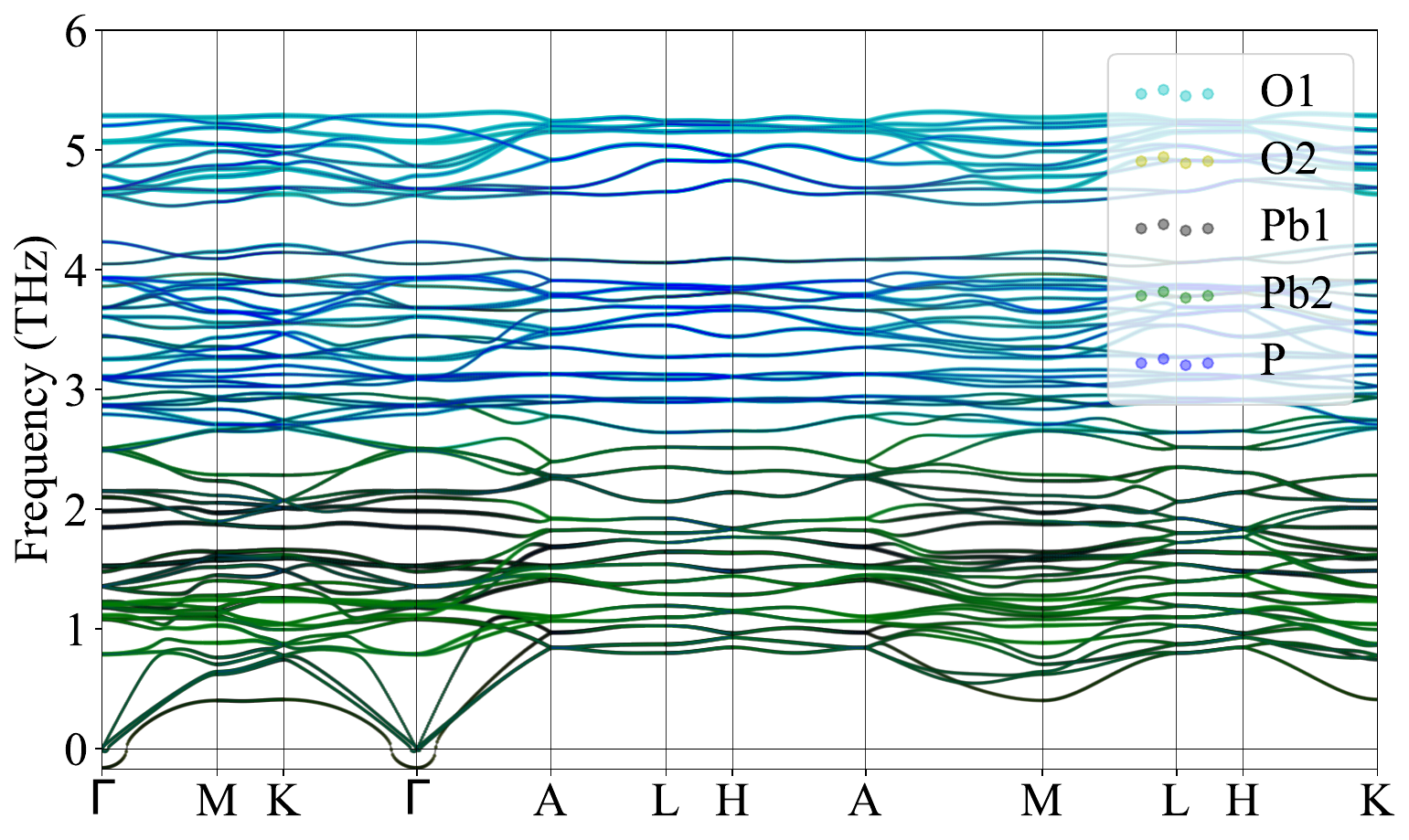}
}
\caption{Phonon spectrum for relaxed \ch{Pb10(PO4)6(OH)2} (without Cu doping) structure at (a) low and (b) high temperature calculated using $1\times 1 \times 1$ supercell, which shows imaginary, unstable phonons in the $k_3 = \pi$ plane. (c) Phonon spectrum calculated using a $1\times 1 \times 2$ supercell at high temperature, which stabilizes the $k_3 = \pi$ modes but softens a single mode at the $\Gamma$ point. In addition, we observe that the soft phonon modes shown in panels (a) and (b) are flat modes in the $k_3=\pi$ plane. A series of flat phonon modes are also presented at the finite frequency in the high-$T$ calculations, such as flat modes around $~$1 THz in panels (c). }
\label{fig:pho_undoped}
\end{figure*}

In this section, we perform the phonon calculations to check the stability of proposed structures in the literature, including \ch{Pb10(PO4)6O}, \ch{Pb10(PO4)6(OH)2}, \ch{Pb9Cu1(PO4)6O} and \ch{Pb9Cu1(PO4)6(OH)2}. We find that, in their nominal 111 unit cells, \emph{all} structures display imaginary phonon frequencies. We focus our discussion below on \ch{Pb9Cu1(PO4)6(OH)2} since powder X-ray diffraction shows a strong similarity between it and LK-99. Supplemental figures for other structures may be found in \App{app:phononplots}. 

For the undoped structures, both the Pb-O and Pb-(OH)$_2$ present negative/imaginary phonon at low-T and high-T with $1\times 1\times 1$ cell, where the phonon of Pb-(OH)$_2$ is shown in Fig.~\ref{fig:pho_undoped}(a,b). Here, the temperature effect is included via electronic smearing in the harmonic approximation level. As the atomic projection presents, the phonon instability is mainly contributed - as theoretically expected - by the heaviest Pb atoms at both Pb1 and Pb2 sites. As the temperature goes higher, the imaginary phonons harden, especially in the $k_z=0$ plane where imaginary modes disappear at high-T. Since the calculation is performed in $1\times 1 \times 1$ cell with a short $c$ compared to $a$, the negative phonon in $k_z=\pi$ plane may be caused by the short cutoff in the $c$ direction. To verify this scenario, we also perform phonon calculation in $1\times 1\times 2$ supercell at high-T. As shown in Fig.~\ref{fig:pho_undoped}(c), the negative branches on the $k_z=\pi$ plane in the $1\times 1 \times 1$ cell calculation become positive, as the cell is enlarged to include the force constants between atoms with longer distance. We expect that the residual imaginary phonon at the $\Gamma$ point can be eliminated by performing phonon calculations in a larger supercell after relaxation. 

We also observe that the phonon spectrum presents a good separation of frequency based on the mass of elements: the heaviest Pb dominates the lowest frequency and gives the imaginary modes, H phonons lie at a much higher frequency, which is not shown in the plot, and P and O phonons lie in the middle. 

For the Cu-doped structure, previous DFT calculations suggest a (ferro/antiferro)magnetic ground state. Therefore, the phonon calculations are performed in both paramagnetic and ferromagnetic phases as presented in \App{app:phononplots} with a $1\times 1\times 1$ cell. Similar to the undoped case, the doped structures show negative phonon modes, but tend to harden in the FM phase. 
Both O and the Cu dopant contribute to the imaginary phonon at low $T$ (see \App{app:phononplots} ) and one should relax the structure with much care for the doping effect to obtain a stable phonon spectrum. The difficulty in obtaining a stable structure even for the undoped parent compounds, which exist in nature, underscores an important challenge for first principles studies of the doped compound, whatever its nature.

\section{Tight-Binding Models}
\label{sec:TBmodels}

In this section, we construct short-range tight-binding models for both \ch{Pb9Cu1(PO4)6O} and \ch{Pb9Cu1(PO4)6(OH)2} compounds and for both considered positions of the Cu dopant. Such models are a prerequisite for studying the many-body phases that LK-99 is conjectured to realize. We emphasize that the precise chemical composition, purity, and structure of the supposed compound are far from being settled, and initial proposals may need to be re-examined. Nevertheless, we study the \emph{proposed} structures here for completeness. We find, in all cases and in agreement with independent calculations, that Cu forms a high-density flat at the Fermi level.  Based on this finding we ask whether a flat band superconductivity scenario is viable. We perform the calculations of the quantum geometry in these bands and find that they are atomically localized and will likely not favor superconductivity.

In all cases, we find that the two bands straddling the Fermi level are atomic and dominated by the Cu $d$-orbitals. In the \ch{Pb9Cu1(PO4)6O} structure, a four-band model is required due to the close proximity of the O bands, which hybridize with the Cu orbitals particularly in the $k_z = \pi$ plane. For this structure, the dispersion of the Cu bands is predominantly due to hybridization with O. In contrast, \ch{Pb9Cu1(PO4)6(OH)2} shows well-isolated Cu bands at the Fermi level, and a two-band model can be constructed. \App{app:SPham} contains a complete discussion of the parameters, symmetry, and quantum geometry of the bands. 

A low-energy model of the flat bands can be constructed from the $d_{yz}, d_{xz}$ orbitals of Cu on the 1b/1c positions for Cu$_1$/Cu$_2$ doping, and the $p_x, p_y$ orbitals of O on the 1a position in space group $P3$ (see \App{app:SPham} for conventions of the lattice). Our DFT calculations show degeneracies at $\Gamma$ and $A$ (which form double Weyl points \cite{HIR23}) that cannot be protected by $C_3$ alone since it forms an abelian group. We check that the separate spin sectors possess a spin-less time-reversal symmetry $\mathcal{T}^2 = +1$ which protects a 2D complex irrep ${}^1E{}^2E$ arising from the $d$ and $p$ orbitals. The preservation of the spin-less $\mathcal{T}$ in the FM phase comes from the fact that the magnetization is taken into account as the local momentum term $M(\bsl{r}) s_z$ (with real scalar $M(\bsl{r})$). The spin-orbital coupling is negligible for the Cu and O atoms.

\subsection{ Pb$_{9}$Cu(PO$_4$)$_6$O }
\begin{figure*}
\centering
\includegraphics[width=8.5cm]{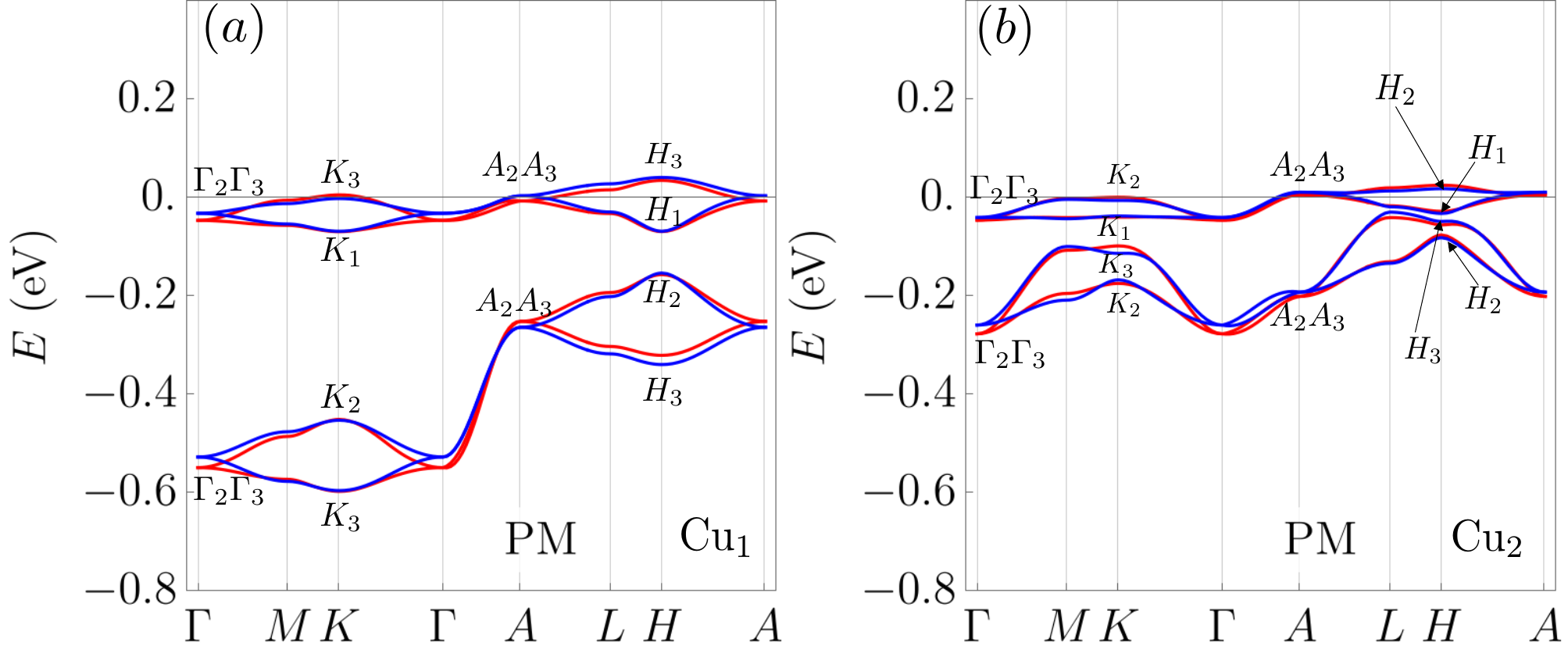} \includegraphics[width=8.5cm]{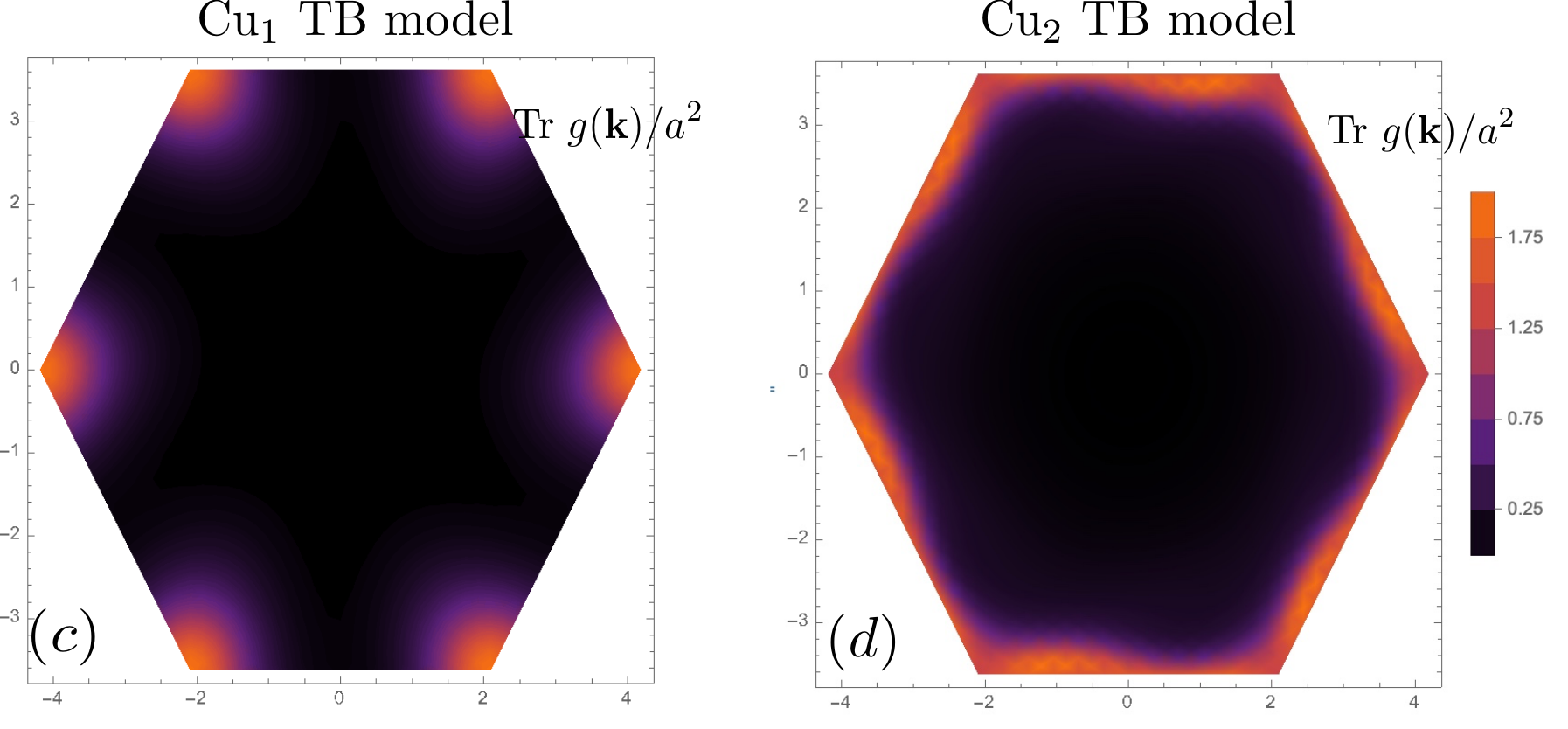} 
\caption{Comparison of DFT and tight-binding model band structures for Pb$_{9}$Cu(PO$_4$)$_6$O. The relaxed DFT (red) and short-range tight-binding model (blue) are shown for Cu$_1$ doping $(a)$ and for Cu$_2$ doping $(b)$. The 2D quantum metric $g(\mbf{k})$ is computed on the plane $k_3 = \pi$ for the Cu$_1$ (c) and Cu$_2$ (d) models, showing peaked features where the Cu and O bands have the smallest direct gap.
} 
\label{fig:bandswan}
\end{figure*}

In the Pb$_{9}$Cu(PO$_4$)$_6$O structure, it is necessary to construct a four-band model due to the O-dominated bands that appear closely below (and hybridize with) the Cu-dominated bands at the Fermi level. The model takes the form
\bea
\label{eq:h4band}
h_4(\mbf{k}) = \bpm h_{C}(\mbf{k}) & h_{CO}(\mbf{k}) \\ h^\dag_{CO}(\mbf{k}) & h_{O}(\mbf{k})\epm
\eea
consistent with these symmetries. For Cu$_1$ doping, it is sufficient to include the three nearest-neighbor Cu-O hoppings ($h_{CO}(\mbf{k})$), six O-O in-plane hoppings for O and Cu, and the two O-O vertical hoppings ($h_{O}(\mbf{k})$). In fact, the Cu hoppings are less than 4meV and can be safely dropped, effectively setting $h_C(\mbf{k}) = 0$. The Cu$_2$ doping structure exhibits a smaller gap between the O and Cu bands, and requires the inclusion of next nearest neighbor hoppings, with full expressions given in \App{app:SPham}. We find excellent agreement with the DFT spectrum and irreps in \Fig{fig:bandswan} within this approximation. 

Our tight-binding model shows that the dispersion of the Cu bands at the Fermi level arises essentially in its entirety from hybridization with O. Nevertheless, symmetry eigenvalues indicate that the bands are topologically trivial: O hybridization does not cause a topological change. To confirm this, we compute the 2D Fubini-Study quantum metric (\Fig{fig:bandswan}(c,d))
\bea
g(\mbf{k}) &= \frac{1}{2} \sum_{i=x,y} \Tr \del_i P(\mbf{k})\del_i P(\mbf{k})
\eea
where $P(\mbf{k}) = U(\mbf{k}) U^\dag(\mbf{k})$ is the projector onto the eigenvector matrix $U(\mbf{k})$ of the two Cu bands, and non-abelian Wilson (\Fig{fig:wilsonloops}) loop 
\bea
W(k_2,k_3) &= U^\dag(2\pi,k_2,k_3) \prod_{k_1}^{2\pi \leftarrow 0} P(\mbf{k}) \, U(0,k_2,k_3)
\eea
computed over the set of two bands at the Fermi level. Both show strongly localized states, consistent with their elementary band representation. 
Furthermore, the basis Wanniers functions are localized: square root of Wannier spread is about $0.20 a\approx 0.26 c$ for Cu and about $0.33 a\approx 0.43 c$ for O).
As we can see, the Wannier function is more localized for Cu than that for O, expalining the fact why the hopping among Cu is smaller than that among O.

\subsection{ Pb$_{9}$Cu(PO$_4$)$_6$(OH)$_2$ }

\begin{figure*}
\centering
 \includegraphics[width=4.25cm]{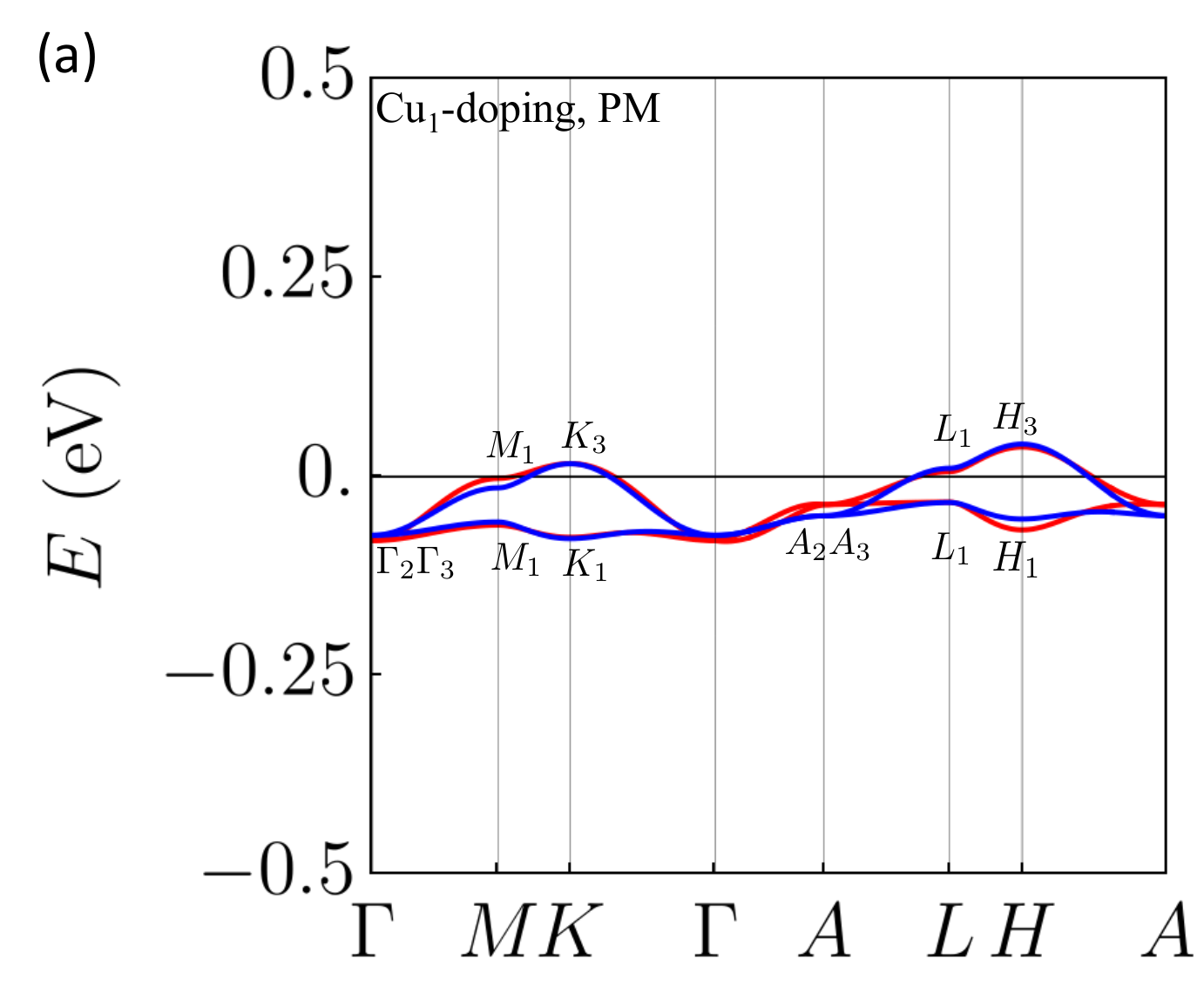}
\includegraphics[width=4.25cm]{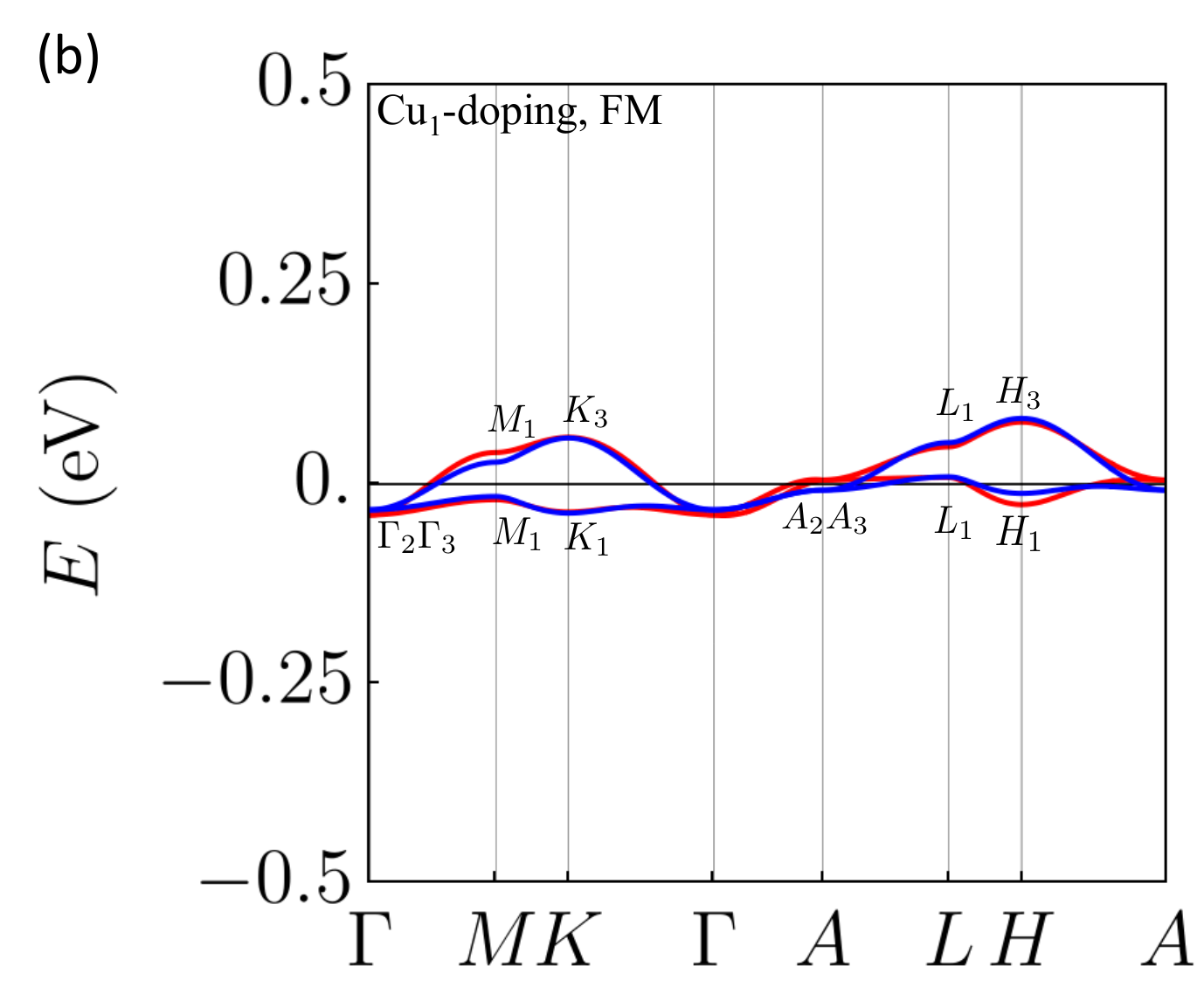}
 \includegraphics[width=4.25cm]{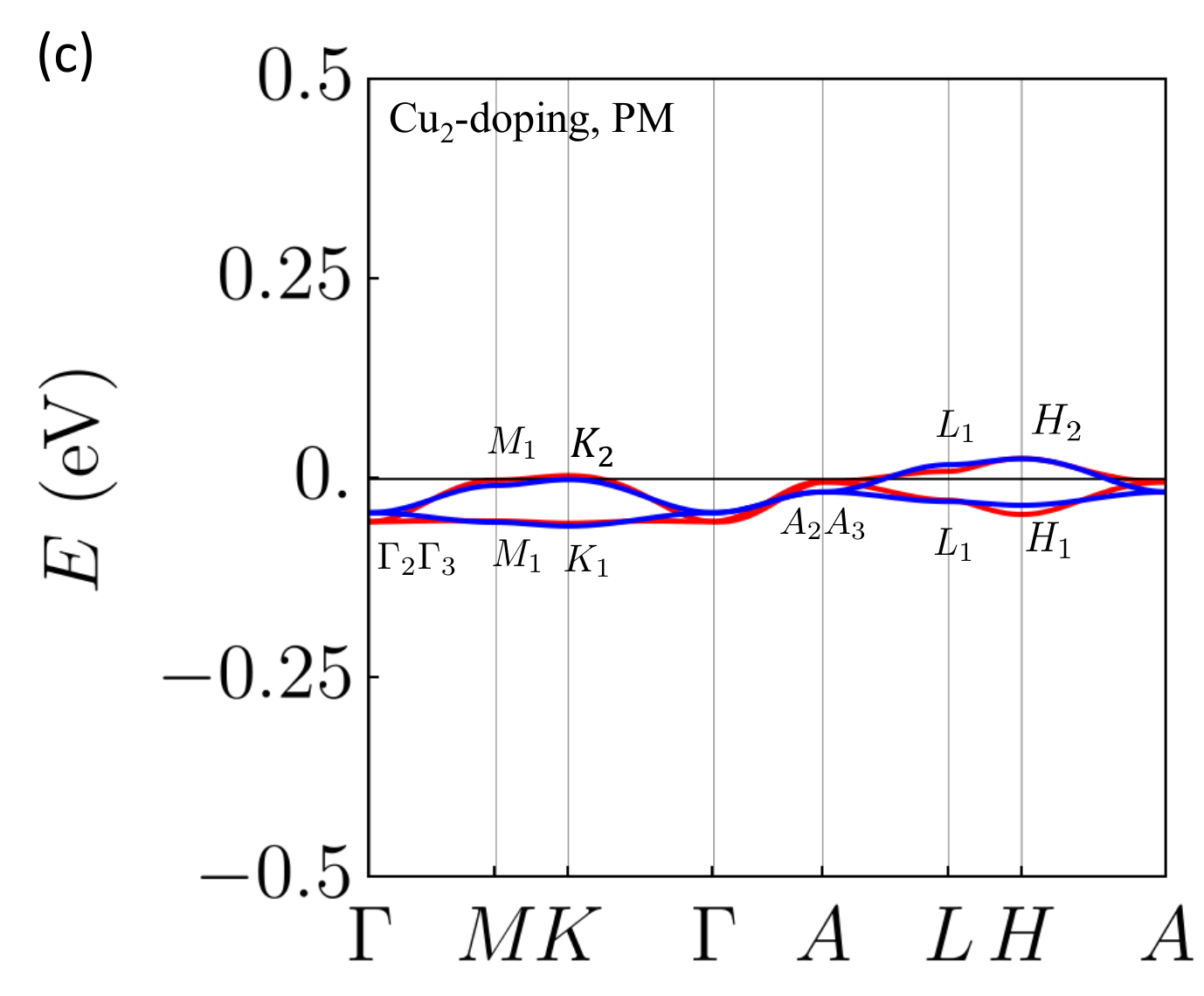}
\includegraphics[width=4.25cm]{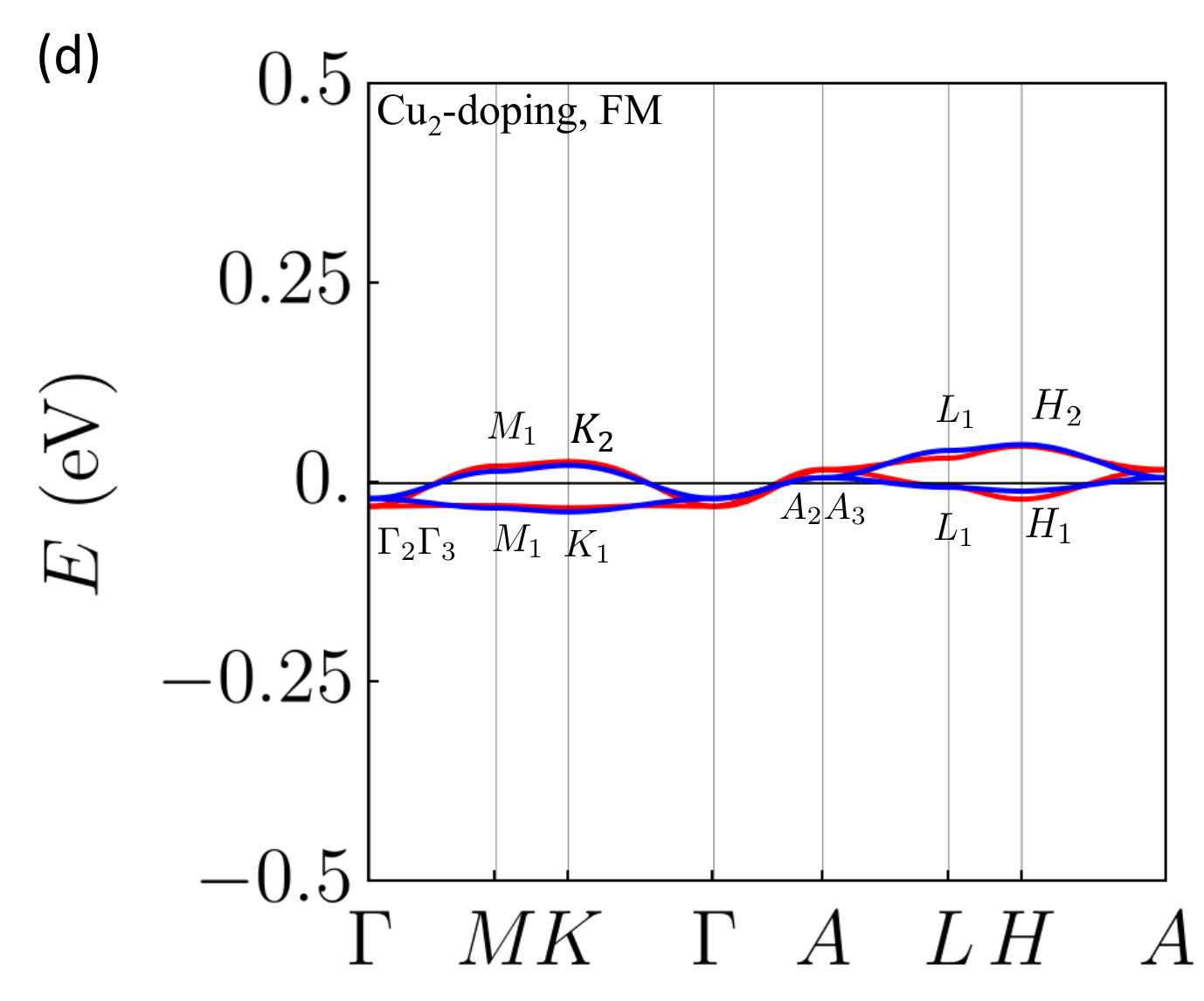}
\caption{
Comparison of the relaxed DFT (red) and 2-band tight-binding model band structures (blue) and irreps for Pb$_{9}$Cu(PO$_{4}$)$_6$(OH)$_2$ in the specified Cu doping and magnetic properties.
The relaxed DFT and the NN-hopping tight-binding model (blue) are in close agreement.
The expression of the model and the values of the model parameters are specified in \appref{app:2bandmodel}. Note that this is a 2-band model, so the quantum geometric tensor of both bands, taken together, vanishes. 
} 
\label{fig:bandswan2band}
\end{figure*}

We construct a 2-band model $h_2(\mbf{k})$ for the two bands near the Fermi level in Pb$_{9}$Cu(PO$_4$)$_6$(OH)$_2$ for both Cu$_1$ doping and Cu$_2$ doping and in both PM and FM phases.
The 2-band model is constructed with $d_{xz}$ and $d_{yz}$ on Cu (at 1b for Cu$_1$ doping and at 1c for Cu$_2$ doping), and it only contains 
NN hopping along all three directions in addition to the onsite energy term.
The form of the model is the same for both doping and for both PM and FM phases, since they all preserve the spinless TR and $C_3$ symmetries for the 2 bands near the Fermi level according to the Wannierization of DFT data.
With parameter values determined from the Wannierization of the DFT data, we can see that the model gives very similar bands as the DFT band structure (\figref{fig:bandswan2band}).
(Detailed expression and parameter values of the 2-band model can be found in \appref{app:2bandmodel}.)

In particular, for one specific Cu doping, we choose the same parameter values for the NN hoppings for PM and FM phases, since the DFT values in the two phases are very close (difference smaller than $0.1meV$); the only non-negligible difference between PM and FM is just an onsite energy shift, indicating that the effect of FM on the states near Fermi level is just uniformly shifting their energy.
The small bandwidth of the bands near the Fermi level comes from the small hopping magnitudes (maximum hopping about 12meV), which comes from the small spread of the Wannier function of the basis (the square root of the Wannier spread of each orbital is about  $0.24 a\approx 0.32 c$). (See \appref{app:2bandmodel} for details.)
We can see the Wannier spread of the 2-band model is larger than that of the Cu in the 4-band model, which is consistent with the larger hopping here compared to the hopping among Cu in the 4-band model.

In the DFT bands structure, the symmetry-protected gapless points at $\Gamma$ and $A$ are double Weyl points with chirality $\pm 2$. (See \appref{app:2bandmodel} for details.)
However, the band splitting along $\Gamma$-A is very small (maximum splitting about $2$meV), and thus in our simplified NN-hopping model, we neglect the band splitting along $\Gamma$-A for simplicity.
Although such simplification in our model will make the two double Weyl points merge into an accidental nodal line along $\Gamma$-A, it will be convenient for future study on the correlated physics, since it makes the eigenvectors of the Hamiltonian independent of $k_z$.

Both \refcite{OH23} and \refcite{HIR23} contain 2-band PM models constructed from $d_{xz}$ and $d_{yz}$ on Cu.
However, the 2-band PM model in \refcite{OH23} has mirror symmetry along $y$, which we do not include in our model since the DFT calculation indicates a considerable breaking of the mirror symmetry. (See \appref{app:2bandmodel} for details.)
The 2-band PM models in both \refcite{OH23} and \refcite{HIR23} have considerably different parameter values than ours.
\refcite{HIR23} shows double Weyl points at $\Gamma$ and A in their 2 band model, for which they choose to include the small band splitting along $\Gamma$-A in their model.

\section{ Interacting Hamiltonian}

We use the constraint random phase approximation (cRPA) method\cite{aryasetiawan2004frequency, solovyev2005screening, aryasetiawan2006calculations, miyake2009ab} to compute the screened Coulomb interaction for the $(d_{xz}, d_{yz})$ orbitals of Cu near $E_f$ for two Cu-doped structure of \ch{Pb9Cu1(PO4)6O} and \ch{Pb9Cu1(PO4)6(OH)2}. In Table. \ref{table:CRPA_U}, we list the values of Hubbard-Kanamori parameters, i.e., the onsite intra-orbital Hubbard $\mathcal{U}$, inter-orbital $\mathcal{U}^\prime$, and onsite exchange $\mathcal{J}$. The interacting Hamiltonian can be constructed as 
\begin{equation}
\begin{aligned}
    \hat{H}_{\text{int}} &=
    \mathcal{U} \sum_{im} n_{im\uparrow}n_{im\downarrow}
    + \mathcal{U}^\prime \sum_{i,m\ne m^\prime} 
    n_{i m\uparrow}n_{i m^\prime\downarrow}\\
    &+  (\mathcal{U}^\prime - \mathcal{J})\sum_{i,m<m^\prime, \sigma} 
    n_{i m \sigma} n_{i m^\prime \sigma}	\\
    & +\mathcal{J} 
    \sum_{i, m\ne m^\prime}
    (-c_{im\uparrow}^\dagger c_{i m \downarrow}  c_{i m^\prime \downarrow}^\dagger c_{i m^\prime \uparrow} 
    + c_{im\uparrow}^\dagger c_{im\downarrow}^\dagger c_{i m^\prime \downarrow} c_{i m^\prime \uparrow}),
\end{aligned}
\end{equation}
where $i$ is the site index and $m$ the orbital index. We remark that the \textit{ab-initio} Hubbard-Kanamori parameters may need to be renormalized as the two quasi-flat bands near $E_f$ also have weights of other orbitals.

\begin{table}[htbp]
\begin{tabular}{c|c|c|c}
\hline
Phase            & $\mathcal{U}$ & $\mathcal{U}^\prime$ & $\mathcal{J}$ \\ \hline
\ch{Pb9Cu1(PO4)6O} Cu$_1$        & 2.75          & 1.71                 & 0.52          \\ \hline
\ch{Pb9Cu1(PO4)6O} Cu$_2$        & 3.53          & 2.38                 & 0.58          \\ \hline
\ch{Pb9Cu1(PO4)6(OH)2} Cu$_1$ & 2.88          & 1.99                 & 0.45          \\ \hline
\ch{Pb9Cu1(PO4)6(OH)2} Cu$_2$ & 1.85          & 0.96                 & 0.14          \\ \hline
\end{tabular}
\caption{The \textit{ab-initio} Hubbard-Kanamori parameters. In the table, $\mathcal{U}$, $\mathcal{U}^\prime$, $\mathcal{J}$ denotes the onsite intra-orbital Hubbard, inter-orbital Hubbard, and onsite exchange interaction. All numbers are in eV.}
\label{table:CRPA_U}
\end{table}

\subsection{Comments on Flat Bands and Interactions}

Flat bands are thought to be beneficial for strongly correlated phases since the interaction strength necessarily dominates over the single-particle bandwidth. However, the quantum geometry of the flat bands is another essential ingredient in determining the nature of the resulting strongly correlated phases. 

For instance in the single-band square lattice Hubbard model at half filling, the flat band limit $t \to 0$ yields a fully decoupled lattice (the atomic limit) which is a perfect paramagnet. It is $O(t^2/U)$ corrections that stabilize an anti-ferromagnetic phase. If, however, the interaction strength is much larger than a set of isolated bands but smaller than the gap between these bands and their complement, a different set of phases can emerge. In the repulsive case, ferromagnetism can be proven \cite{PhysRevLett.62.1201,1993CMaPh.158..341M}, and superconductivity (or phase separation) for attractive case \cite{PhysRevLett.62.1201,2022arXiv220900007H,2016PhRvB..94x5149T}. Features of the many-body phase, for instance the mass of the Cooper pair, stiffness of the spin wave, and a lower bound on the electron-phonon coupling are determined by quantum geometry \cite{2022arXiv220900007H,2016PhRvB..94x5149T,2023arXiv230502340Y}.

To entertain the possibility of superconductivity in flat bands, we recall that the mean-field critical temperature in flat bands will be proportional to the interaction strength \cite{2015NatCo...6.8944P}. Although this scenario is exponentially improved over one-band BCS theory, it still requires an attractive interaction of roughly 100 meV. This order of magnitude is larger than the bandwidth of the two-band model in the OH structure, and roughly equal to the bandwidth of the upper Cu bands in the four-band model for the O structure. (We note in this case that the band gap is small, and projecting the interaction to the flat bands may not be justified.) The repulsive Coulomb interaction  we computed is eV scale, and is much larger than the total bandwidth of both tight-binding models. 

Two recent papers \cite{HIR23,TAV23} have proposed models where one of the Cu bands is perfectly flat. Although the two Cu bands together form an indecomposable elementary band representation with trivial topology, analyzing only one of the two bands  (per spin) away from their degeneracy points at $\Gamma$ and $A$ can show strong quantum geometry. It is only appropriate to consider strong coupling groundstate a single flat (gapless) band in the limit where the interaction strength is much smaller than the bandwidth. This limit, while interesting and worthy of study, places an upper bound on the interaction strength which limits its applicability to a high-temperature phase.

\section{Further Verification and Tests}

\subsection{Immediate Experiments}
As we experimentally found that samples are multi-phase, short-term experiments should focus on isolating the different phases and characterizing their compositions and structures. Also several more synthesis should be performed to confirm that the outcomes are repeatable. Once we know reliably the outcome of the synthesis the sample needs to be thoroughly characterized with a wide range of methods. SEM/EDX will be a good first attempt to analyze how many different phases are in the materials and what their respective compositions are. It would be ideal if those phases can be separated either manually of in different synthesis attempts targeting the respective compositions found via SEM/EDX. As we showed it is possible to pick single crystals from at least one phase.

\subsection{Chemical Structure Verification}

Once phase-pure materials are obtained, they can be thoroughly characterized. If single crystals are obtained, SXRD is ideal in combination with chemical analysis such as SEM/EDX or, if enough samples can be separated, ICP-OES. Otherwise PXRD in combination with Rietveld refinement will be necessary. If the phases are not crystalline, the structural characterization becomes more complicated and chemical analysis is a more important first step, likely to be followed by high-resolution electron microscopy. 

Characterization of physical properties is most meaningful on single-phase materials and can be performed after through characterization of all components of the samples. Similarly, theoretical analysis of the electronic structures is most meaningful if the final crystal structures are known.

\subsection{Theoretical Analysis}

Once the chemical structure is firmly established and the set of bands at the Fermi level is settled, the bands must be fully analyzed based on the principles outlined in this paper, including orbital, quantum geometric, topological, and localization content in all the multiple phases that result from the reaction. The interacting Hamiltonian must then be derived and with it the values of the Hubbard $U$. Then the ground state of the system must be determined. Even if not superconducting, one must still explain the levitation properties shown in initial experiments - either large diamagnetism or some flavor of ferromagnetism. The phonons, and more importantly the electron-phonon interaction need to be obtained; an understanding is needed of the rather flat Pb phonon bands on the $k_z=\pi$ plane that seem to become negative at low temperature.  If indeed, however, the sample does turn out to be superconducting, short of being a fundamental discovery, it would also point out the limitation of our theoretical understanding of the mechanisms that create it.

\section{Conclusions}

Assuming the reported structure and the location of the Cu dopants, we have performed first principle calculations of the electronic structure of \ch{Pb9Cu1(PO4)6O} and \ch{Pb9Cu1(PO4)6(OH)2}. We find that the bands are almost flat, localized on the Cu atoms, and have weak quantum geometry and trivial topology. Hence in a  
"flat-band superconductivity" scenario, already alluded to previously,  our calculation of the quantum geometric properties of the active bands shows that such a phase would have extremely small superfluid stiffness even at zero temperature. Instead, their flatness and the strong Coulomb repulsion is compatible with the (anti/)ferromagnetic ground states found in calculations. The phonon spectra seem to contain imaginary phonons for both high and low-temperature paramagnetic phases (with the reported unrelaxed structure). The experimental findings also do not, as of yet, strongly suggest the presence of superconductivity. Hence either the reported structure is incorrect, or the ab-initio calculations and the structure are correct but then furthemore give rise to a non-superconducting ground-state, or the ab-initio calculations are incorrect due to large correlations or other factors. In fact, our preliminary experiments and structural solutions point to LK-99 being a multi-phase material, where the part that structurally agrees with a doped version of Pb apatite is transparent and thus probably not superconducting. Either way, this points to the difficulty of predicting and/or explaining superconductors by ab initio methods, even when they exist. It also suggests that extra cases be taken in the literature, both experimental and theoretical. Experimentally, samples need to be much more carefully analyzed with a wide range of diffraction and spectroscopic methods. The individual phases should be isolated and their properties should be studied separately. Theoretically, one must at the very minimum check the correct symmetry of the states, their localization and topology, and try to obtain superconductivity from the first principle Hamiltonian, rather than introducing it by hand in a BdG formalism. These represent hard challenges that only serious investigations can overcome. 

\section{Acknowledgements}
H.H. and Y.J. were supported by the European Research Council (ERC) under the European Union’s Horizon 2020 research and innovation program (Grant Agreement No. 101020833). D.C. acknowledges the hospitality of the Donostia International Physics Center, at which this work was carried out. J.H-A. is supported by a Hertz Fellowship. D.C. and B.A.B. were supported by the Simons Investigator Grant No. 404513, the Gordon and Betty Moore Foundation through Grant No. GBMF8685 towards the Princeton theory program, the Gordon and Betty Moore Foundation’s EPiQS Initiative (Grant No. GBMF11070), Office of Naval Research (ONR Grant No. N00014-20-1-2303),  BSF Israel US foundation No. 2018226, NSF-MERSEC (Grant No. MERSEC DMR 2011750). J.Y. is supported by the Gordon and Betty Moore Foundation through Grant No. GBMF8685 towards the Princeton theory program. B.A.B. and C.F. are also part of the SuperC collaboration. Y.J. and S.B-C. acknowledge financial support from the MINECO of Spain through the project PID2021- 122609NB-C21 and by MCIN and by the European Union Next Generation EU/PRTR-C17.I1, as well as by IKUR Strategy under the collaboration agreement between Ikerbasque Foundation and DIPC on behalf of the Department of Education of the Basque Government. BAB also acknowledges support  the European Research Council (ERC) under the European Union’s Horizon 2020 research and innovation program (grant agreement no. 101020833). J.H. and D.C. are supported by DOE Grant No. DE-SC0016239. J.H is also supported by a Hertz Fellowship. C.H.H, P.S.B, and E.L.G are supported by DMR-1956403, and the Research Corporation for Science Advancement (Cottrell Scholar Program) for non-tenured faculty. We are grateful to T. T. Debela, A. Walsh, D. Scanlon, A. Rosen, and C. Musgrave for helpful discussions related to defect formation enthalpy calculations. The authors acknowledge the use of Princeton’s Imaging and Analysis Center, which is partially supported by the Princeton Center for Complex Materials, a National Science Foundation (NSF) – MRSEC program (DMR-2011750). The authors would also like to acknowledge G. Cheng for helping with carbon coating. LMS is supported by the Gordon and Betty Moore Foundation’s EPIQS initiative through Grant No. GBMF9064, as well as the David and Lucille Packard foundation. SBL is supported by the National Science Foundation Graduate Research Fellowship Program under Grant No. (DGE-2039656). Any opinions, findings, and conclusions or recommendations expressed in this material are those of the author(s) and do not necessarily reflect the views of the National Science Foundation.
\onecolumngrid
\tableofcontents

\appendix

\section{Additional DFT results}

\subsection{Crystal structure}\label{Sec:appendix_relaxed_crystal_struct}
In Table. \ref{table:relaxed_latt_const}, we list the experimental and DFT relaxed lattice constants in undoped and Cu-doped phases of \ch{Pb10(PO4)6O} and \ch{Pb10(PO4)6(OH)2}.
In Table. \ref{table:relaxed_atom_positions}, we give the relaxed atomic positions.

\begin{table}[htbp]
\begin{tabular}{c|c|c|c|c}
\hline\hline
Compound & Phase & $a$ & $c$  &  Volume  \\ \hline
\multirow{4}{*}{Pb-O} & Experiment\cite{krivovichev2003crystal} & 9.865 & 7.431 & 626.251 \\ \cline{2-5}
& undoped  & 10.018 & 7.485 & 650.585 \\ \cline{2-5}
& Cu$_1$-doped & 9.795 & 7.339 & 609.820 \\ \cline{2-5}
& Cu$_2$-doped & 9.868 & 7.410 & 624.970 \\ \hline
\multirow{4}{*}{Pb-(OH)$_2$} &  Experiment\cite{bruckner1995crystal} & 9.866 & 7.426 & 625.991 \\ \cline{2-5}
& undoped & 9.866 & 7.426 & 625.991 \\ \cline{2-5}
& Cu$_1$-doped & 9.731 & 7.296 & 598.352 \\ \cline{2-5}
& Cu$_2$-doped & 9.718 & 7.301 & 597.151 \\ 
\hline\hline
\end{tabular}
\caption{The experimental and relaxed lattice constants in undoped and Cu-doped phases \ch{Pb10(PO4)6O} and \ch{Pb10(PO4)6(OH)2} (written as Pb-O and Pb-(OH)$_2$ for short). The experimental structures are given by Ref.\cite{krivovichev2003crystal, bruckner1995crystal}.
}
\label{table:relaxed_latt_const}
\end{table}

\begin{table}[htbp]
\begin{tabular}{c|c|c|c}
\hline\hline
Compound & Phase & Atom   & Position  \\ \hline
\multirow{5}{*}{Pb-O} & undoped 
& tri-O & $(0,0,0.726)$ \\ \cline{2-4}
& \multirow{2}{*}{Cu$_1$-doped}
& tri-O & $(0,0,0.821)$ \\ 
& & Cu$_1$ & $(\frac{1}{3}, \frac{2}{3}, 0.498)$ \\ \cline{2-4}
& \multirow{2}{*}{Cu$_2$-doped}
& tri-O & $(0,0,0.746)$ \\ 
& & Cu$_2$ & $(\frac{2}{3}, \frac{1}{3}, 0.502)$ \\ \hline
\multirow{8}{*}{Pb-(OH)$_2$} & \multirow{2}{*}{undoped} 
& tri-O & $(0,0,0.040),(0,0,0.540)$ \\ 
& & tri-H & $(0,0,0.900),(0,0,0.400)$ \\ \cline{2-4}
& \multirow{3}{*}{Cu$_1$-doped}
& tri-O & $(0,0,0.154),(0,0,0.762)$ \\ 
& & tri-H & $(0,0,0.019),(0,0,0.628)$ \\ 
& & Cu$_1$ & $(\frac{1}{3}, \frac{2}{3}, 0.473)$ \\ \cline{2-4}
& \multirow{3}{*}{Cu$_2$-doped}
& tri-O & $(0,0,0.154),(0,0,0.630)$ \\ 
& & tri-H & $(0,0,0.021),(0,0,0.497)$
\\ 
& & Cu$_2$ & $(\frac{2}{3}, \frac{1}{3}, 0.498)$ \\ \hline\hline
\end{tabular}
\caption{The relaxed atomic positions in undoped and Cu-doped phases \ch{Pb10(PO4)6O} and \ch{Pb10(PO4)6(OH)2} (written as Pb-O and Pb-(OH)$_2$ for short). 
}
\label{table:relaxed_atom_positions}
\end{table}

\subsection{Orbital projections}\label{Sec:appendix_orb_proj}

\begin{figure}[htbp]
    \centering
    \includegraphics[width=\columnwidth]{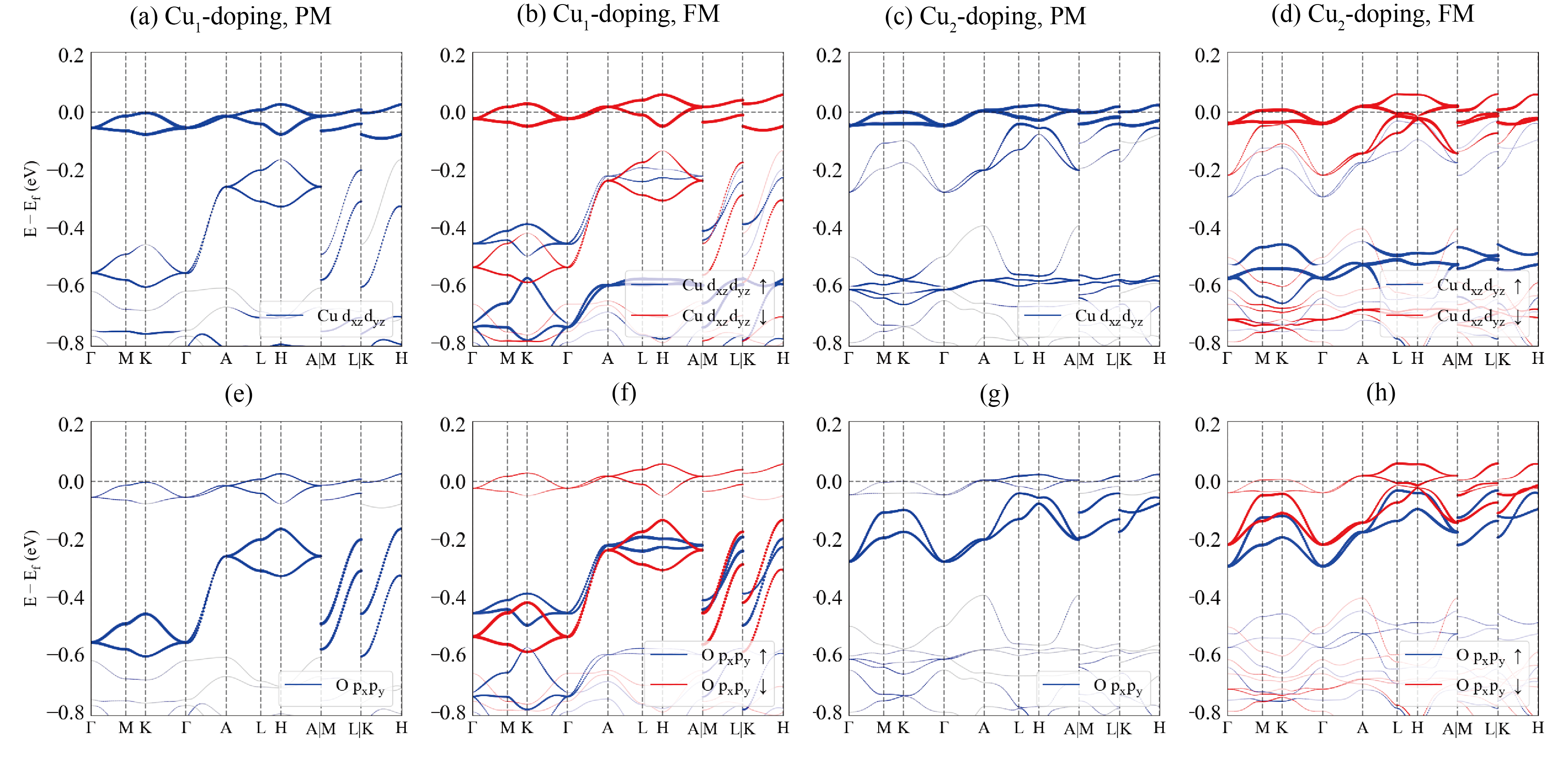}
    \caption{Orbital projections of \ch{Pb9Cu1(PO4)6O} in Cu$_1$- and Cu$_2$-doping, PM and FM phases, where the first row is the projection of $(d_{xz}, d_{yz})$ orbitals of Cu, and the second row is the $(p_x,p_y)$ orbitals of trigonal-O.
    }
    \label{fig:1O_orbital_proj}
\end{figure}

\begin{figure}[htbp]
    \centering
    \includegraphics[width=\columnwidth]{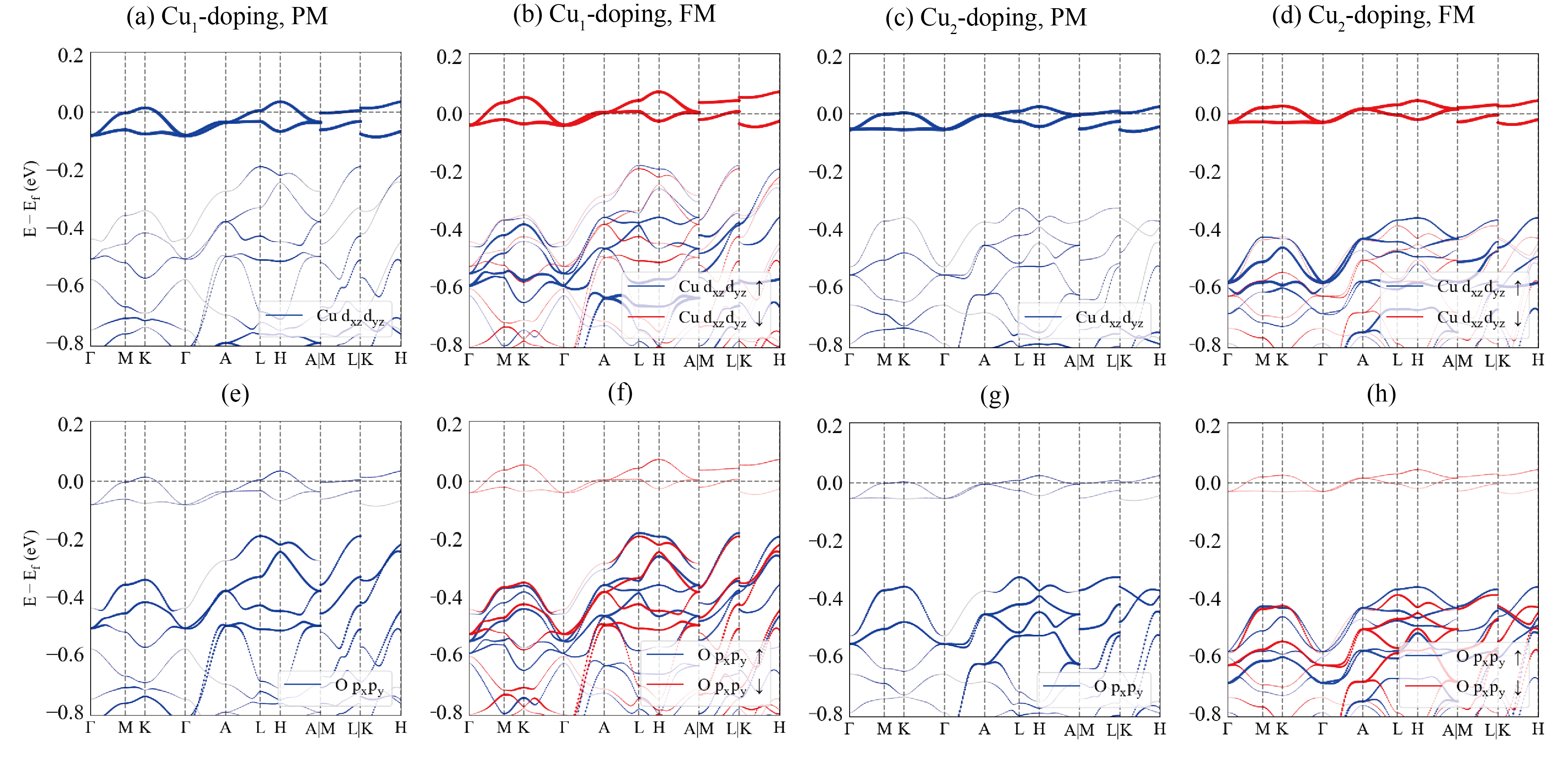}
    \caption{Orbital projections of \ch{Pb9Cu1(PO4)6(OH)2} in Cu$_1$- and Cu$_2$-doping, PM and FM phases, where the first row is the projection of $(d_{xz}, d_{yz})$ orbitals of Cu, and the second row is the $(p_x,p_y)$ orbitals of trigonal-O. 
    }
    \label{fig:2OH_orbital_proj}
\end{figure}

In Fig. \ref{fig:1O_orbital_proj} and Fig. \ref{fig:2OH_orbital_proj}, we show the orbital projections of \ch{Pb10(PO4)6O} and \ch{Pb10(PO4)6(OH)2}, respectively. 

\subsection{Bands of unrelaxed structures}\label{Sec:appendix_unrelaxed_struct}

\begin{figure}[htbp]
    \centering
    \includegraphics[width=\columnwidth]{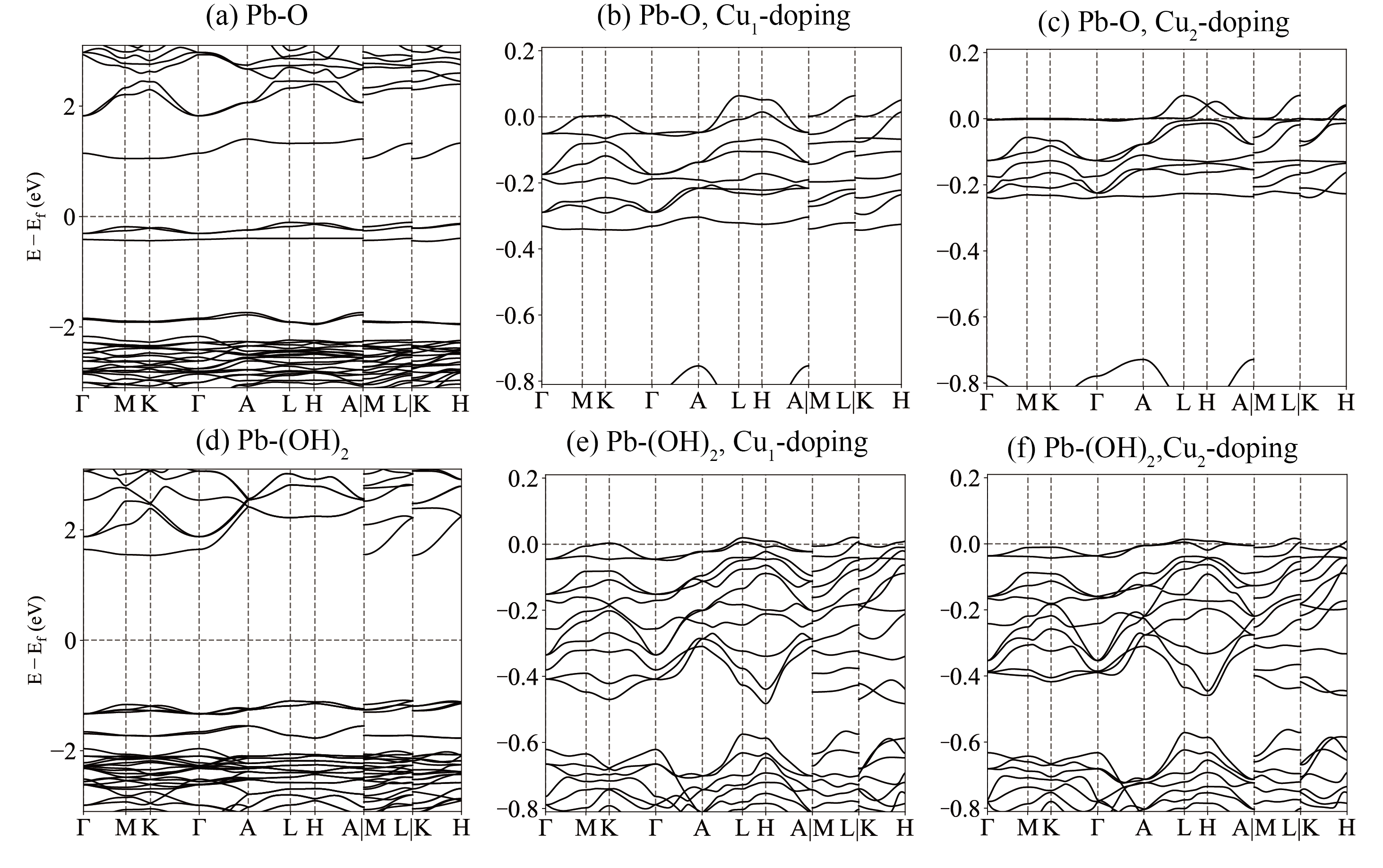}
    \caption{The DFT bands of unrelaxed structures. (a)-(c) are undoped, Cu$_1$-doped, and Cu$_2$-doped bands for \ch{Pb10(PO4)6O}, and (d)-(f) are for \ch{Pb10(PO4)6(OH)2}.
    }
    \label{fig:unrelaxed_bands}
\end{figure}

As mentioned in the main text, the original experimental structures have fractional occupancy for O or (OH)$_2$. Hence the unrelaxed structures used here have fixed the position for O and (OH)$_2$ and remove the fractional occupancy. This means that these unrelaxed structures are not experimental and need further relaxation in order to obtain stable structures for DFT. The Cu-doped phases also do not have experimental structures and need further relaxation. For completeness, we also show in Fig. \ref{fig:unrelaxed_bands} the bands of unrelaxed structure for \ch{Pb10(PO4)6O} and \ch{Pb10(PO4)6(OH)2}, in undoped, Cu$_1$-doped, and Cu$_2$-doped phases. 

\begin{figure}[htbp]
\centering
\subfloat[Pb-O at low-T.]{
\label{fig:O1LowT}
\includegraphics[width=0.44\textwidth]{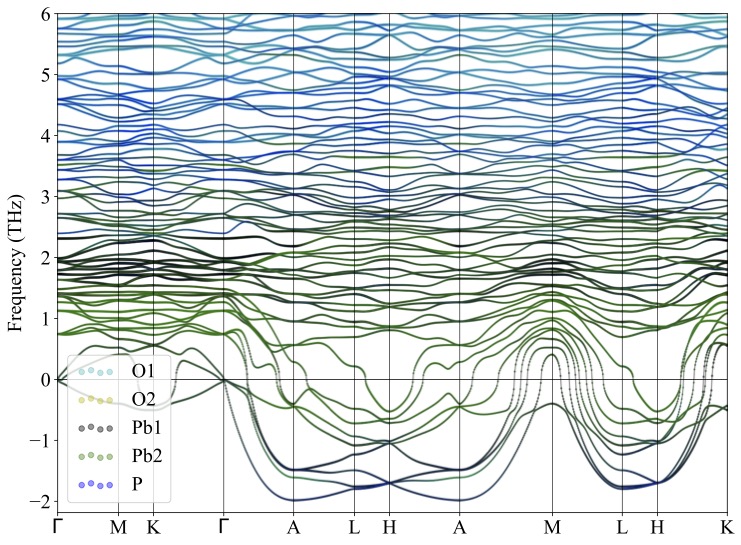}
}
\subfloat[Pb-O at high-T.]{
\label{fig:O1HighT}
\includegraphics[width=0.44\textwidth]{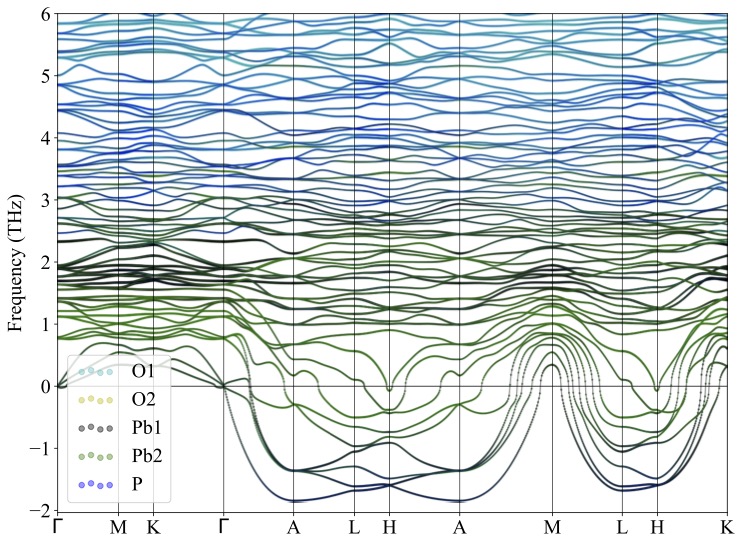}
}
\caption{Phonon spectrum for relaxed \ch{Pb10(PO4)6O} without doping Cu. The negative branches get harden at $k_3=0$ plane as the temperature goes higher. Instead, the imaginary phonon at $k_3=\pi$ are still soft, which may be caused by the short cutoff in $c$-direction.}
\label{fig:1Opho_undoped}
\end{figure}

\begin{figure}[htbp]
\centering
\subfloat[Cu$_1$-doped \ch{Pb9Cu1(PO4)6O} in PM phase at low-T.]{
\label{fig:1OCu1_PM_lowT}
\includegraphics[width=0.42\textwidth]{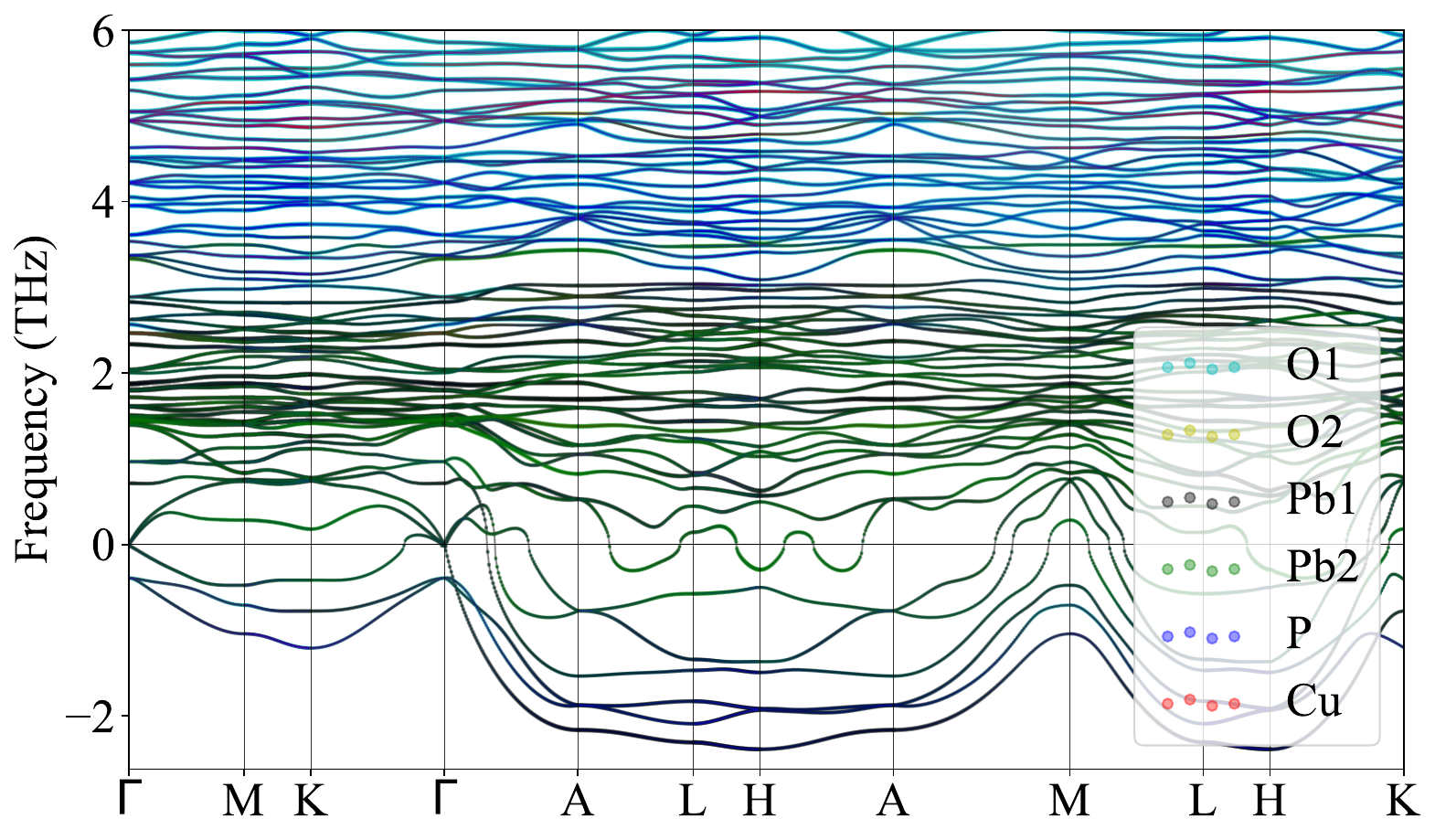}
}
\subfloat[Cu$_1$-doped \ch{Pb9Cu1(PO4)6O} in PM phase at high-T.]{
\label{fig:1OCu1_PM_highT}
\includegraphics[width=0.42\textwidth]{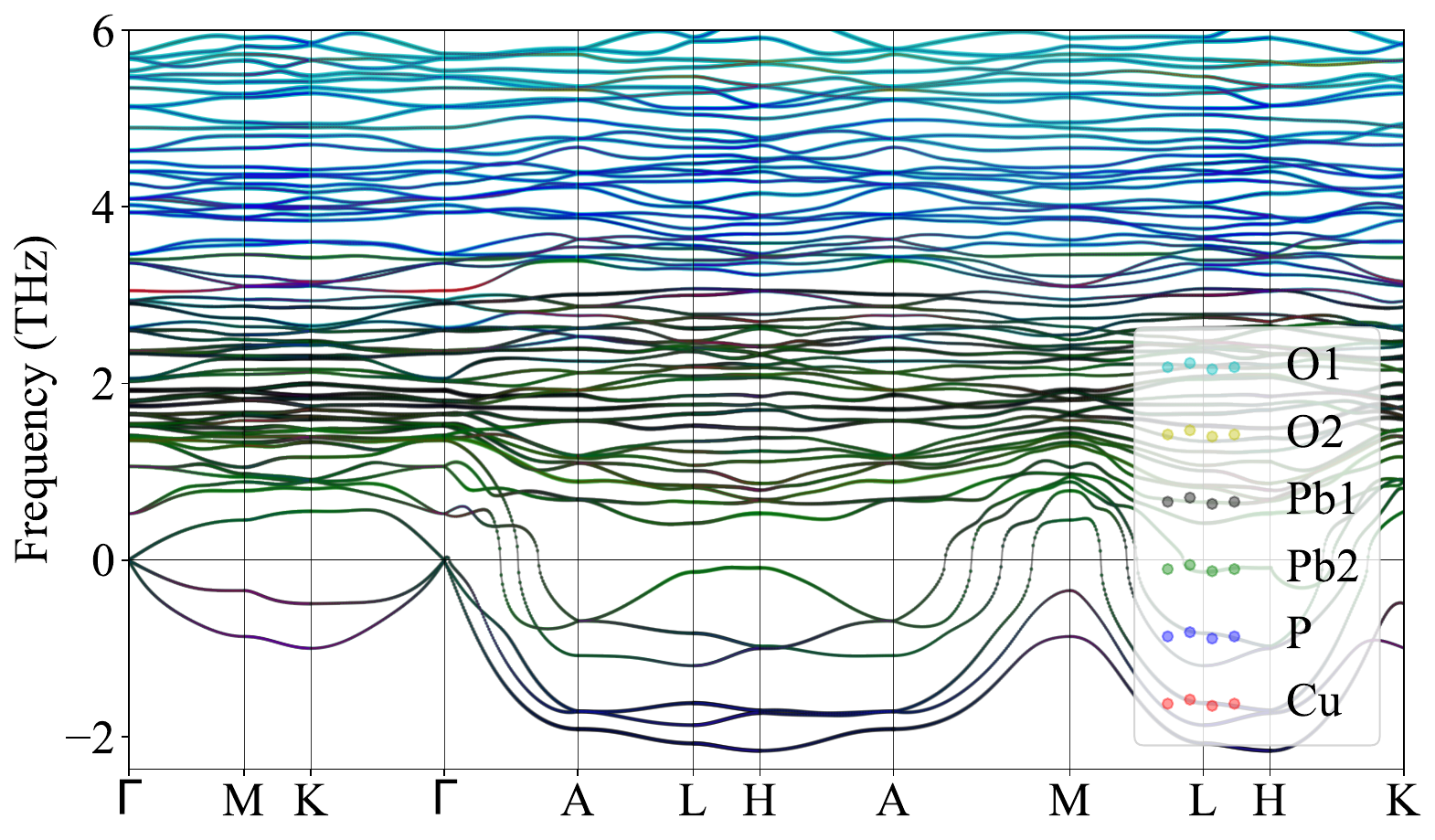}
}
\\
\subfloat[Cu$_1$-doped \ch{Pb9Cu1(PO4)6O} in FM phase at low-T.]{
\label{fig:1OCu1_FM_lowT}
\includegraphics[width=0.42\textwidth]{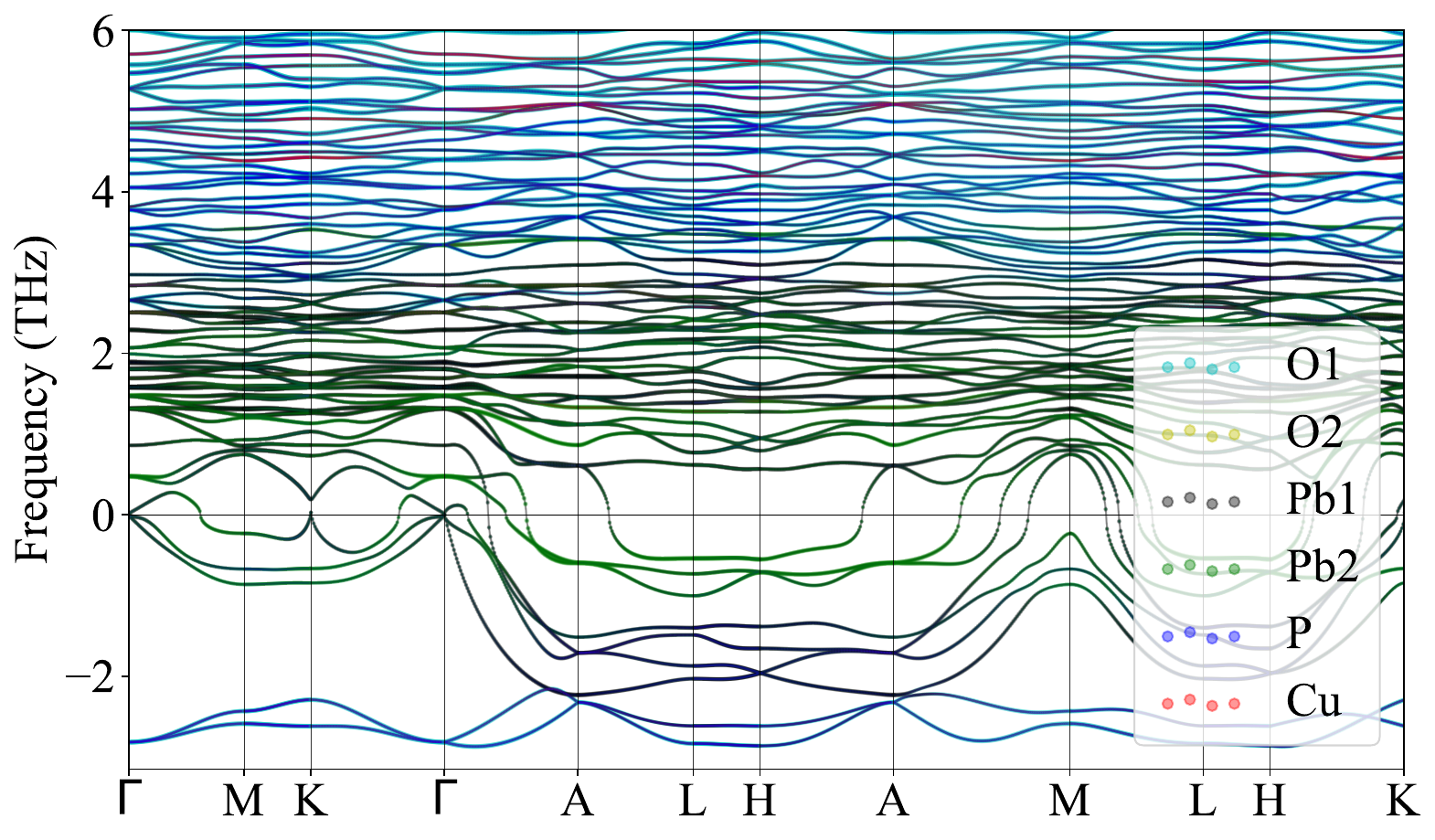}
}
\subfloat[Cu$_1$-doped \ch{Pb9Cu1(PO4)6O} in FM phase at high-T.]{
\label{fig:1OCu1_FM_highT}
\includegraphics[width=0.42\textwidth]{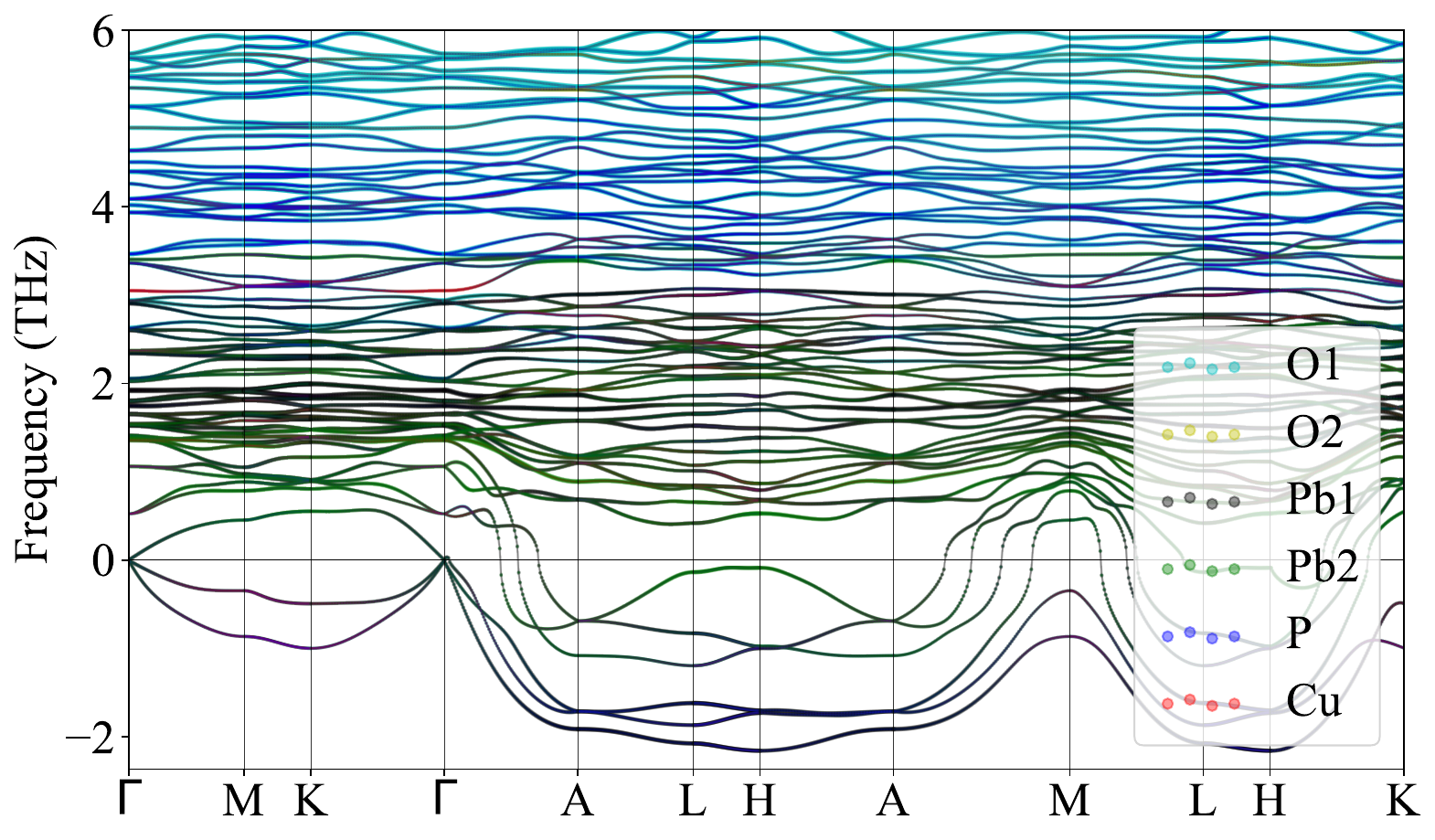}
}
\\
\subfloat[Cu$_2$-doped \ch{Pb9Cu1(PO4)6O} in PM phase at low-T.]{
\label{fig:1OCu2_PM_LowT}
\includegraphics[width=0.42\textwidth]{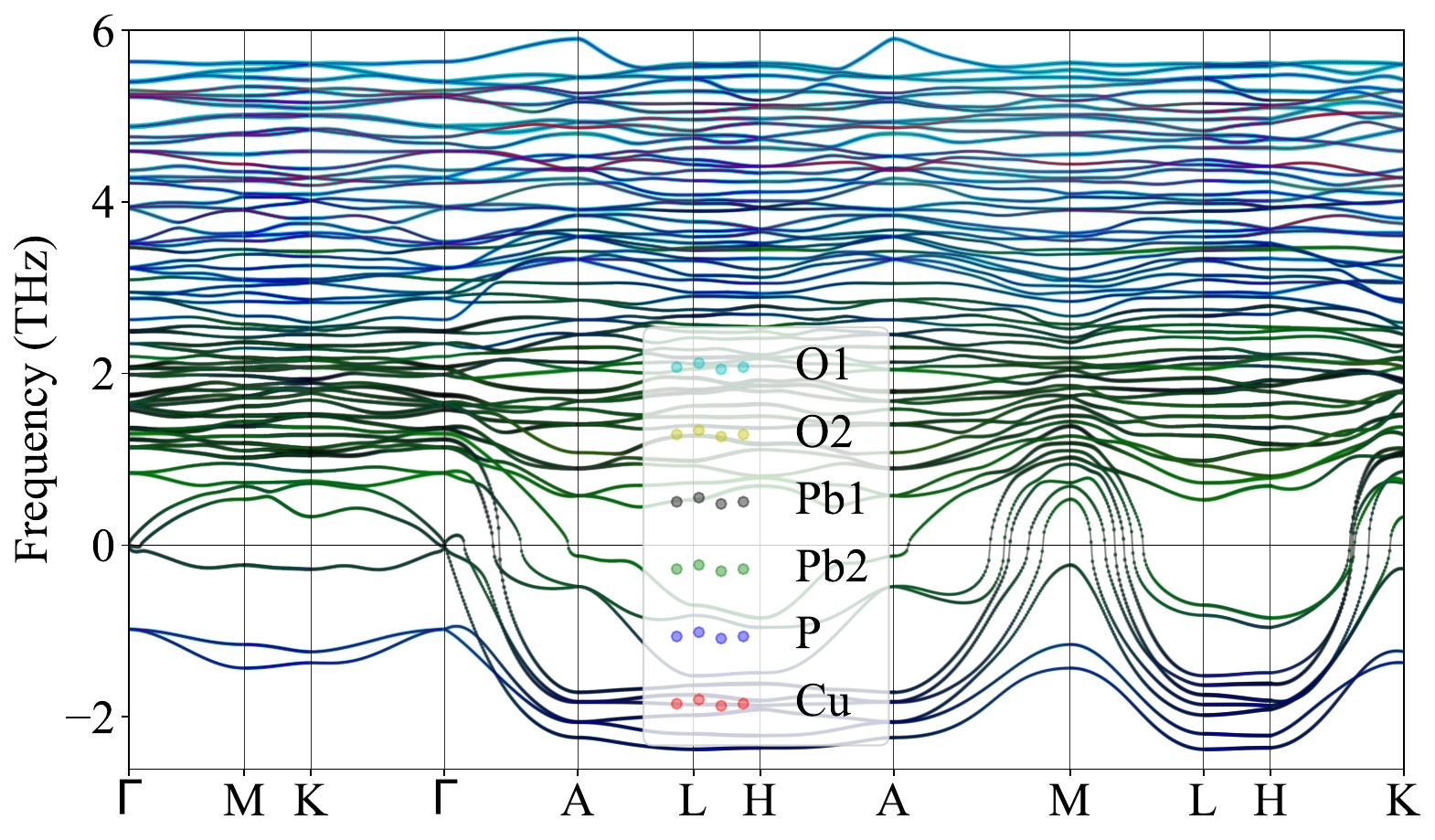}
}
\subfloat[Cu$_2$-doped \ch{Pb9Cu1(PO4)6O} in PM phase at high-T.]{
\label{fig:1OCu2_PM_highT}
\includegraphics[width=0.42\textwidth]{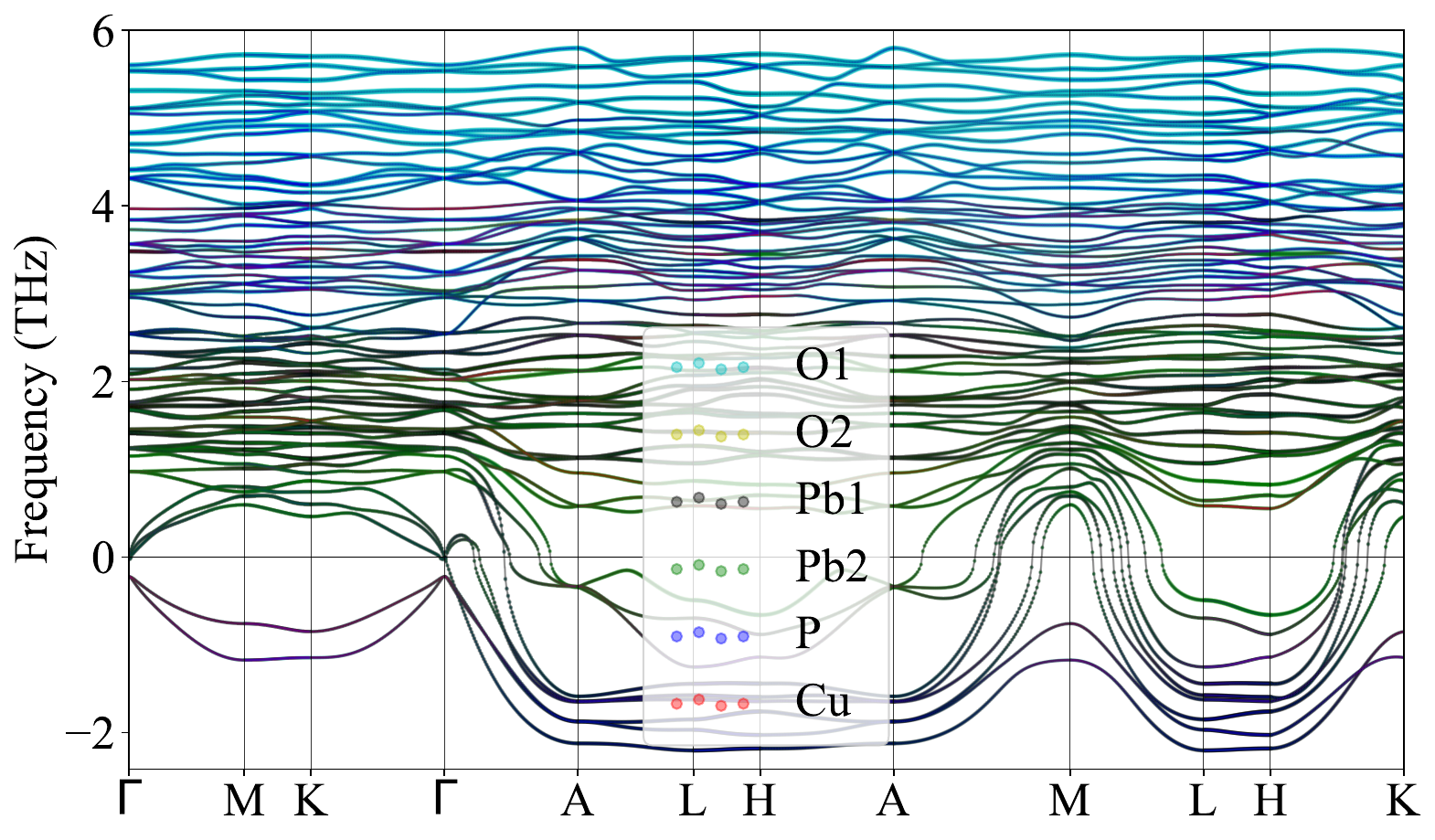}
}
\\
\subfloat[Cu$_2$-doped \ch{Pb9Cu1(PO4)6O} in FM phase at low-T.]{
\label{fig:1OCu2_FM_LowT}
\includegraphics[width=0.42\textwidth]{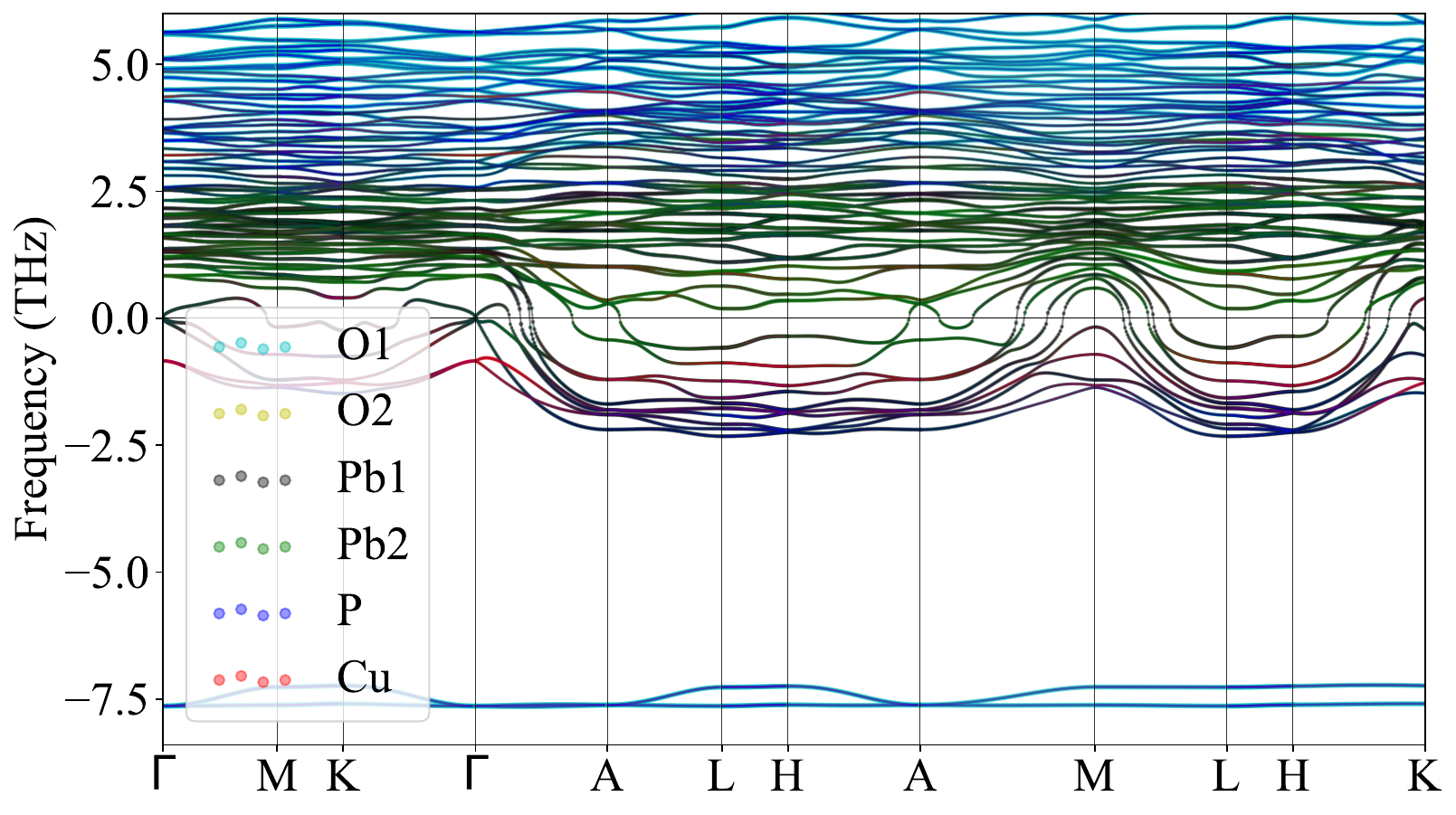}
}
\subfloat[Cu$_2$-doped \ch{Pb9Cu1(PO4)6O} in FM phase at high-T.]{
\label{fig:1OCu2_FM_highT}
\includegraphics[width=0.42\textwidth]{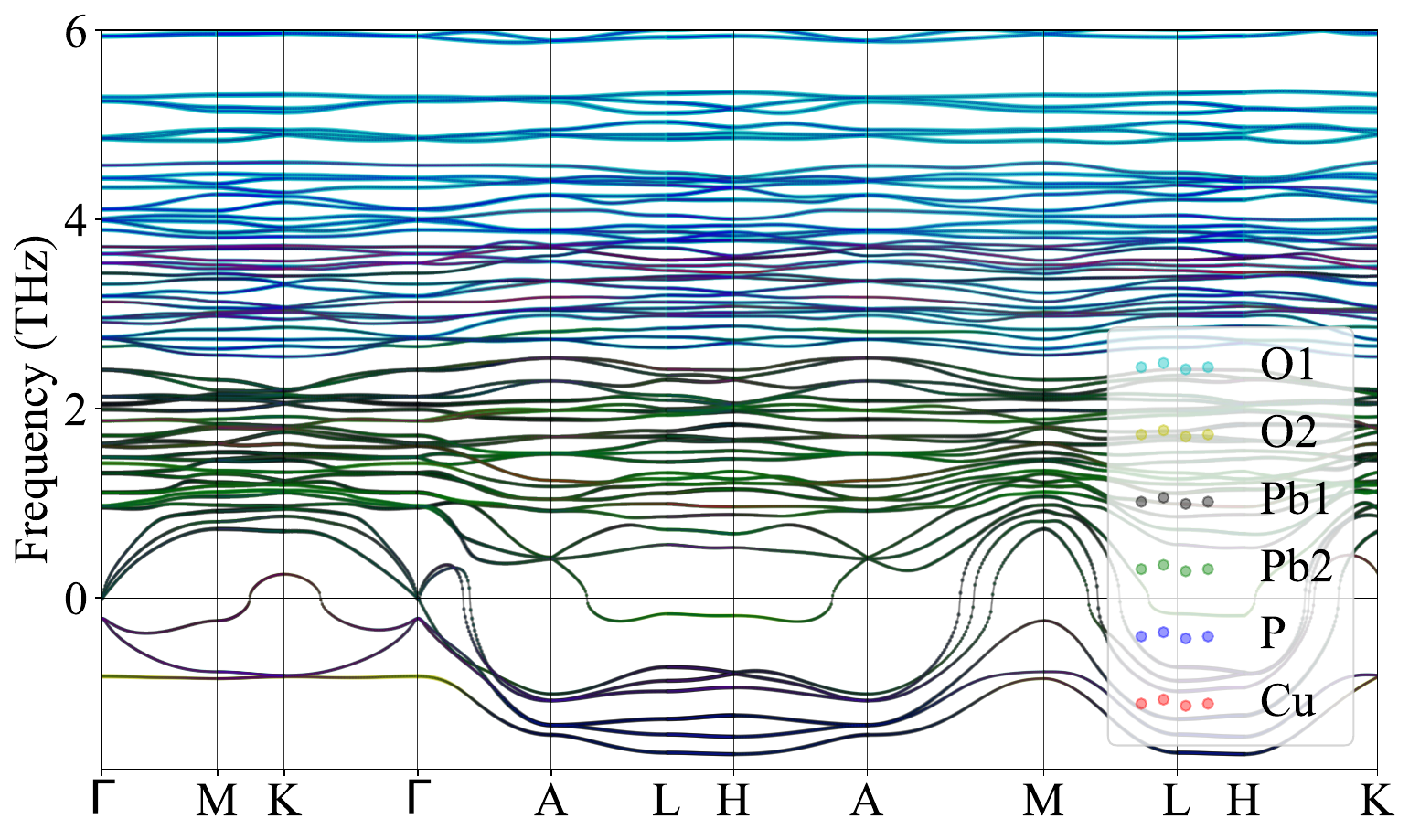}
}
\caption{Phonon spectrum for relaxed \ch{Pb9Cu1(PO4)6O} in Cu$_1$- and Cu$_2$-doping state. Different from the undoped structure, the doped ones present imaginary phonon contributed by O and Cu at low-T, while their contributions are eliminated at high-T.}
\label{fig:1O_doped}
\end{figure}

\subsection{Phonon Spectra}
\label{app:phononplots}

Here, we present the phonon spectrum of relaxed Pb-O, the doped Pb-O, and Pb-(OH)$_2$ compounds. The temperature effect is included via the smearing method in harmonic approximation level. Without specific indication, these calculations are performed within 111 cells. As shown below, all these phonon spectrums present imaginary modes in the 111-cell results. Similar to the phonon of undoped Pb-(OH)$_2$ in Sec.~\ref{ssec:pho}, the phonon presents a hardening trend as the temperature goes higher. For the imaginary modes at $k_3=\pi$ plane, we expect they can be eased in larger supercell calculation as distant force constants are included. In the doped structure, the O1 and Cu sites also contribute to the imaginary branches at low temperatures, which are absent in the high-temperature results. This is much more evident in the FM phonon spectrum. Since O and Cu are much lighter than Pb, the current imaginary phonons may not be physical, and further relaxation and larger supercell calculations may eliminate them, which we leave for future study. We also emphasize that current instabilities don't determine the stability issues of these compounds since larger supercell takes longer time.

\begin{figure}[htbp]
\centering
\subfloat[Cu$_1$-doped Pb-(OH)$_2$ in PM phase at low-T.]{
\label{fig:2OHCu1_PM_lowT}
\includegraphics[width=0.4\textwidth]{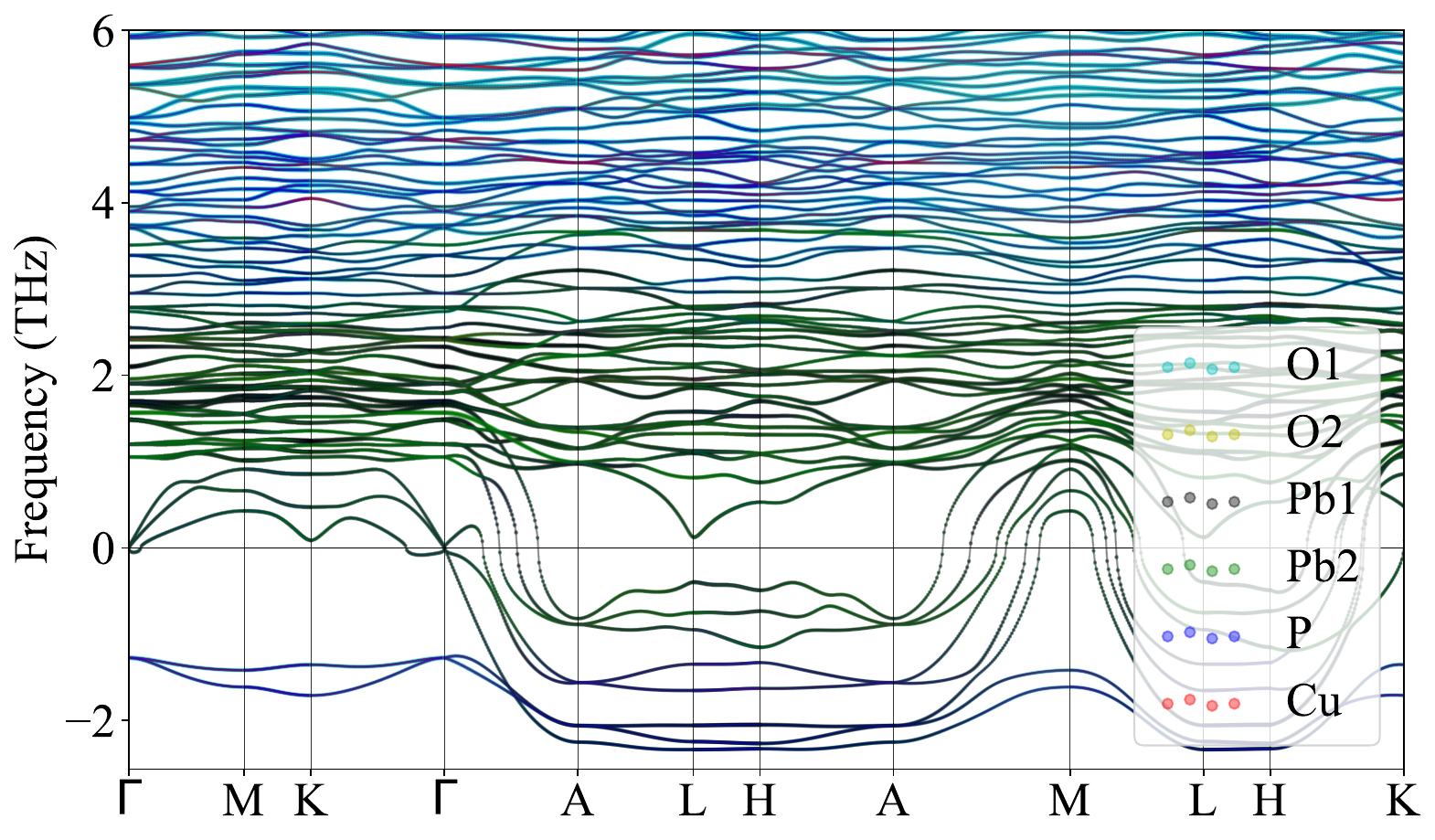}
}
\hspace{5pt}
\subfloat[Cu$_1$-doped Pb-(OH)$_2$ in PM phase at high-T.]{
\label{fig:2OHCu1_PM_highT}
\includegraphics[width=0.4\textwidth]{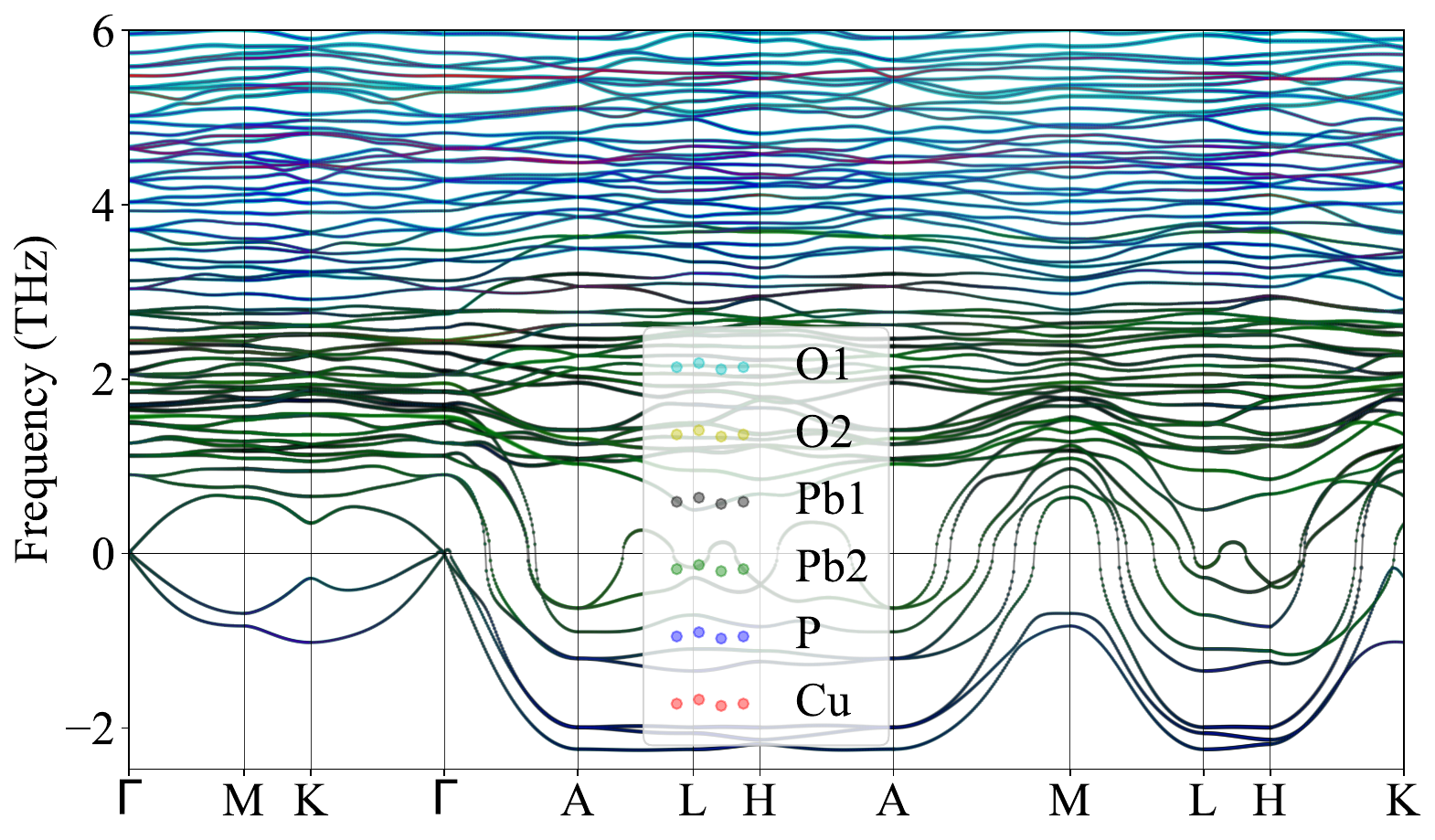}
}
\\
\subfloat[Cu$_1$-doped Pb-(OH)$_2$ in FM phase at low-T.]{
\label{fig:2OHCu1_FM_lowT}
\includegraphics[width=0.4\textwidth]{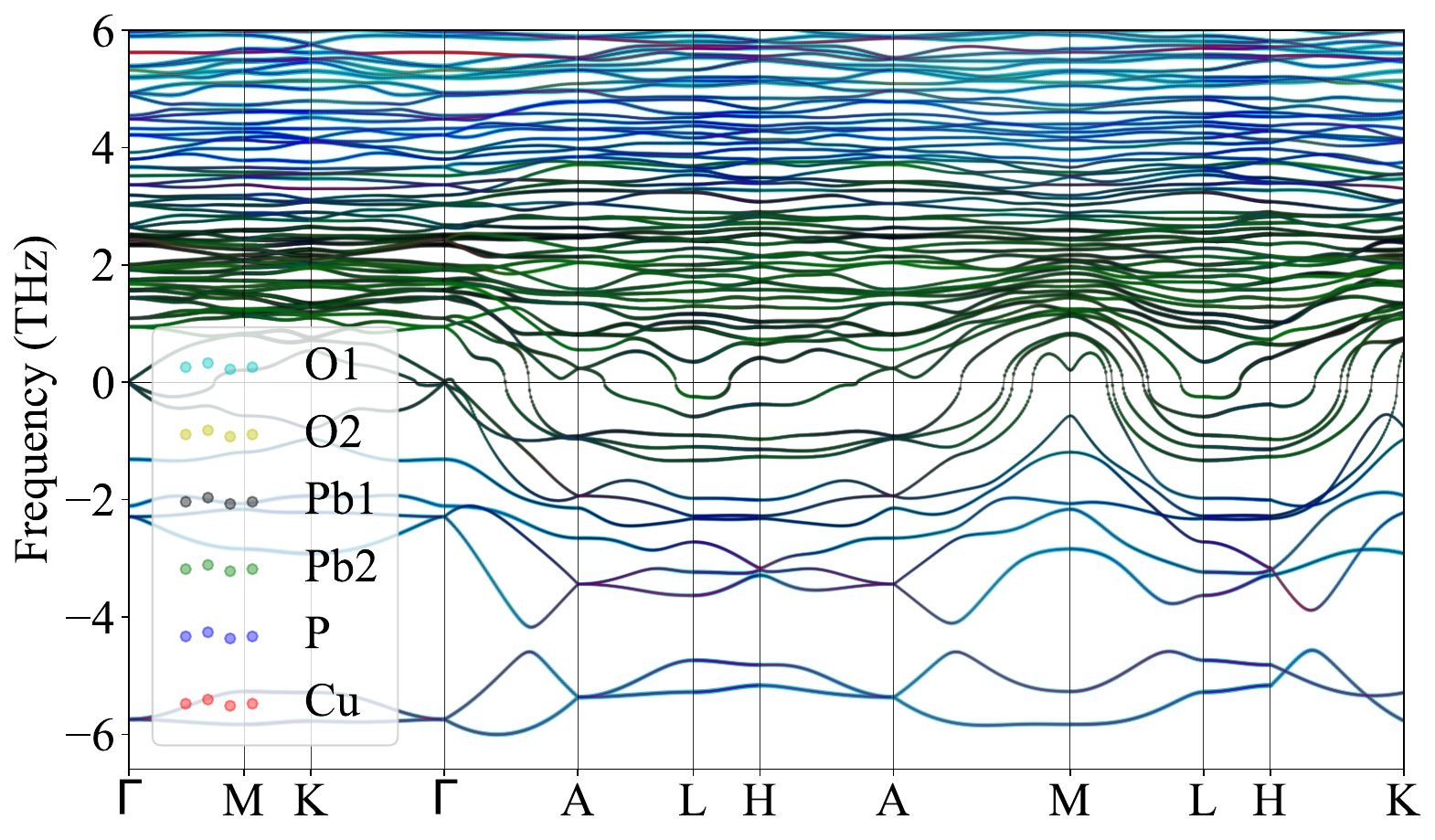}
}
\hspace{5pt}
\subfloat[Cu$_1$-doped Pb-(OH)$_2$ in FM phase at high-T.]{
\label{fig:2OHCu1_FM_highT}
\includegraphics[width=0.4\textwidth]{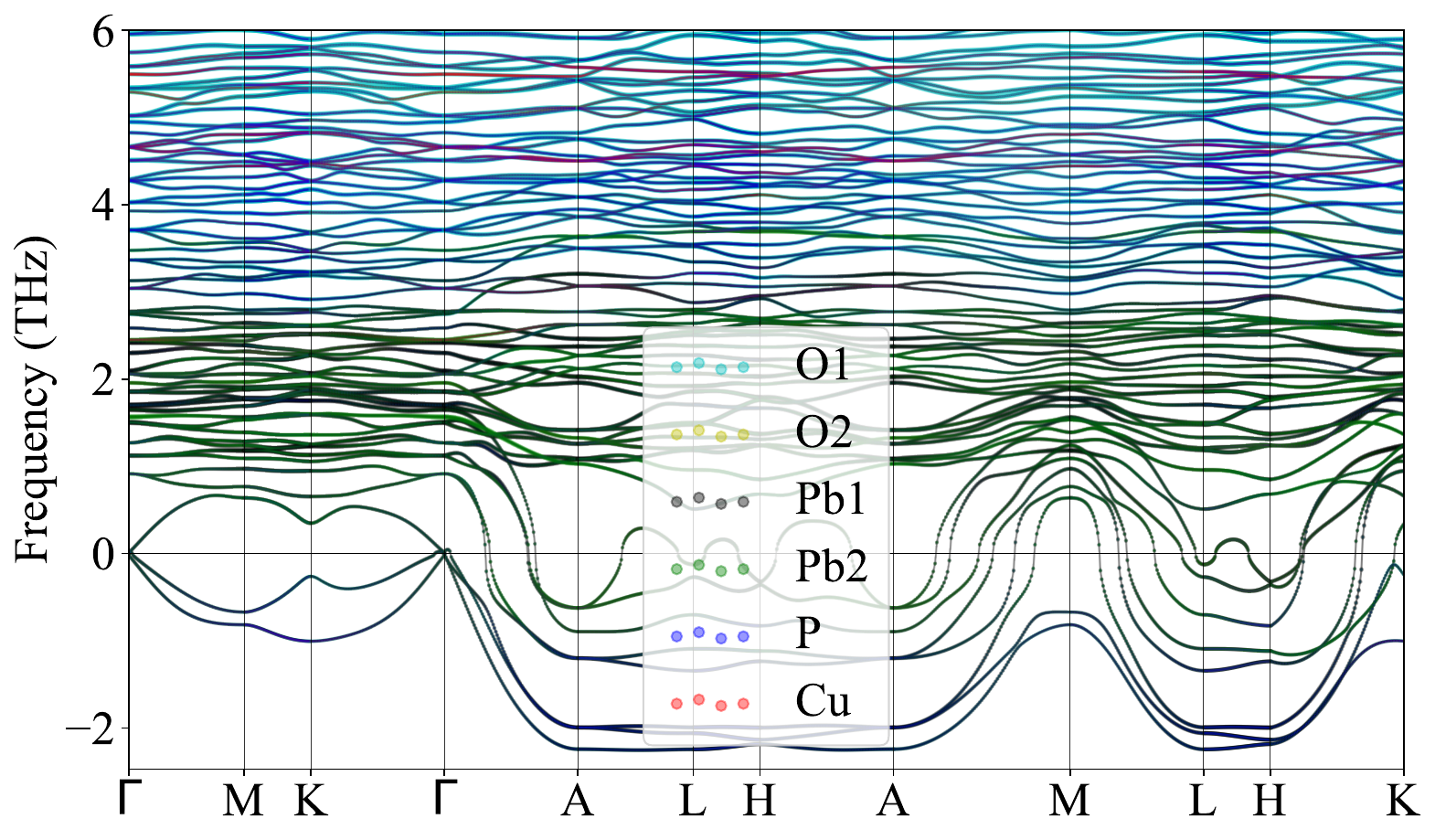}
}
\\
\subfloat[Cu$_2$-doped Pb-(OH)$_2$ in PM phase at low-T.]{
\label{fig:2OHCu2_PM_lowT}
\includegraphics[width=0.4\textwidth]{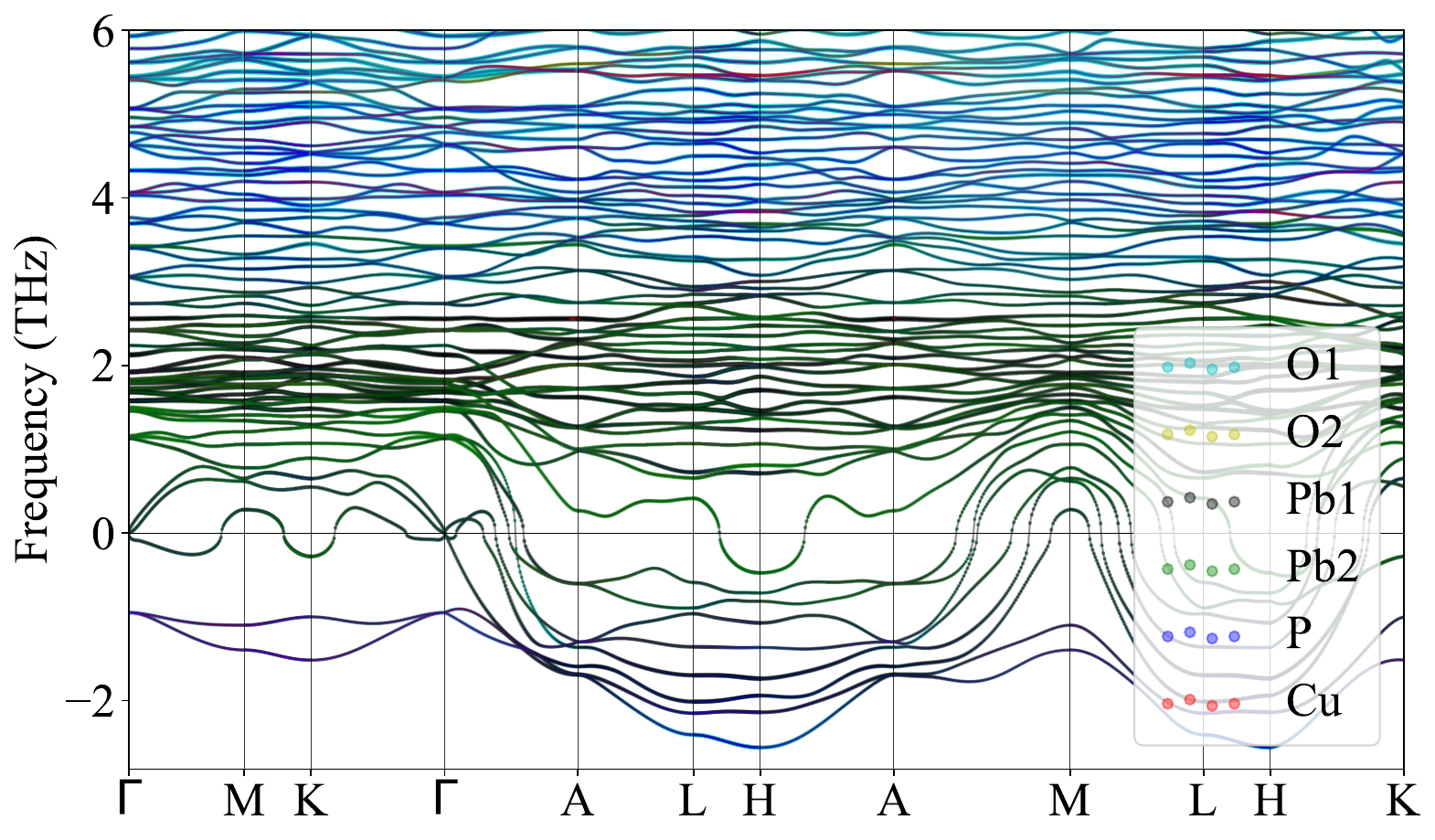}
}
\hspace{5pt}
\subfloat[Cu$_2$-doped Pb-(OH)$_2$ in PM phase at high-T.]{
\label{fig:2OHCu2_PM_highT}
\includegraphics[width=0.4\textwidth]{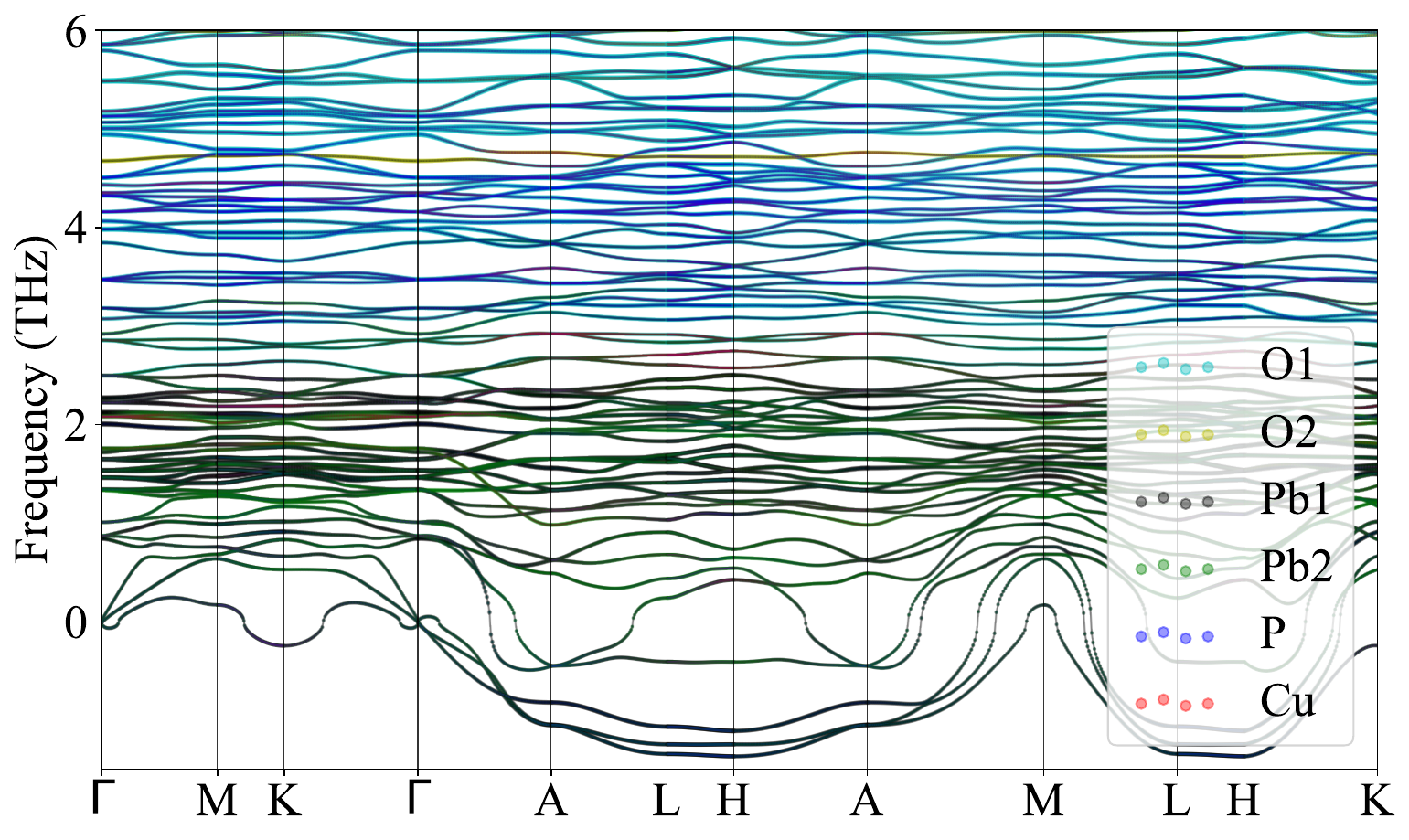}
}
\caption{Phonon spectrum for relaxed Pb-(OH)$_2$ in Cu$_1$- and Cu$_2$-doping state. As in Cu-doped Pb-O compounds, O also contributes to the imaginary phonon modes at low-T, which are absent at high-T.}
\label{fig:2OH_doped}
\end{figure}

\subsection{DFT computational details}
We use the Vienna \textit{ab-initio} Simulation Package (VASP)\cite{kresse1996efficiency, kresse1993ab1, kresse1993ab2, kresse1994ab, kresse1996efficient} to perform the \textit{ab-initio} computations. The generalized gradient approximation (GGA) with Perbew-Burke-Ernzerhof (PBE) exchange-correlation potential\cite{perdew1996generalized} is adopted. A cutoff energy of 500 eV is adopted and no extra Hubbard U correction is applied in the calculation of band structures. 
The maximally localized Wannier functions (MLWFs) are constructed using Wannier90\cite{marzari1997maximally, souza2001maximally, marzari2012maximally, pizzi2020wannier90} to obtain onsite energies and hoppings parameters of orbitals. The phonon spectrum is calculated through density functional perturbation theory (DFPT) assisted by Phonopy package~\cite{phonopy2008}.

Defect formation energy calculations were performed using a 2x2x2 supercell, computed with the same base function, a 500 eV cutoff, and a 4 eV Hubbard U correction for Cu. Bulk Cu, Cu$_2$S, and Cu$_3$P were used for the competing chemical phases.

\section{Single-Particle Hamiltonians}
\label{app:SPham}

In this Appendix, we give a detailed, symmetry-based construction of the single-particle bands for two possibles structures of Cu doped lead apatite. Both models show partially occupied topologically trivial bands dominated by Cu $d$-orbitals at the Fermi level. We then fit our parameters to ab initio Wannier calculations. 

\subsection{ Pb$_{9}$Cu(PO$_4$)$_6$O}

We first provide the details of the single-particle Hamiltonians in the PB$_{9}$Cu(PO$_4$)$_6$O structure. The location of the Cu dopant, which is proposed to replace Pb (although experimental characterization of the system is crucial to test this hypothesis), can occur either at the 1b (Cu$_1$ doping) or 1c position (Cu$_2$ doping). Note that Cu doping and structural relaxation reduce the space group of the original Pb$_{9}$Cu(PO$_4$)$_6$O compound to $P31'$ generated by translations, $C_3$, and spin-less time-reversal $\mathcal{T}$.

\begin{figure*}
   (a)\includegraphics[width=0.3\columnwidth]{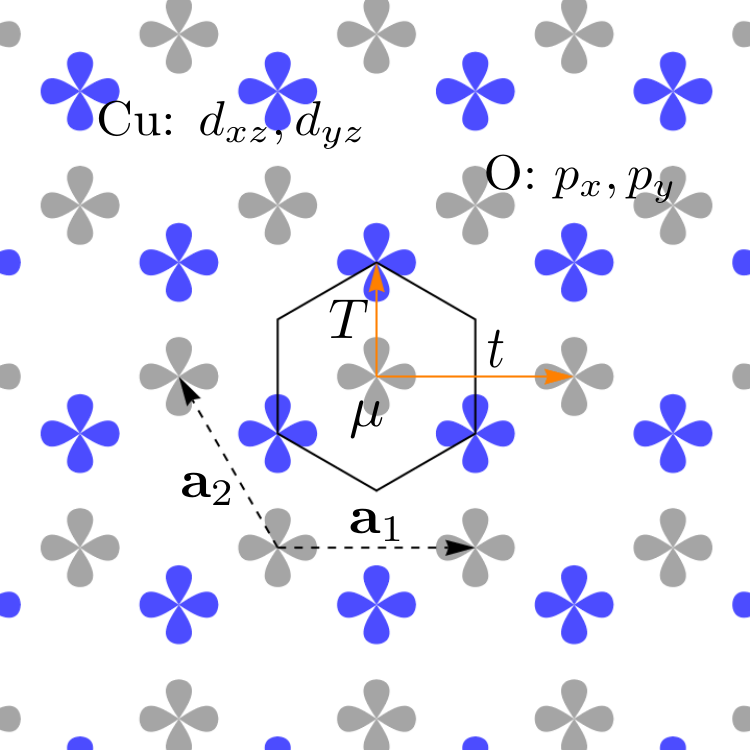} \qquad 
   (b)\includegraphics[width=0.3\columnwidth]{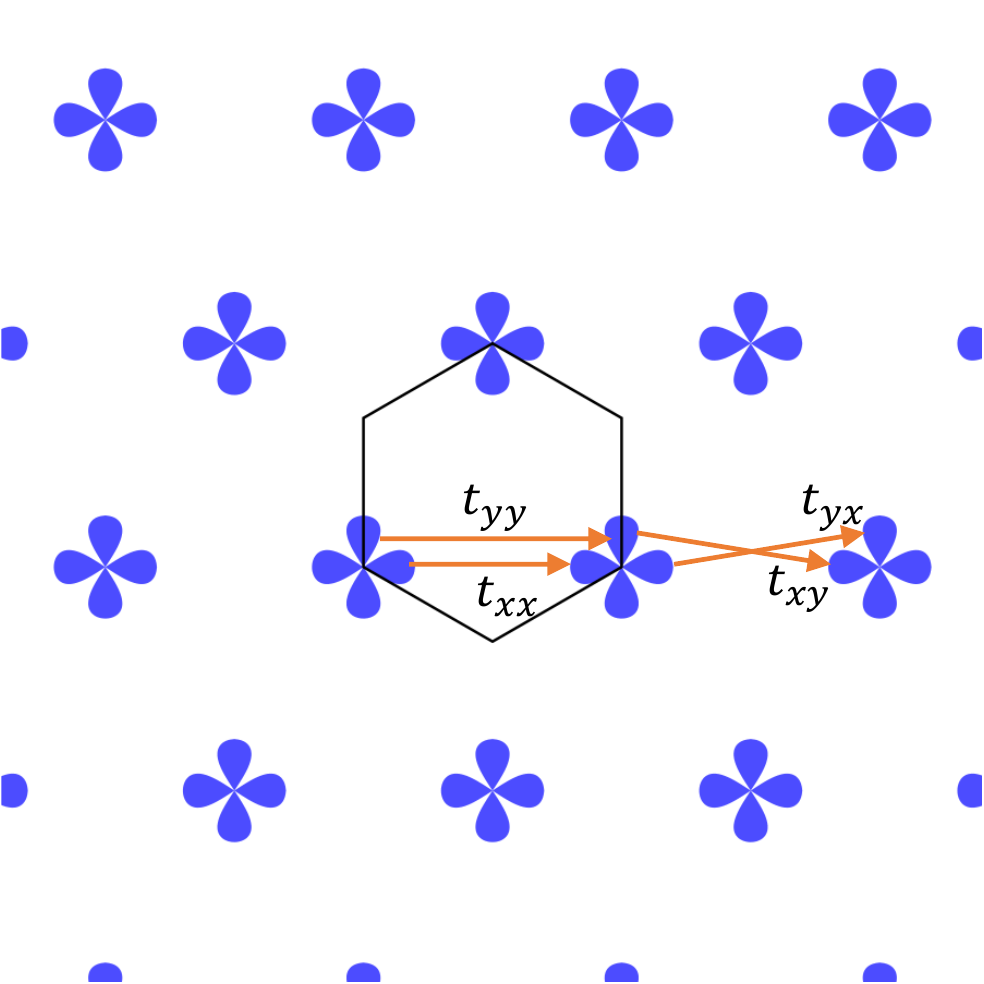}
   \caption{Real Space lattice and hoppings. (a) Hopping structure of the 4-band model in 2D. The only relevant 3D term is an interlayer $t'$ hopping between Cu orbitals. (b) Hopping structure of the 2-band model in 2D.}
   \label{fig:hoppings}
\end{figure*}

In both cases, DFT orbital projections indicate that the dominant orbitals in the bands near the Fermi surface are the $d$-orbitals of the Cu atom at the 1b/1c position and the $p$ orbitals of the O atom at the 1a position (shown in \Fig{fig:hoppings}).  We use lattice vectors $\mbf{a}_1 = a (1,0,0), \mbf{a}_2 = C_3 \mbf{a}_1, \mbf{a}_3 = c (0,0,0)$ where $C_3$ is a three-fold rotation. The Wannier centers/locations of Cu atoms are $\mbf{r}_{1b} = (\mbf{a}_1+ 2\mbf{a}_2)/3 + z \mbf{a}_3$ and $\mbf{r}_{1c} = (2\mbf{a}_1+ \mbf{a}_2)/3 + z' \mbf{a}_3$. The realistic structures show a small vertical displacement between O and Cu atoms (which are not fixed by symmetry since $C_3$ is in-plane) given by $z,z' \sim .25$. 

The Cu $d$ orbitals and O $p$ orbitals both transform in the 2D irrep ${}^1E{}^2E$ with angular momentum $\pm 1 \mod 3$. Topological quantum chemistry \cite{bradlyn2017topological} provides the following momentum space irreps when these local orbitals are induced to the space group:
\bea
{}^1E{}^2E_{1a} \uparrow P31' &=  \Gamma_2\Gamma_3+A_2A_3 + K_2+K_3 + H_2+H_3, \\
{}^1E{}^2E_{1b} \uparrow P31' &=  \Gamma_2\Gamma_3+A_2A_3 + K_1+K_3 + H_1+H_3 \\
{}^1E{}^2E_{1c} \uparrow P31' &=  \Gamma_2\Gamma_3+A_2A_3 + K_1+K_2 + H_1+H_2 \\
\eea
and the irrep notation is contained in the character tables below.
\bea
\begin{array}{c|ccc}
31' &1& C_3  \\
\hline
A &1&1 \\
{}^1E{}^2E &2 & -1  \\
\end{array} \qquad \begin{array}{c|ccc}
31' &1& C_3  \\
\hline
\Gamma_1 &1&1 \\
\Gamma_2\Gamma_3 &2 & -1  \\
\end{array}\qquad  \begin{array}{c|ccc}
31' &1& C_3  \\
\hline
A_1 &1&1 \\
A_2A_3 &2 & -1  \\
\end{array}  \qquad
\begin{array}{c|ccc}
31' &1& C_3  \\
\hline
K_1 &1&1 \\
K_2 &1&e^{\frac{2\pi i}{3}}  \\
K_3 &1&e^{-\frac{2\pi i}{3}}  \\
\end{array} \qquad
\begin{array}{c|ccc}
31' &1& C_3  \\
\hline
H_1 &1&1 \\
H_2 &1&e^{\frac{2\pi i}{3}}  \\
H_3 &1&e^{-\frac{2\pi i}{3}}  \\
\end{array}
\eea
Note that the $K'$ and $H'$ points can be obtained by time-reversal, and the only irreps at $M$ and $L$ are the trivial irrep of $\mathcal{T}$. More details can be found on the Bilbao Crystallographic Server \url{https://www.cryst.ehu.es/}. We see that degeneracies occur at the $\Gamma$ point and $A$, and the rest of the bands are split. 

We now build a short ranged tight-binding model from these orbitals. We will primarily discuss the para-magnetic (spin-unpolarized) DFT results which provide a single-particle band structure. Including Coulomb repulsion favors a ferromagnetic groundstate which is expected due to the flat bands obtained in the tight-binding models. 

For both dopant positions, we write the Hamiltonian as 
\bea
h_{4}(\mbf{k}) = \bpm h_{C}(\mbf{k}) & h_{CO}(\mbf{k}) \\ h^\dag_{CO}(\mbf{k}) & h_{O}(\mbf{k}) \epm
\eea
describing the two Wannier orbitals dominated by Cu and O respectively, and their coupling. The symmetries of the model have representations
\bea
D[C_3] h(\mbf{k}) D^\dag[C_3] &= h(C_3 \mbf{k}), \qquad h^*(\mbf{k}) = h(-\mbf{k}), \qquad h(\mbf{k}+\mbf{G}) = V[\mbf{G}] h(\mbf{k})V^\dag[\mbf{G}] 
\eea
where $D[C_3] = R_{2\pi/3} \oplus R_{2\pi/3}$, $V[\mbf{G}] = e^{-i \mbf{G} \cdot \mbf{r}_{1b}} \sigma_0 \oplus \sigma_0$ and $R_{2\pi/3} = e^{-i \frac{2\pi}{3} \sigma_2}$ is the 2D rotation matrix. 

The O block is dominated by a relative potential $\mu_O$, an in-plane $\mbf{a}_1$ hopping matrix $t_O$, and an out-of-plane  $\mbf{a}_3$ hopping matrix $t'_O$. For higher accuracy, we also include  $\mbf{a}_1 \pm \mbf{a}_3$ hoppings denoted $t^\pm_{O}$. The O-O Hamiltonian is 
\bea
h_{O}(\mbf{k}) = \mu_O \sigma_0 +  \lp t'_O e^{- i \mbf{k} \cdot \mbf{a}_3}+ \sum_{n=0}^2 R_{2\pi/3}^n (t_O e^{- i \mbf{k} \cdot C_3^n \mbf{a}_1} + t^+_O e^{- i \mbf{k} \cdot C_3^n (\mbf{a}_1+\mbf{a}_3)} + t^-_O e^{- i \mbf{k} \cdot C_3^n (\mbf{a}_1-\mbf{a}_3)} )R^n_{-2\pi/3}  +h.c. \rp \ .
\eea
The Cu $d_{xz}, d_{yz}$ orbitals have the same $C_3$ representations, and their block has the analogous form (although we shall see that the hopping strengths are much weaker):
\bea
h_{C}(\mbf{k}) = \mu_{Cu} \sigma_0 +  \lp t'_{Cu} e^{- i \mbf{k} \cdot \mbf{a}_3}+ \sum_{n=0}^2 R_{2\pi/3}^n (t_{Cu} e^{- i \mbf{k} \cdot C_3^n \mbf{a}_1} + t^+_{Cu} e^{- i \mbf{k} \cdot C_3^n (\mbf{a}_1+\mbf{a}_3)} + t^-_{Cu} e^{- i \mbf{k} \cdot C_3^n (\mbf{a}_1-\mbf{a}_3)} )R^n_{-2\pi/3}  +h.c. \rp \ .
\eea
Lastly, the hybridization term contains the nearest-neighbor Cu-O hopping $T$, as well as out-of-plane couplings $T_\pm$:
\bea
h_{CO}(\mbf{k}) = \sum_{n=0}^2 R^n_{2\pi/3} (T e^{- i \mbf{k} \cdot C_3^n \pmb{\delta}}+T_+ e^{- i \mbf{k} \cdot C_3^n (\pmb{\delta}+\mbf{a}_3)}+T_-e^{- i \mbf{k} \cdot C_3^n (\pmb{\delta}-\mbf{a}_3)} )
 R^n_{-2\pi/3} \ .
\eea
Here $\pmb{\delta} = -\mbf{r}_1b/\mbf{r}_1c$ for the Cu$_1$ doping model and Cu$_2$ doping model respectively. 

Incorporating the dominant terms from DFT, we use the following hopping matrices for Cu$_1$ doping:
\bea
\label{eq:cu1dopingparamO}
\mu_O &= 3.803, \quad t_O = -0.018 (\sigma_1 - i \sigma_2)/2, \quad t'_O = -0.074 \sigma_0, \quad \mu_C = 4.094, \quad t_C = -0.0035 (\sigma_1 - i \sigma_2)/2\\
T &= \bpm 0.056 & 0.0190 \\ -0.056 & -0.045 \epm, \qquad \text{all other terms} = 0  
\eea
and Cu$_2$ doping:
\bea
\mu_O &= 3.624, \quad t_O = -0.025 (\sigma_0 +\sigma_3)/2, \quad t'_O = -0.021 \sigma_0, \quad t_O^+ = .007 (\sigma_1 - i \sigma_2)/2 , \quad  t_O^- = .003 (\sigma_1 - i \sigma_2)/2 \\
\mu_C &= 3.736, \quad t_C = R_{-2\pi/3} \bpm 0 & 0 \\ .006 & -.002 \epm R_{2\pi/3}, \quad t_C' = -.004 \sigma_0, \quad t_C^+ = 0, \quad t_C^- = -.003 (\sigma_0 + \sigma_3)/2 \\
T &= \left(
\begin{array}{cc}
 -0.005 & 0.021 \\
 -0.012 & 0.004 \\
\end{array}
\right), \quad T_+ = -.004 (\sigma_1 + i \sigma_2)/2, \quad T_- = -.006 (\sigma_1 + i \sigma_2)/2 \ .
\eea
 The single-particle second-quantized Hamiltonian is
\bea
H_4 &= \sum_{\mbf{k}, \al \be, \sigma = \uparrow,\d} c^\dag_{\mbf{k} \al,\sigma} [h_{4}(\mbf{k})]_{\al \be} c^\dag_{\mbf{k} \be,\sigma} 
\eea
using $SU(2)$ spin symmetry due to the small spin-orbit coupling. The Fermi level of the single-particle model occurs at $3/4$ filling of the upper two bands in each spin sector, leading to a total of one hole per unit cell in the spinful model. 

\begin{figure}[H]
    \centering
    \includegraphics[width=.5\columnwidth]{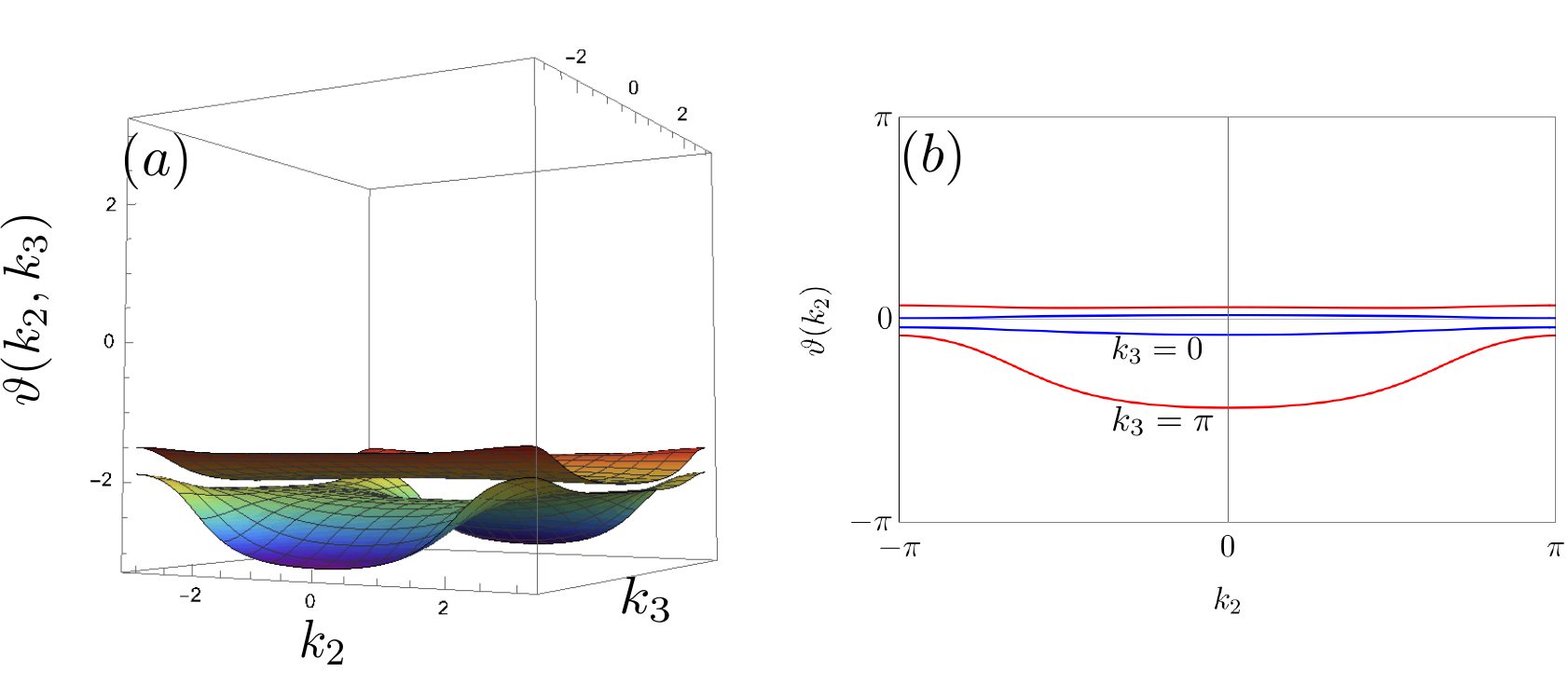}\includegraphics[width=.5\columnwidth]{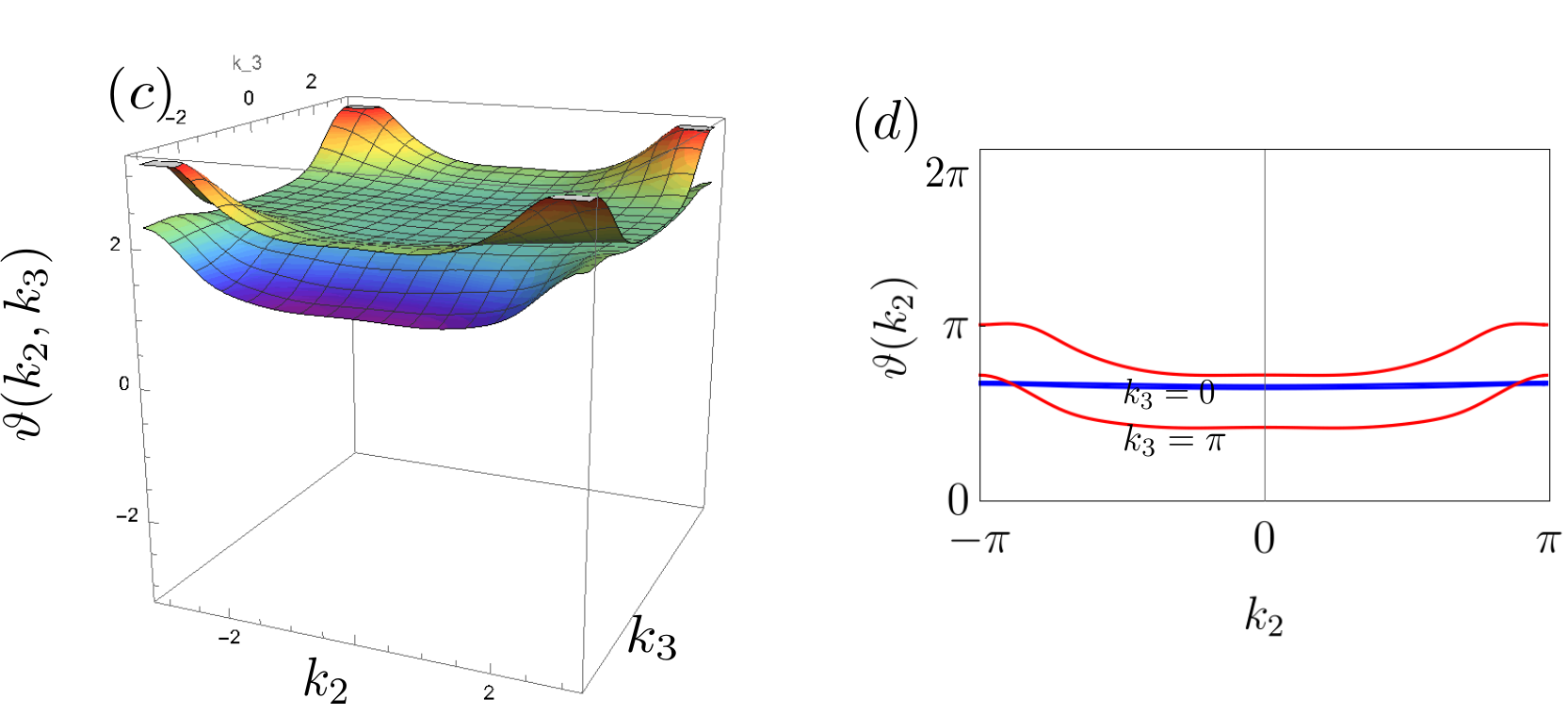}
    \caption{Wilson loops. We compute the non-abelian Wilson loop $W(k_2,k_3) = \exp i \oint dk_1 A_1(\mbf{k})$ for the Cu$_1$ phase (a) and Cu$_2$ phase (c). The Wilson loop phases/Wannier centers $\vartheta(k_2,k_3)$ show no winding, but show greater variation in the $k_z = \pi$ plane. Direct comparison of the Wilson loop spectra at $k_3 = 0$ (blue) and $k_3 = \pi)$ (red) for Cu$_1$ (b) and Cu$_2$ (d). Stronger hybridization with the O bands at $k_3 = \pi$ is responsible for the dispersion of the Wilson loop.
    }
    \label{fig:wilsonloops}
\end{figure}

\begin{figure}[H]
    \centering
    \includegraphics[width=.5\columnwidth]{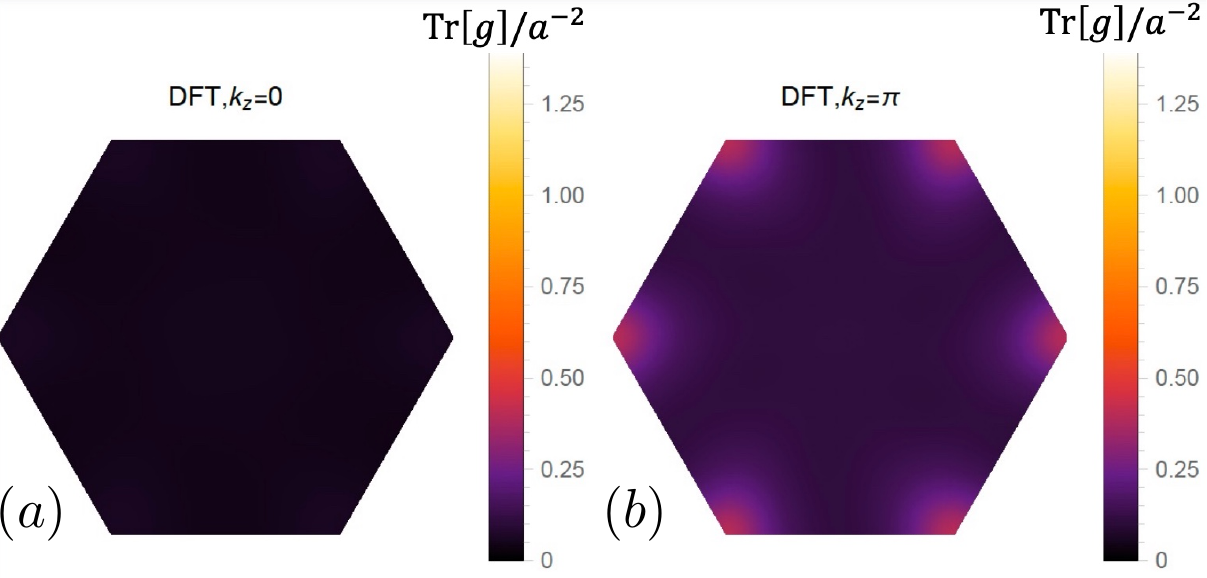}\includegraphics[width=.5\columnwidth]{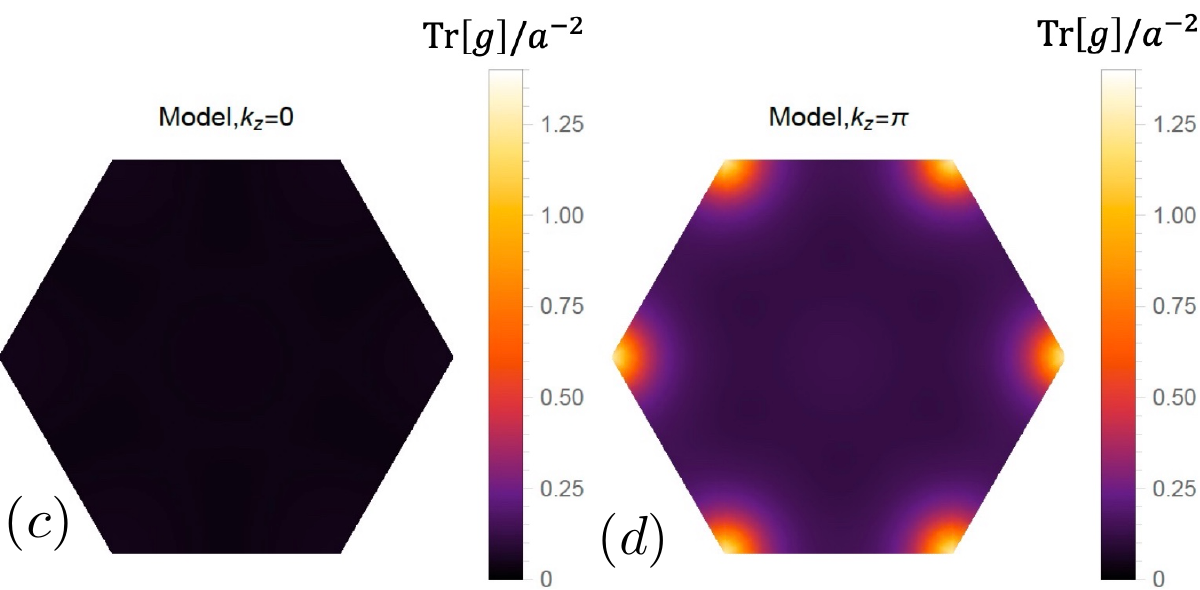}
    \caption{The 2D Fubini-Study metric $g(\mbf{k}) = \frac{1}{2} \Tr (\del_x P\del_x P + \del_y P\del_y P)$ (in unit of $a^2$ of the upper two bands for the DFT calculation and the tight-binding model \Eq{eq:h4band}.
    }
    \label{fig:FSMplotapp}
\end{figure}

In addition to an accurate modeling of the spectrum, it is crucial to study the localization and topological properties of the bands, whose effect on the many-body physics is magnified by a large density of states. To do so, we compute the two-band non-abelian Wilson loop in \Fig{fig:wilsonloops} and the two-band Fubini-Study metric at $k_z = 0,\pi$ in \Fig{fig:FSMplotapp}. Our results are consistent with tight localization, as expected from the atomic band representation formed by the Cu $d$-orbitals. 

We emphasize that, although the two Cu bands together are trivial and form an EBR, the double degeneracy enforced by $C_3$ and $\mathcal{T}$ (the $\Gamma_2\Gamma)3$ and $A_2A_3$ irreps) creates non-trivial quantum geometry within a \emph{single} band. In particular, the Fubini-Study metric will diverge at the touching, as required by the Berry curvature monopole (pointed out in Ref.~\cite{HIR23}) or double Weyl point.

\begin{figure}[H]
    \centering
    \includegraphics[width=.66\columnwidth]{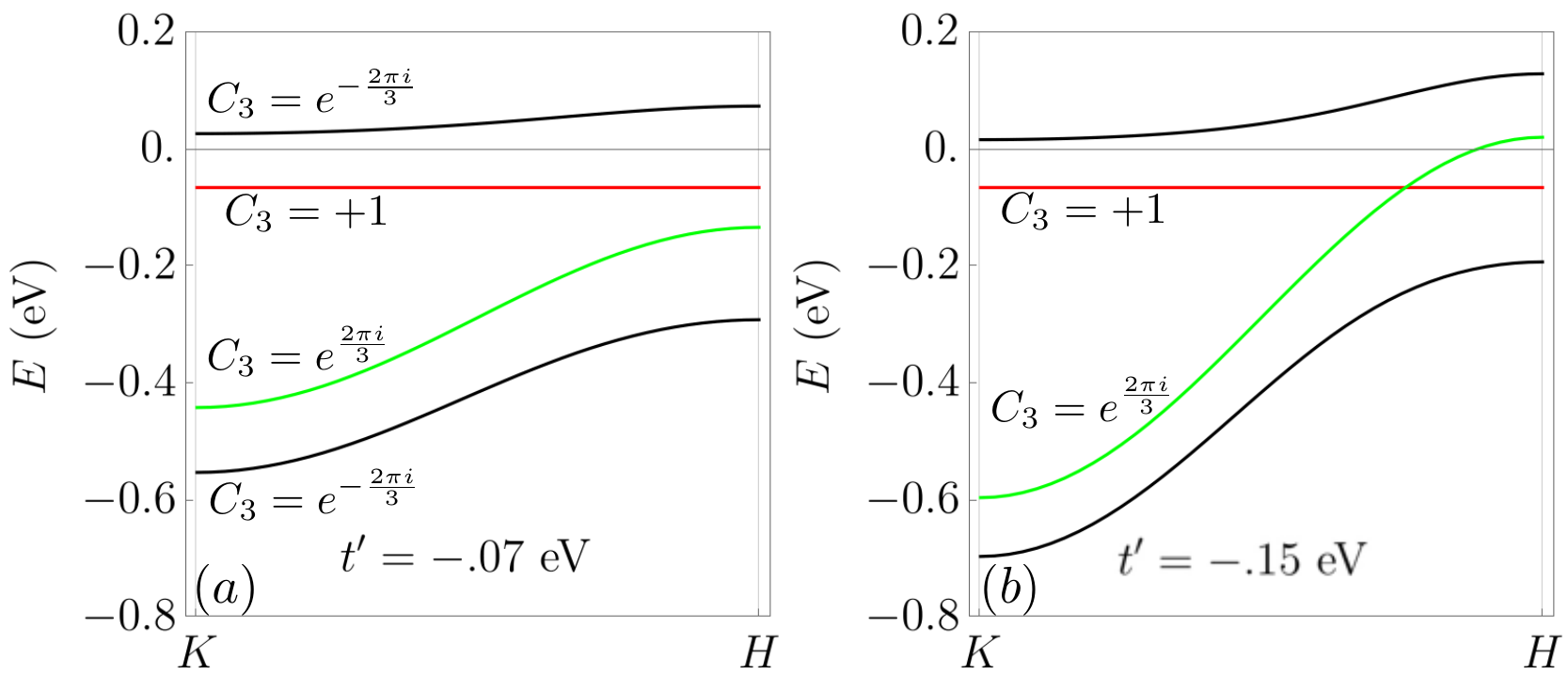} \includegraphics[width=.33\columnwidth]{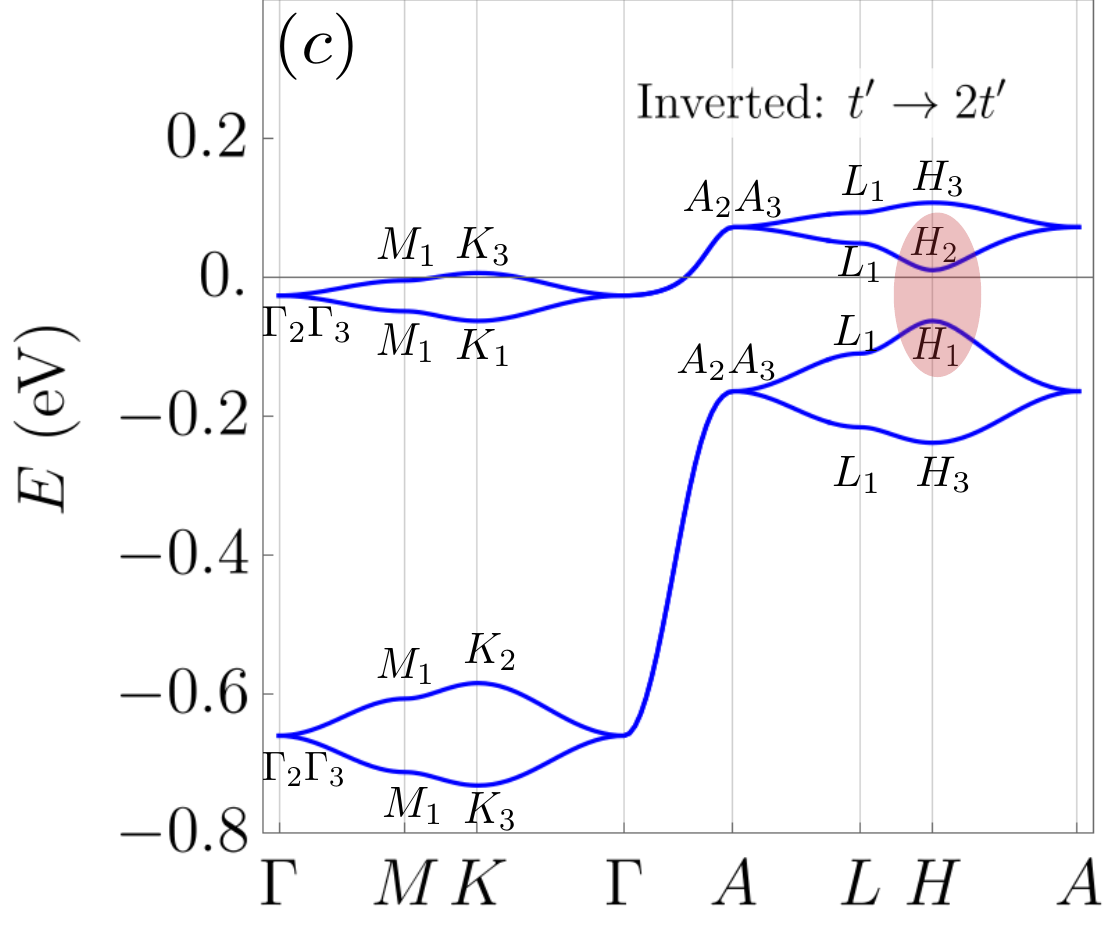}
    \caption{Inverted Semimetal phase obtained for the Cu$_1$ structure by increasing the O out-of-plane hopping $t'_O$ in $h_4(\mbf{k})$. We compare the dispersions of the $C_3$ irreps along the $KH$ line in the normal (a) and inverted (b) phases. The inverted phase is a topological semimetal due to the protected crossing of the different $C_3$ eigenvalues. Note $C_3$ is preserved along all points in $KH$. (c) show the band structure in the inverted phase.
    }
    \label{fig:inverted}
\end{figure}

\Fig{fig:bandswan} shows that Cu and O DFT bands nearly touch at the $H$ point, indicating close proximity to a band inversion. We check that tuning the $t'$ vertical hopping of the O orbitals realixes this band crossing, resulting in a topological phase transition into a semi-metal (see \Fig{fig:inverted}).

\begin{figure}[H]
    \centering
    \includegraphics[width=.75\columnwidth]{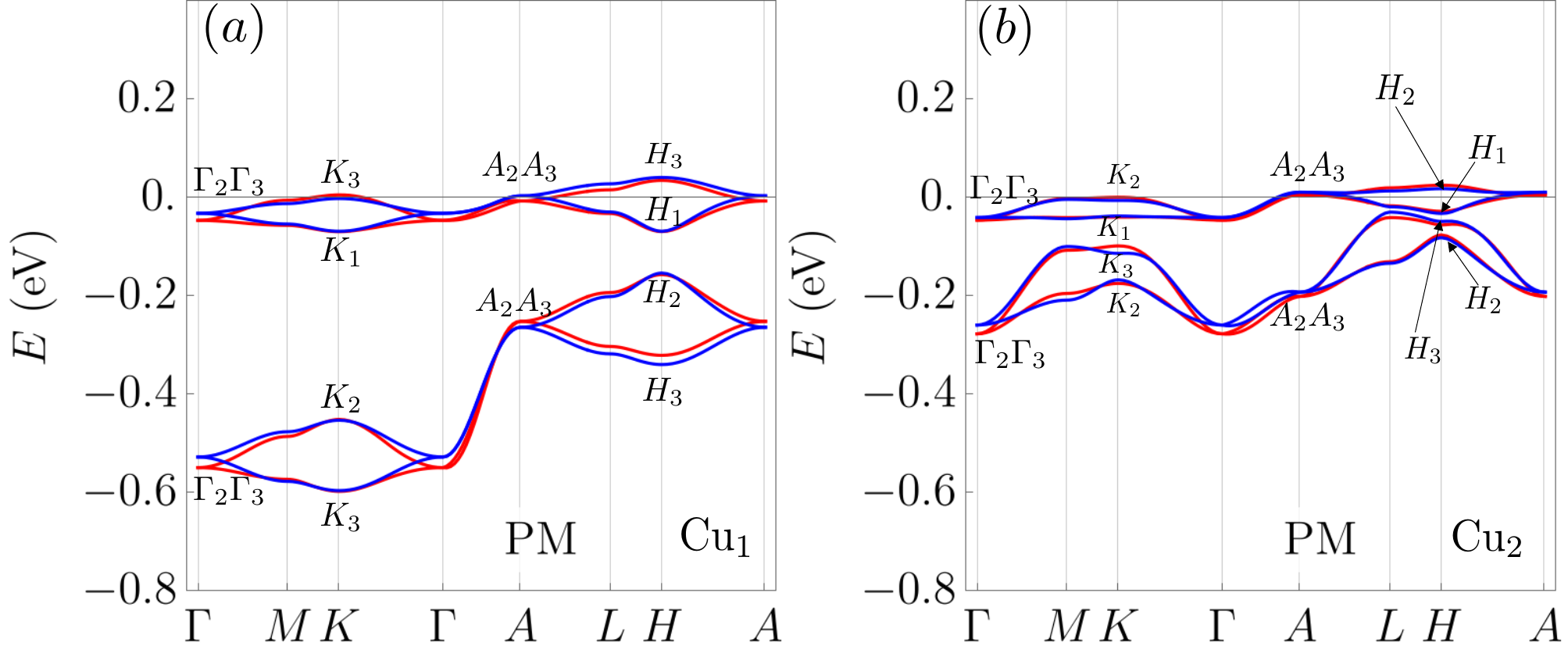} 
    \caption{The ferromagnetic bands are very similar to the paramagnetic bands (see Main Text). The essential difference between the two cases is the Fermi level, which fills $3/4$ of the spin-degenerate Cu bands in the paramagnetic case, but fills $1/2$ of the spin-polarized Cu bands in the ferromagnetic case.
    }
    \label{fig_O_Cu12}
\end{figure}

DFT shows that ferromagnetism (and also a degenerate anti-ferromagnetism) is favored over the paramagnetic (spin-degenerate) single-particle bands given above. We find that the spin-polarized bands at the Fermi surface have essentially identical parameters as \Eq{eq:cu1dopingparamO}, whereas small adjustments in the parameters are required for Cu$_2$ doping:
\bea
\mu_O &= 3.695, \quad t_O = -0.025 (\sigma_0 +\sigma_3)/2, \quad t'_O = -0.022 \sigma_0, \quad t_O^+ = .006 (\sigma_1 - i \sigma_2)/2 , \quad  t_O^- = .003 (\sigma_1 - i \sigma_2)/2 \\
\mu_C &= 3.751, \quad t_C = R_{-2\pi/3} \bpm 0 & 0 \\ .006 & -.002 \epm R_{2\pi/3}, \quad t_C' = -.004 \sigma_0, \quad t_C^+ = 0, \quad t_C^- = -.003 (\sigma_0 + \sigma_3)/2 \\
T &= \left(
\begin{array}{cc}
 -0.005 & 0.021 \\
 -0.013 & 0.005 \\
\end{array}
\right), \quad T_+ =0, \quad T_- = -.006 (\sigma_1 + i \sigma_2)/2 \ .
\eea
Their main effect is to further shrink the gap between the O and Cu bands, as can be seen in \Fig{fig_O_Cu12}

\subsection{ Pb$_{9}$Cu(PO$_4$)$_6$(OH)$_2$ }
\label{app:2bandmodel}

We provide the details of the single-particle Hamiltonians in the Pb$_{9}$Cu(PO$_4$)$_6$(OH)$_2$ structure for both Cu$_1$ and Cu$_2$ doping and in both PM and FM phases.
For this structure, we have space group P3.
In all the cases, we only have two bands near the Fermi level~\figref{fig:2OH_doped_bands}, they are trivial atomic bands given by $d_{xz}$ and $d_{yz}$ orbitals of Cu. (Cu atoms are at 1b for Cu$_1$ and at 1c for Cu$_2$.)
Wanneirization of the DFT bands would directly generate a DFT 2-band tight-binding model, which can generate band structure that perfectly matches the DFT ones.
The DFT 2-band tight-binding model shows that the hoppings longer than NN are small (\ie, less than 1meV), and thus we will build a 2-band NN-hopping model for the two bands in each case. 
Briefly, the form of the model is the same for all cases due to the same symmetries.
The hopping parameter values in the model are the same for the PM and FM phases for one specific Cu doping, though the onsite energies are different.
The hopping values do change across different Cu doping, as indicated by the different bands in \figref{fig:2OH_doped_bands} for different Cu doping.
Details are discussed below.

The 2-band NN hopping model that we build with $d_{xz}$ and $d_{yz}$ at the same position in each unit cell reads
\begin{equation}
\label{eq:H_2band}
H_{2band} = \sum_{\bsl{k}} c^\dagger_{\bsl{k},Cu} \left[ E_{Cu} + 2 t_{z} \cos(k_z c) + \left( \sum_{n=0,1,2} R_{2\pi/3}^n t_{\bsl{a}_1} R_{-2\pi/3}^{n} e^{-\ii (C_3^n \bsl{a}_1) \cdot \bsl{k}} + h.c. \right) \right] c_{\bsl{k},Cu}\ ,
\end{equation}
where 
e\eq{
t_{\bsl{a}_1} = \mat{ t_{xx} & t_{xy} \\ t_{yx} & t_{yy} } 
}
is the hopping matrix along $\bsl{a}_1$, $c^\dagger_{\bsl{k},Cu} = (c^\dagger_{\bsl{k},Cu,d_{xz}},  c^\dagger_{\bsl{k},Cu,d_{yz}} )$.

For Cu$_1$ doping, the parameter values are
\eq{
\label{eq:paramagnetic_2band_parameters}
E_{Cu}= -0.0326\eV\ , \ t_z = -0.0062 \eV\ ,\ t_{xx} = -0.0094 \eV \ ,\ t_{xy} = 0.0066 \eV \ ,\  t_{yx} = -0.0115 \eV, t_{yy} = 0 \ .
}
The in-plane hopping is shown in \figref{fig:hoppings}(b).Compared to the most general symmetry-allowed NN hoppings, we directly neglect the off-diagonal part of the $\bsl{a}_3$ hopping (as it is about 0.5meV, equivalent to maximum band splitting about $2$meV along $\Gamma$-A). We also set $t_{yy}=0$ since it is smaller than 2meV. The band structure from the model \eqnref{eq:H_2band} is shown in \figref{fig:bandswan2band}(a), which has a good agreement with the DFT band structure. The flatness of the bands, which form an indecomposablt atomic representation of the Cu orbitals, is merely due to the small hopping amplitudes/localized Wannier functions. To be specific, the square root of the Wannier spread of each orbital is $2.402 \AA = 0.2469 a = 0.3293 c$, indicating that the Wannier function is very localized. 
Owing to $t_{yx}/t_{xy}=-1.74\neq -1$, the mirror symmetry along $y$ is broken.

\begin{figure}
    \centering
    \includegraphics[width=0.7\columnwidth]{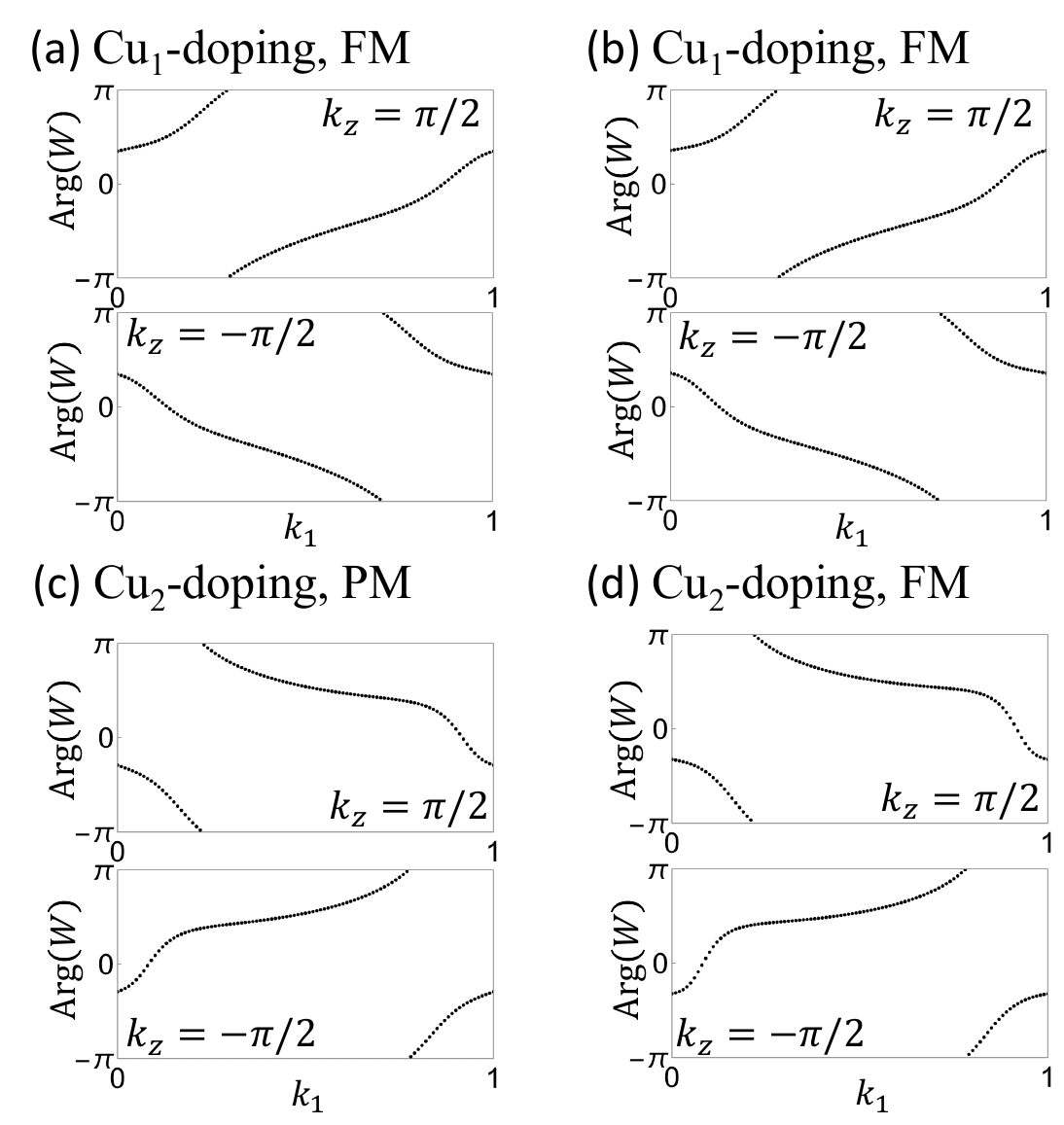}
    \caption{The Wilson loop spetrum of the lower band of the DFT 2-band model in each specified case on $k_z=\pi/2$.
    The Wilson loop $W$ is calcuated along $\bsl{b}_2$, and is ploted as a function of $k_1$ is along $\bsl{b}_1$.
    }
    \label{fig:onebandWLofthe2bandmodel}
\end{figure}

Since this is a 2-band model, the geometric properties (\eg, Fubini-Study metric) of the two bands, taken together, vanishes.
However, we may look at the Wilson loop of the one of the two bands, for which we choose lower band.
The Wilson loop is ill-defiend for $k_z=0$ and $k_z=\pi$ planes, since the two bands touch on the two planes at $\Gamma$ or A (enforced by symmetries), and thus we look at $k_z=\pm \pi/2$ for the DFT 2-band tight-binding model, which is shown in \figref{fig:onebandWLofthe2bandmodel}(a).
Clearly, the lower band of the DFT 2-band model has nonzero and opposite Chern numbers at $k_z=\pm \pi/2$, showing that the gapless points at $\Gamma$ and A are double Weyl points with chirality $\pm 2$.
However, in our simplified model (\eqnref{eq:H_2band}), we neglect the band splitting along $\Gamma$-A, since it is very small. 
Although such simplfiication merge the two double Weyl points into an accidental nodal line along $\Gamma$-A, it would be convenience for later study of correlated phases based on our model, since it make the eigenvectors of the Hamiltonian independent of $k_z$.

Now we discuss the ferromagnetic case for Cu$_1$ doping. With ferromagnetism, only the spin-down bands remain near the Fermi level~\figref{fig:2OH_doped_bands}.
The DFT spin-down bands are very close to the DFT paramagnetic bands as shown in \figref{fig:bandswan2band}(a,b). The effect of FM is approximately just a shift of the onsite energy for bands near Fermi level, as the hopping values for the DFT FM spin-down two bands are extremely similar to those for the DFT PM two bands (difference smaller $0.1meV$) except of a considerable shift of the onsite energy.
Therefore, we can build a 2-band NN-hopping model that has the same form as \eqnref{eq:H_2band}, and choose the same hopping parameter values as \eqnref{eq:paramagnetic_2band_parameters}, while shifting onsite energy to 
\eq{
\label{eq:paramagnetic_2band_parameters_FM}
E_{Cu}= 0.008907\eV\ .
}
Again, the band structure from the model has a good agreement with the DFT band structure, as shown in \figref{fig:bandswan2band}.
Again, the DFT 2-band tight-binding model has double Weyl points with chirality $\pm 2$ at $\Gamma$ and A (\figref{fig:onebandWLofthe2bandmodel}(b)), which we choose to merge into an accidental nodal line along $\Gamma$-A in our simplified model.

We now discuss the Cu$_2$ doping parameters, which are very similar to the Cu$_1$ doping. A key difference is that the Cu is now at 1c, but this is simply a choice of origin in the two-band model we build and does not affect the parameterization of the Hamiltonian.
In the PM phase, the hopping parameter values read
\eqa{
\label{eq:paramagnetic_2band_parameters_Cu_2}
& E_{Cu}=-0.0207\eV,\ t_{z} = -0.0065\eV,\ t_{xx} =  \frac{1}{4} \left(3 t_{2}-\sqrt{3} t_{1}\right),\ t_{xy} = \frac{1}{4} \left(-3 t_{1}-\sqrt{3} t_{2}\right) \\ 
& t_{yx} = \frac{1}{4} \left(t_{1}-\sqrt{3} t_{2}\right),t_{yy} = \frac{1}{4} \left(\sqrt{3} t_{1}+t_{2}\right)
}
with 
\eq{
t_1 =  0.0112\eV\ , \ t_2 = -0.0027\eV\ .
}
The reason for us to parameterize the in-plane hoppings by only $t_1$ and $t_2$, because along $\bsl{a}_2$ direction the hopping matrix approximately reads
\eq{
R_{2\pi/3}t_{\bsl{a}_1} R_{-2\pi/3} = t_{\bsl{a}_2} = \mat{ 0 & 0 \\ t_1 & t_2 }
}
according to the Wannierization of the DFT data, where the neglected elements are smaller than 1meV. 
In particular, $t_{xy}= -0.0072\eV$ and $t_{yx} = 0.0040\eV$ indicate the breaking of the mirror as $y$ since $t_{xy}/t_{yx}= -1.81\neq -1$.
The band structure from the model \eqnref{eq:H_2band} again has a good agreement with the DFT band structure, as shown in \figref{fig:bandswan2band}(b).
Small hopping again the small spread of the Wannier function of the basis: the square root of the Wannier spread of each orbital is $2.271 \AA = 0.2334 a = 0.3113 c$.
Again, the DFT 2-band tight-binding model has double Weyl points with chirality $\pm 2$ at $\Gamma$ and A (\figref{fig:onebandWLofthe2bandmodel}(c)), which we choose to merge into an accidental nodal line along $\Gamma$-A in our simplified model.

In the ferromagnetic case, the model parameters again only defer from \eqnref{eq:paramagnetic_2band_parameters_Cu_2} by an onsite energy shift:
\eq{
\label{eq:paramagnetic_2band_parameters_Cu_2_FM}
E_{Cu}= 0.0023\eV\ .
}
Again, the band structure from the model has a good agreement with the DFT band structure, as shown in \figref{fig:bandswan2band}(d).
Again, the DFT 2-band tight-binding model has double Weyl points with chirality $\pm 2$ at $\Gamma$ and A (\figref{fig:onebandWLofthe2bandmodel}(d)), which we choose to merge into an accidental nodal line along $\Gamma$-A in our simplified model.

\section{Experimental Details}
\label{app:ExpDet}

In this Appendix, we give detailed information on synthesis and characterization of lead apatites.
\subsection{Synthesis}
\begin{figure}[H]
    \centering
    \includegraphics[width=\columnwidth]{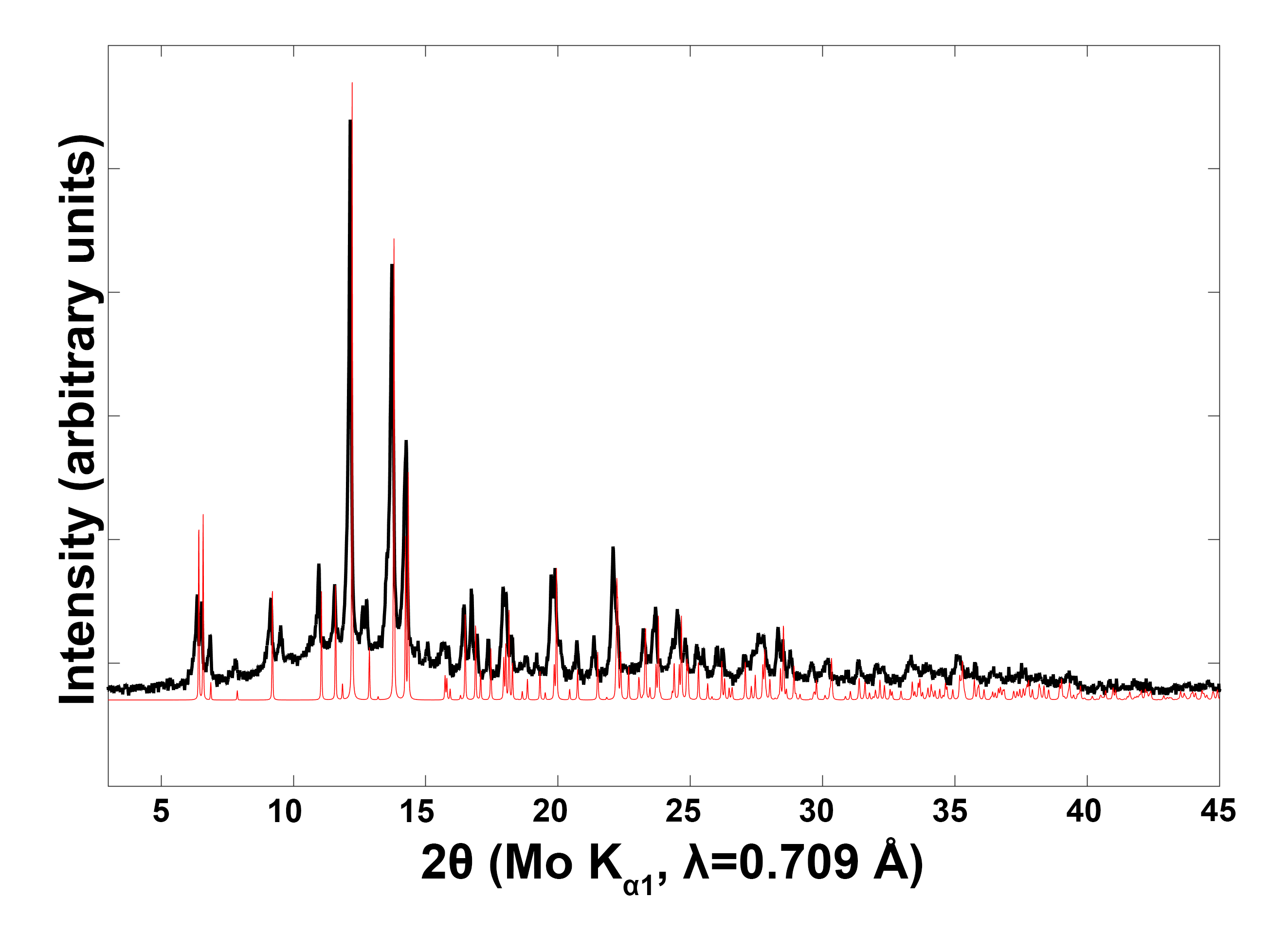}
    \caption{Comparisons of experimental (black) and calculated powder X-ray diffraction patters of Pb$_2$(SO$_4$)O (red).}
    \label{fig:Pb2SO5}
\end{figure}

Well homogenized powders of Pb$_2$(SO$_4$)O were synthesized by grinding stoichiometric amounts of PbO (Sigma Aldrich, $>$99.0 percent) and PbSO$_4$ (Sigma Aldrich, $>$99.0 percent) in an agate mortar and pestle. The mixture was loaded in alumina crucibles in 3g amounts and then placed in a quartz tube and sealed under dynamic vacuum. Experiments found that sealing with oxygen present (i.e. not backfilling with Ar) resulted in the highest purity of product. Sample purity was confirmed via powder X-ray diffraction using a STOE Stadi P powder X-ray diffractometer equipped with a Mo K$\alpha$ ($\lambda$$=0.71073$ \AA) sealed-tube X-ray source and graphite monochromator at room temperature in transmission geometry \Fig{fig:Pb2SO5}

\begin{figure}[H]
    \centering
    \includegraphics[width=\columnwidth]{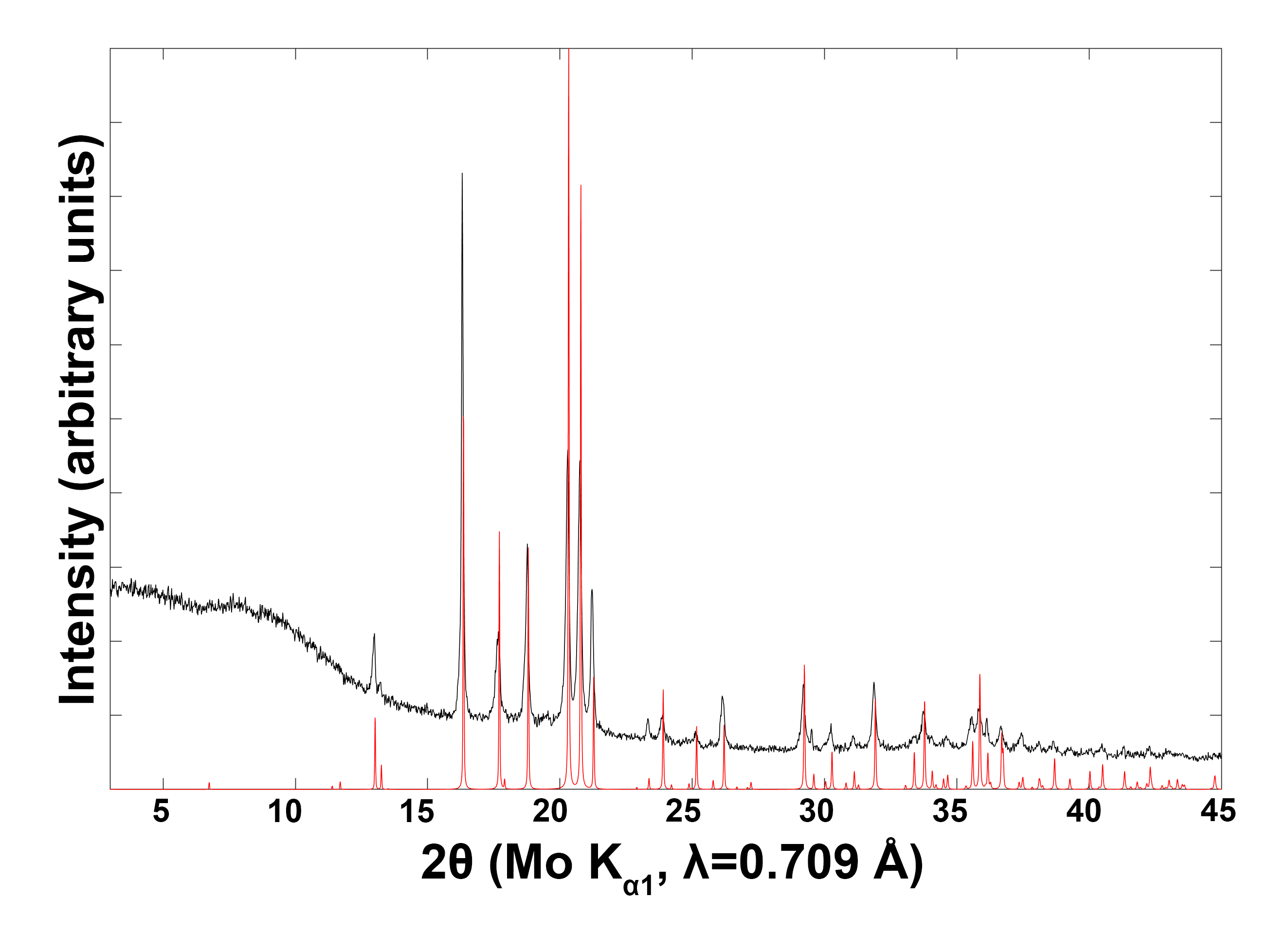}
    \caption{Comparisons of experimental (black) and calculated powder X-ray diffraction patters of Cu$_3$P (red).}
    \label{fig:Cu3P}
\end{figure}

Cu$_3$P was grown phase pure by mixing elemental Cu (Alfa Aesar, -170+270 mesh, 99.9 percent) and elemental red Phosphorus (Alfa Aesar, 99.999+ percent) in a 3:1 stoichiometric ratio. The sample was placed in an alumina crucible and subsequently sealed in a quartz tube. For this synthetic step the quartz tube was evacuated three times and backfilled with argon to prevent any oxidation before sealing. The tube was then loaded in a furnace heated first to 450 $^{\circ}$C over 2 hours and held there for 3 hours to prereact the phosphorus. The furnace was then ramped to 1050 $^{\circ}$C over 4 hours, kept at this temperature for 8 hours, and then shut off to cool quickly. This saved nearly 24 hours in synthesis time compared to the original synthesis and results in a phase pure product of the target compound, collected in transmission geometry on the same instrument \Fig{fig:Cu3P}.

Finally, a 1:1 molar ratio of Cu$_3$P and Pb$_2$(SO$_4$)O were ground into a fine powder in mortar and pestle. The powders were once again loaded into an alumina crucible, placed in a quartz tube, and sealed under dynamic vacuum. The sample was then heated to 950 $^{\circ}$C, kept at this temperature for 6hrs, and then shut off to cool quickly.

\begin{figure}[H]
    \centering
    \includegraphics[width=\columnwidth]{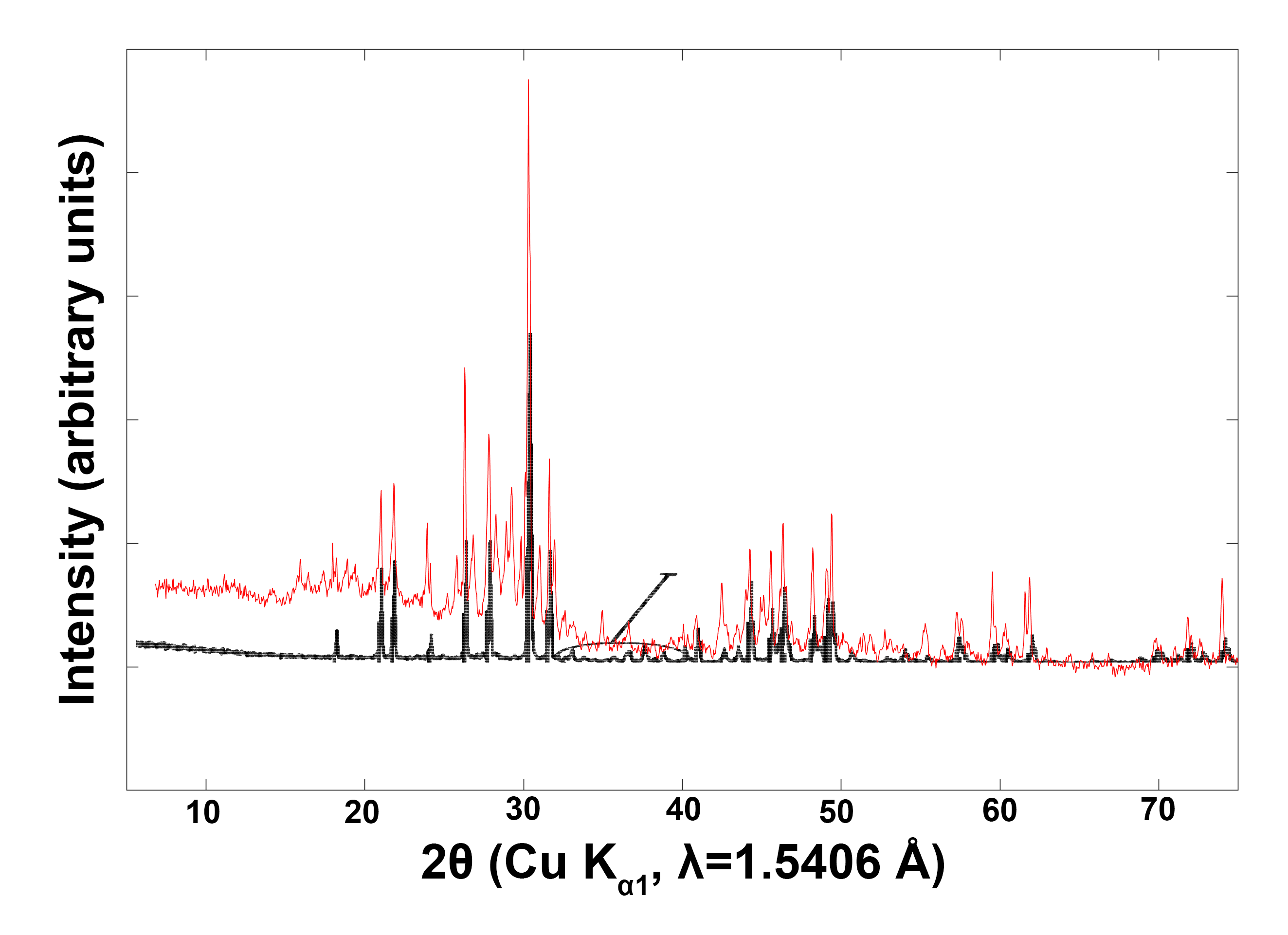}
    \caption{Comparisons of previously reported (black) and collected powder X-ray data for proposed Pb$_{10-x}$Cu$_x$(PO$_4$)$_6$O (red).}
    \label{fig:ExOver}
\end{figure}

\begin{figure}[H]
    \centering
    \includegraphics[width=0.5\columnwidth]{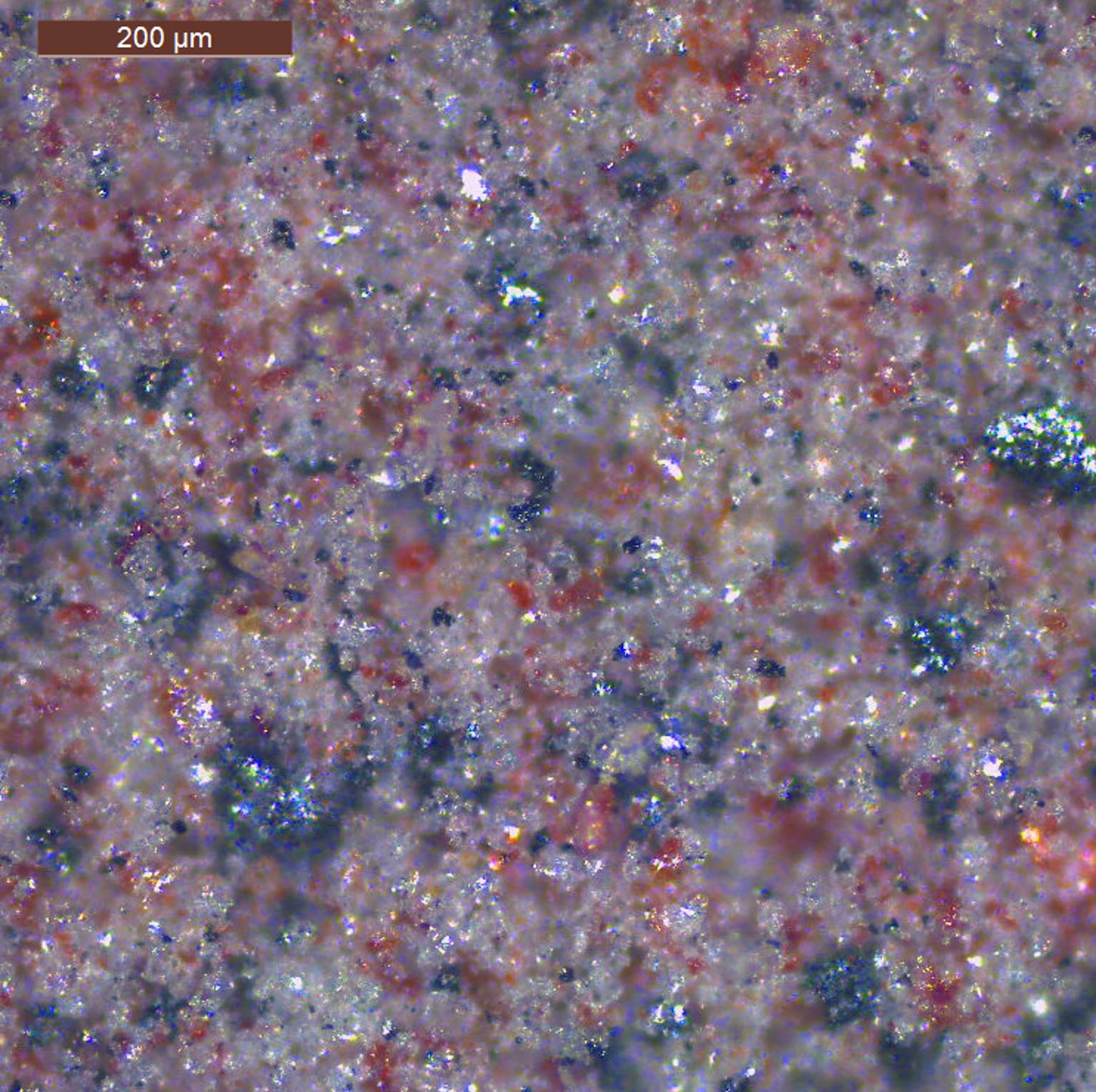}
    \caption{A representative sample of the final product phase, ground into a powder. We observe at least 3 phases with seemingly different properties.}
    \label{fig:Multiphase}
\end{figure}

\begin{figure}[H]
    \centering
    \includegraphics[width=\columnwidth]{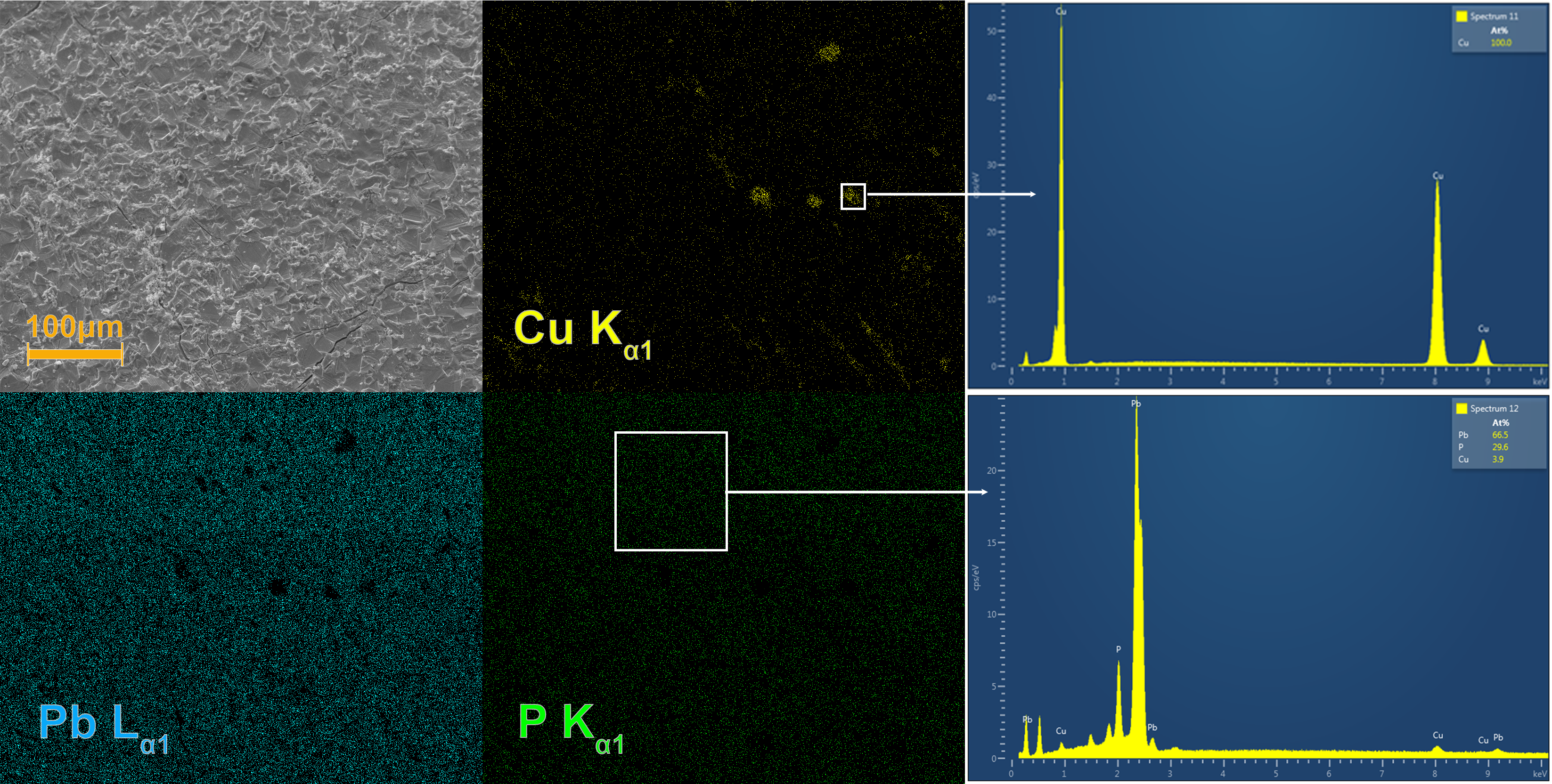}
    \caption{EDS of the red and white transparent phase showing pockets of Cu embedded in a phase consistent with Pb$_{10-x}$Cu$_x$(PO$_4$)$_6$(OH)(SH).}
    \label{fig:redandwhite}
\end{figure}

\begin{figure}[H]
    \centering
    \includegraphics[width=\columnwidth]{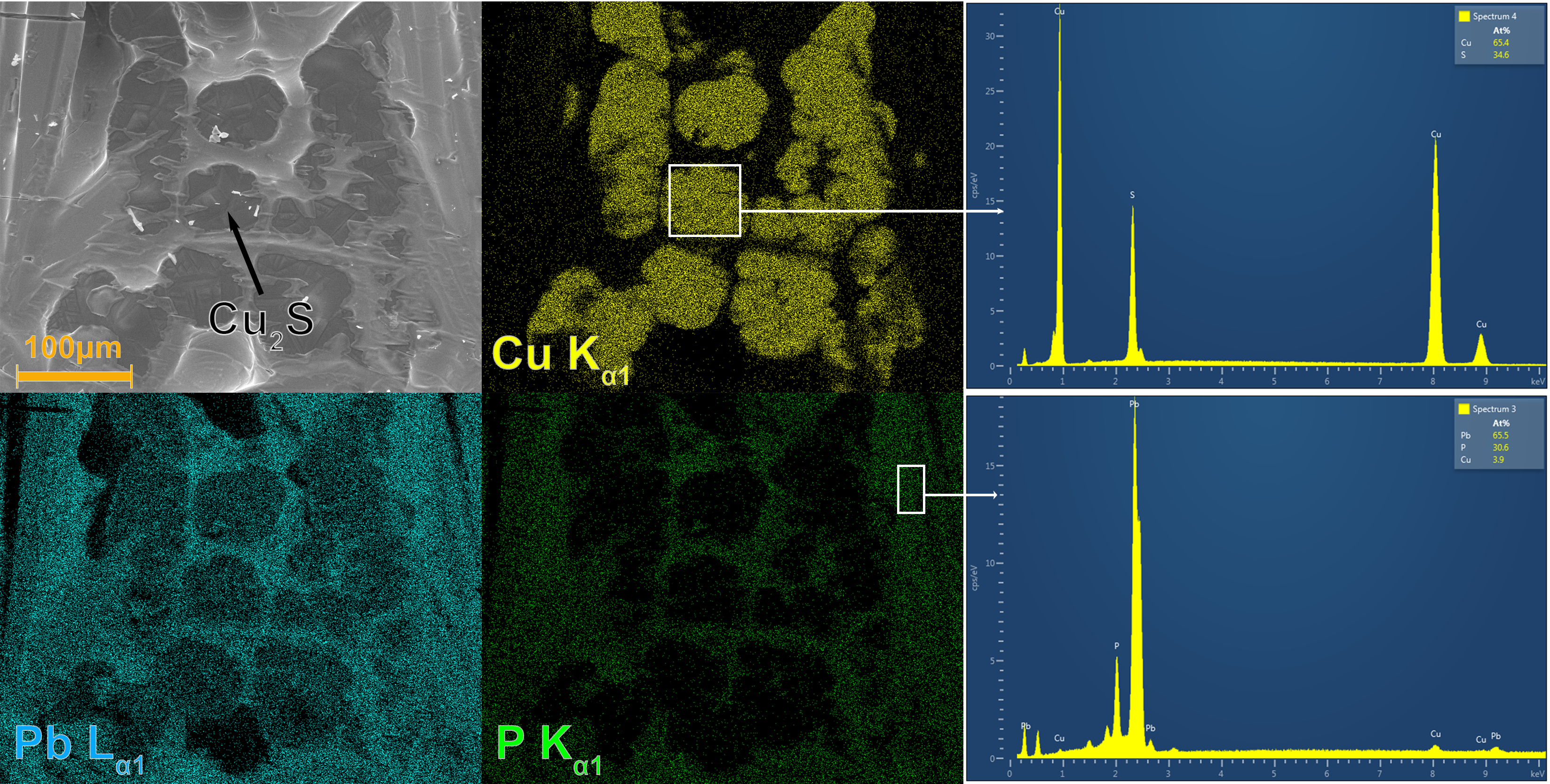}
    \caption{EDS of the metallic gray phase showing Cu$_2$S embedded in a phase consistent with Pb$_{10-x}$Cu$_x$(PO$_4$)$_6$(OH)(SH).}
    \label{fig:metallic}
\end{figure}

\subsection{Refinement}
Calculated precession images for data in refined herein are shown below.
\begin{figure}[H]
    \centering
    \includegraphics[width=\columnwidth]{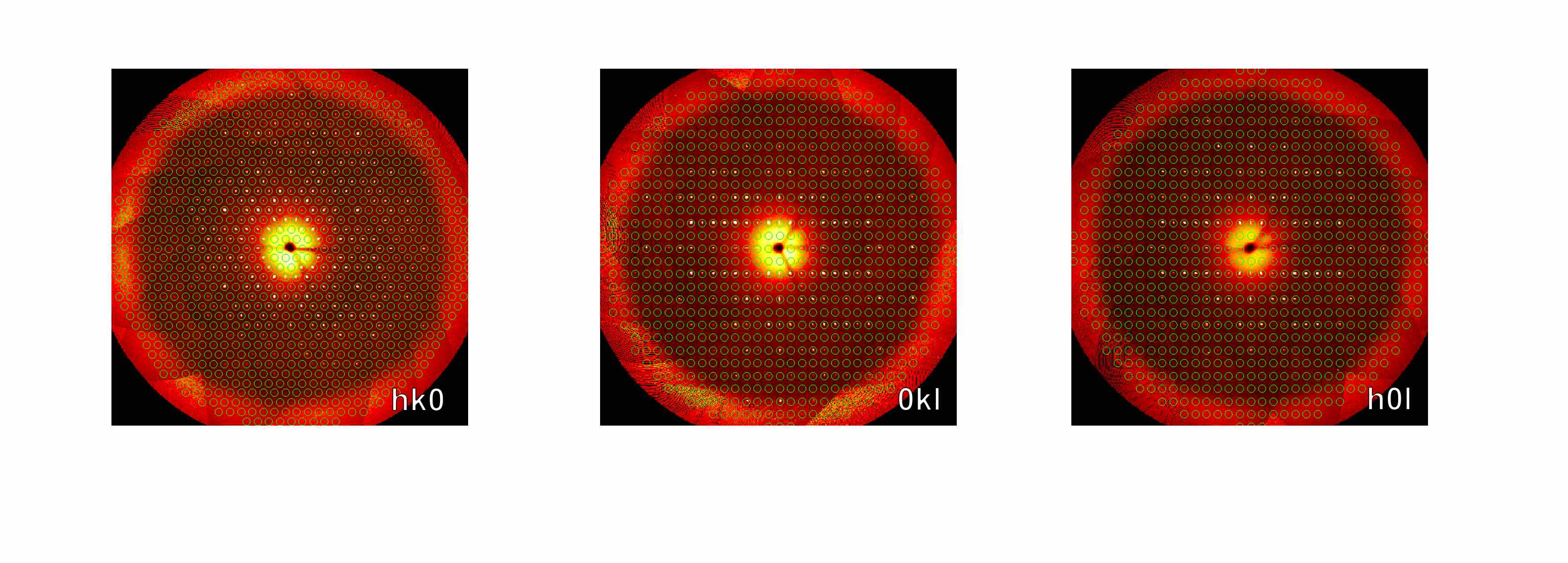}
    \caption{Precession images for a thin transparent crystal. The only systematic absences are [00l]=2n agreeing with the space group P6$_3$/$m$.}
    \label{fig:Prec}
\end{figure}
Collection and integration statistics for the full hemisphere data set are shown below.

\begin{center}
\begin{table*}
\renewcommand{\tablename}{Table S}
\caption{Collection and integration statistics for lead appetite structures.}
\begin{tabular}{@{}lllll@{}}
\hline
Crystal Dimension (mm)                                          & 0.55 $\times$ 0.092 $\times$ 0.176 mm      &  &  &  \\
Temperature (K)                                                 & 298(2)                                   &  &  &  \\
Radiation Source, $\lambda$ (\AA)                               &  Mo K$\alpha$, $\lambda$=0.71073 \AA  &  &  &  \\
Absorption Correction                                           & analytical                            &  &  &  \\
Space Group                                                     & P6$_3$/$m$                                  &  &  &  \\
$a$ (\AA)                                                       & 9.8508(1)                            &  &  &  \\
$c$ (\AA)                                                       & 7.4395(2)                            &  &  &  \\
Cell Volume (\AA$^{3}$)                                         &  625.198(19)                          &  &  &  \\
Absorption Coefficient (mm$^{-1}$)                              &  67.952                                &  &  &  \\
$\theta_{min}$ , $\theta_{max}$                                 & 2.39, 36.34                           &  &  &  \\
Number of Reflections                                           & 32013                                 &  &  &  \\
Unique Reflections (I \textgreater 3$\sigma$, all)              & 763, 1066                              &  &  &  \\
Rint(I\textgreater{}3$\sigma$, all)                             & 6.54, 7.40                            &  &  &  \\
\hline
\end{tabular}
\end{table*}

\end{center}

Freely refining occupancies for both an O at [0 0 $1/4$] and S located at [0 0 0] results in site occupancy factors of 0.54 and 0.56, respectively. The freely refined occupancies having a summation very close to 1 seems suggestive that this could be a physically real interpretation of the structural solution. After refining the thermal parameters anisotropically, a final refinement converges with a goodness of fit (GoF) parameter of 1.22 and R value of 3.94 compared to all reflection. Here, our site occupancies for O (1.08) and S (1.12) sum to above 1. If we decide to restrict the occupancies of the 2$b$ S and 2$a$ O to sum to a site occupancy factor of 1, we end refinement with a final composition Pb$_{10}$(PO$_4$)$_6$(OH)$_{0.94}$(SH)$_{1.06}$. This refinement has nominally the same refinement statistics (GoF(all) $= 1.22$, R(all) $=3.95$). 

\begin{center}

\begin{table*}
\renewcommand{\tablename}{Table S}
\caption{Refinement Statistics for lead appetite structures without Cu doping.}
\begin{tabular}{@{}lllll@{}}
\hline
Refined   Composition     & Pb$_{10}$(PO$_4$)$_6$(OH)$_{0.94}$(SH)$_{1.06}$   &&& Pb$_{10}$(PO$_4$)$_6$(OH) $_{1.12}$ (SH) $_{1.08}$ \\ 
Refinement Method                                          & { F$^{2}$} &&&  { F$^{2}$} \\

Number of Parameters                                            & 43                                   &  &  &  44 \\
R(I\textgreater{}$3\sigma$), R$_{w}$(I\textgreater{}$3\sigma$)  & 2.48,5.43                            &&& 2.47, 5.43 	\\
R(all), R$_{w}$(all)                                            & 3.95, 5.70                           &&& 	 3.94, 5.69	  \\
S(I\textgreater{}$3\sigma$), S(all)                             & 1.39, 1.22                    &&&  1.39, 1.22	\\
$\Delta\rho_{max}$ , $\Delta\rho_{min}$ (e \AA$^{-3}$)          & 3.38, -3.19                          &&& 	3.35, -3.21	 \\ 
\hline
\end{tabular}
\end{table*}

\end{center}

\begin{center}
\begin{table*}
\renewcommand{\tablename}{Table S}
\caption{Refined atomic coordinates for Pb$_{10}$(PO$_4$)$_6$(OH)$_{0.94}$(SH)$_{1.06}$.}
\begin{tabular}{@{}lllllllllllll@{}}
\hline
Site && Wyckoff Position && x   && y    && z          && Occupancy &&  \\  
\hline
Pb1  && 6h       && 0.24486(4) && 0.24667(4) && $1/4$ && 1         &&  \\
Pb2  && 4f               && $2/3$ && $1/3$ && -0.00339(4)        && 1         &&  \\
P1  && 6h               &&  0.3729(2)  && 0.4020(2) && $3/4$ && 1         &&  \\ 
O1  && 6h               &&  0.4851(7)  && 0.3366(8) && $3/4$ && 1         &&  \\ 
O2  && 12i               &&  0.2650(6)  && 0.3480(6) && 0.5834(6) && 1         &&  \\ 
O3  && 6h               &&  0.5337(8)  && 0.4166(7) && $1/4$ && 1         &&  \\ 
S1  && 2b               &&  0  && 0 && 0 && 0.53         &&  \\ 
O4  && 2a              &&  0  && 0 && $1/4$ && 0.47         &&  \\ 
\hline
\end{tabular}
\end{table*}
\end{center}
\begin{center}

\begin{table*}
\renewcommand{\tablename}{Table S}
\caption{Refined anisotropic displacement parameters for Pb$_{10}$(PO$_4$)$_6$(OH)$_{0.94}$(SH)$_{1.06}$.}
\begin{tabular}{lllllllllllllll}
\hline
Site && U$_{11}$ && U$_{22}$   && U$_{33}$    && U$_{12}$ && U$_{13}$         && U$_{23}$ &&  \\
\hline
Pb1 && 0.02052(15) && 0.01996(15) && 0.02682(15) && 0.01447(12) && 0.00000 && 0.00000 && \\ 
Pb2 && 0.01890(12) && 0.01890(12) && 0.01231(13) && 0.00945(6) && 0.00000 && 0.00000  && \\
P1 && 0.0088(7) && 0.0096(7) && 0.0125(7) && 0.0055(6) && 0.00000 && 0.00000 && \\ 
O1 && 0.022(3) && 0.028(3) && 0.024(3) && 0.020(3) && 0.00000 && 0.00000 && \\ 
O2 && 0.022(2) && 0.040(3) && 0.0212(19) && 0.018(2) && -0.0064(17) && -0.0101(19) && \\ 
O3 && 0.029(3) && 0.010(2) && 0.034(3) && 0.006(2) && 0.00000 && 0.00000 && \\ 
S1 && 0.015(2) && 0.015(2) && 0.044(5) && 0.0077(11) && 0.00000 && 0.00000 && \\ 
O4 && 0.014(6) && 0.014(6) && 0.054(16) && 0.007(3) && 0.00000 && 0.00000 &&\\
 \hline 
\end{tabular}
\end{table*}
\end{center}

\begin{center}
\begin{table*}
\renewcommand{\tablename}{Table S}
\caption{Refined atomic coordinates for Pb$_{10}$(PO$_4$)$_6$(OH)$_{1.12}$(SH)$_{1.08}.$}
\begin{tabular}{@{}lllllllllllll@{}}
\hline
Site && Wyckoff Position && x   && y    && z          && Occupancy &&  \\  
\hline
Pb1  && 6h       && 0.24486(4) && 0.24667(4) && $1/4$ && 1         &&  \\
Pb2  && 4f               && $2/3$ && $1/3$ && -0.00339(4)        && 1         &&  \\
P1  && 6h               &&  0.3729(2)  && 0.4020(2) && $3/4$ && 1         &&  \\ 
O1  && 6h               &&  0.4851(7)  && 0.3366(8) && $3/4$ && 1         &&  \\ 
O2  && 12i               &&  0.2650(6)  && 0.3480(6) && 0.5834(6) && 1         &&  \\ 
O3  && 6h               &&  0.5337(8)  && 0.4166(7) && $1/4$ && 1         &&  \\ 
S1  && 2b               &&  0  && 0 && 0 && 0.54         &&  \\ 
O4  && 2a              &&  0  && 0 && $1/4$ && 0.56         &&  \\ 

\hline
\end{tabular}
\end{table*}
\end{center}

\begin{center}
\begin{table*}
\renewcommand{\tablename}{Table S}
\caption{Refined anisotropic displacement parameters for Pb$_{10}$(PO$_4$)$_6$(OH)$_{1.12}$(SH)$_{1.08}.$}
\begin{tabular}{lllllllllllllll}
\hline
Site && U$_{11}$ && U$_{22}$   && U$_{33}$    && U$_{12}$ && U$_{13}$         && U$_{23}$ &&  \\
\hline
Pb1   && 0.02052(15)               && 0.01996(8) && 0.00796(9) && 0.00000 && 0.00000         && 0.00000 && \\
Pb2   && 0.00687(11)               && 0.00721(10) && 0.00817(11) && 0.00000        && 0.00000         && 0.00000 && \\
P1  && 0.00854(10)               && 0.00664(10)   && 0.00948(11) && 0.00000 && 0.00000         && 0.00000 && \\
O1   && 0.00854(10)               && 0.00664(10)   && 0.00948(11) && 0.00000 && 0.00000         && 0.00000 && \\
O2   && 0.00854(10)               && 0.00664(10)   && 0.00948(11) && 0.00000 && 0.00000         && 0.00000 && \\
O3   && 0.00854(10)               && 0.00664(10)   && 0.00948(11) && 0.00000 && 0.00000         && 0.00000 && \\
S1   && 0.00854(10)               && 0.00664(10)   && 0.00948(11) && 0.00000 && 0.00000         && 0.00000 && \\
O4   && 0.00854(10)               && 0.00664(10)   && 0.00948(11) && 0.00000 && 0.00000         && 0.00000 && \\
\hline 
\end{tabular}
\end{table*}

\end{center}

We also investigated two ways of possible Cu doping within our refinement: one in which the Cu atom substitutes on the Pb lattice site, and another which the Cu atom is inserted along the chain.  Attempts to dope both Pb sites with Cu results in a refined composition of Pb$_{9.55}$Cu$_{0.45}$(PO$_4$)$_6$(OH)$_{0.94}$(SH)$_{1.06}$ with similar statistics (GoF$= 1.21$, R $= 3.93$). We found the Cu has to be added in as a split site after Pb position has been refined completely. The last two steps of occupancy and anisotropic parameter refinement had to be done by restraining the Pb/Cu split site position with automatic refining keys switched off. 

We can also obtain a reasonable refinement to the data if we place Cu into the channel with composition Pb$_{10}$(PO$_4$)$_6$(OH) $_{1.11}$ Cu$_{0.49}$  (GoF $= 1.22$, R $= 3.93$). Still, as the crystals are transparent, charge balance needs to be maintained and thus this last solution is not chemically reasonable.
\begin{center}

\begin{table*}
\renewcommand{\tablename}{Table S}
\caption{Refinement Statistics for lead appetite structures with Cu doping.}
\begin{tabular}{@{}lllll@{}}
\hline
Refined   Composition     & Pb$_{9.55}$Cu$_{0.45}$(PO$_4$)$_6$(OH)$_{0.94}$(SH)$_{1.06}$  &&&  Pb$_{10}$(PO$_4$)$_6$(OH) $_{1.11}$ Cu$_{0.49}$ \\ 
Refinement Method                                          & { F$^{2}$} &&&  { F$^{2}$} \\

Number of Parameters                                            & 44                                    &  &  &  42  \\
R(I\textgreater{}$3\sigma$), R$_{w}$(I\textgreater{}$3\sigma$)  	&   	2.45, 5.41  &&&   2.46, 5.42 \\
R(all), R$_{w}$(all)                                            	&  3.93, 5.67	&&&  3.94, 5.68 \\
S(I\textgreater{}$3\sigma$), S(all)                            	& 1.38, 1.21	 &&& 1.39, 1.22  \\
$\Delta\rho_{max}$ , $\Delta\rho_{min}$ (e \AA$^{-3}$)           & 3.34,  -3.09	  &&&  3.38,  -3.17 \\ 
\hline
\end{tabular}
\end{table*}

\end{center}

\begin{center}
\begin{table*}
\renewcommand{\tablename}{Table S}
\caption{Refined atomic coordinates for Pb$_{9.55}$Cu$_{0.45}$(PO$_4$)$_6$(OH)$_{0.94}$(SH)$_{1.06}$ .}
\begin{tabular}{@{}lllllllllllll@{}}
\hline
Site && Wyckoff Position && x   && y    && z          && Occupancy &&  \\  
\hline
Pb1  && 6h       && 0.24487 && 0.24674 && $1/4$ && 0.956         &&  \\
Cu1'  && 6h       && 0.24487 && 0.24674 && $1/4$ && 0.044         &&  \\
Pb2  && 4f               && $2/3$ && $1/3$ && -0.00344        && 0.953        &&  \\
Cu2'  && 4f               && $2/3$ && $1/3$ && -0.00344        && 0.047         &&  \\
P1  && 6h               &&  0.3728(2)  && 0.4019(2) && $3/4$ && 1         &&  \\ 
O1  && 6h               &&  0.4853(7)  && 0.3370(8) && $3/4$ && 1         &&  \\ 
O2  && 12i               &&  0.2652(6)  && 0.3480(6) && 0.5834(6) && 1         &&  \\ 
O3  && 6h               &&  0.5341(8)  && 0.4168(7) && $1/4$ && 1         &&  \\ 
S1  && 2b               &&  0  && 0 && 0 && 0.53         &&  \\ 
O4  && 2a              &&  0  && 0 && $1/4$ && 0.47         &&  \\ 
\hline
\end{tabular}
\end{table*}
\end{center}
\begin{center}

\begin{table*}
\renewcommand{\tablename}{Table S}
\caption{Refined anisotropic displacement parameters for Pb$_{9.55}$Cu$_{0.45}$(PO$_4$)$_6$(OH)$_{0.94}$(SH)$_{1.06}$.}
\begin{tabular}{lllllllllllllll}
\hline
Site && U$_{11}$ && U$_{22}$   && U$_{33}$    && U$_{12}$ && U$_{13}$         && U$_{23}$ &&  \\
\hline
Pb1 &&  0.02041(17) &&  0.01986(17)  && 0.02671(17)  && 0.01442(13) && 0.00000  && 0.00000  && \\
Cu1' &&  0.02041(17)  && 0.01986(17)  && 0.02671(17)  && 0.01442(13) &&  0.00000  && 0.00000  && \\
Pb2  && 0.01871(14) &&  0.01871(14)  && 0.01213(15)  && 0.00936(7)  && 0.00000  && 0.00000  && \\
Cu2' &&  0.01871(14) &&  0.01871(14) &&  0.01213(15)  && 0.00936(7) &&  0.00000  && 0.00000  && \\
P1  && 0.0100(8)  && 0.0107(8)  && 0.0138(7)  && 0.0060(6)  && 0.00000 && 0.00000  && \\
O1  && 0.024(3) &&  0.029(3) &&  0.026(3) &&  0.021(3)  && 0.00000  && 0.00000 && \\ 
O2  && 0.024(2)  && 0.042(3) &&  0.023(2)  &&  0.019(2)  && -0.0067(17)  && -0.0102(19)  && \\
O3  && 0.031(3) &&  0.011(2)  && 0.035(3) &&  0.006(2)  && 0.00000  && 0.00000  && \\
S1 &&  0.017(2) &&  0.017(2) &&  0.045(5)  && 0.0084(12)  && 0.00000  && 0.00000 && \\ 
O4  && 0.013(6)  && 0.013(6)  && 0.061(17) &&  0.007(3) &&  0.00000 &&  0.00000 && \\
\hline 
\end{tabular}
\end{table*}
\end{center}

\begin{center}
\begin{table*}
\renewcommand{\tablename}{Table S}
\caption{Refined atomic coordinates for Pb$_{10}$(PO$_4$)$_6$(OH) $_{1.11}$ Cu$_{0.49}$.}
\begin{tabular}{@{}lllllllllllll@{}}
\hline
Site && Wyckoff Position && x   && y    && z          && Occupancy &&  \\  
\hline
Pb1  && 6h       && 0.24486(4) && 0.24667(4) && $1/4$ && 1         &&  \\
Pb2  && 4f               && $2/3$ && $1/3$ && -0.00339(4)        && 1         &&  \\
P1  && 6h               &&  0.3729(2)  && 0.4020(2) && $3/4$ && 1         &&  \\ 
O1  && 6h               &&  0.4851(7)  && 0.3366(8) && $3/4$ && 1         &&  \\ 
O2  && 12i               &&  0.2650(6)  && 0.3480(6) && 0.5834(6) && 1         &&  \\ 
O3  && 6h               &&  0.5336(8)  && 0.4165(7) && $1/4$ && 1         &&  \\ 
Cu1  && 2b               &&  0  && 0 && 0 && 0.247         &&  \\ 
O4  && 2a              &&  0  && 0 && $1/4$ && 0.55         &&  \\ 

\hline
\end{tabular}
\end{table*}
\end{center}

\begin{center}
\begin{table*}
\renewcommand{\tablename}{Table S}
\caption{Refined anisotropic displacement parameters for Pb$_{10}$(PO$_4$)$_6$(OH) $_{1.11}$ Cu$_{0.49}$.}
\begin{tabular}{lllllllllllllll}
\hline
Site && U$_{11}$ && U$_{22}$   && U$_{33}$    && U$_{12}$ && U$_{13}$         && U$_{23}$ &&  \\
\hline
Pb1 &&  0.02052(15)  && 0.01995(15) &&  0.02683(15) &&  0.01447(12)  && 0.00000  && 0.00000  && \\
Pb2 &&  0.01891(12)  && 0.01891(12)  && 0.01231(13)  && 0.00946(6)  && 0.00000  && 0.00000  && \\
P1  && 0.0088(7)  && 0.0095(7) &&  0.0125(7)  && 0.0054(6)  && 0.00000  && 0.00000  && \\
O1  && 0.022(3)  && 0.028(3) &&  0.024(3) &&  0.020(3)  && 0.00000  && 0.00000  &&  \\
O2  && 0.022(2)  && 0.040(3)  && 0.0211(19) &&  0.019(2)  && -0.0065(17)  && -0.0102(19) && \\ 
O3  && 0.029(3)  && 0.010(2)  && 0.034(3) &&  0.006(2) &&  0.00000 &&  0.00000  && \\
Cu1  && 0.012(2) &&  0.012(2)  && 0.041(5) &&  0.0062(12) &&  0.00000  && 0.00000  && \\
O4  && 0.020(8)  && 0.020(8)  && 0.07(2) &&  0.010(4) &&  0.00000  && 0.00000  && \\

\hline 
\end{tabular}
\end{table*}

\end{center}

\subsection{Powder Diffraction Comparisons}

\begin{figure*}[t]
    \centering
    \includegraphics[width=0.8\columnwidth]{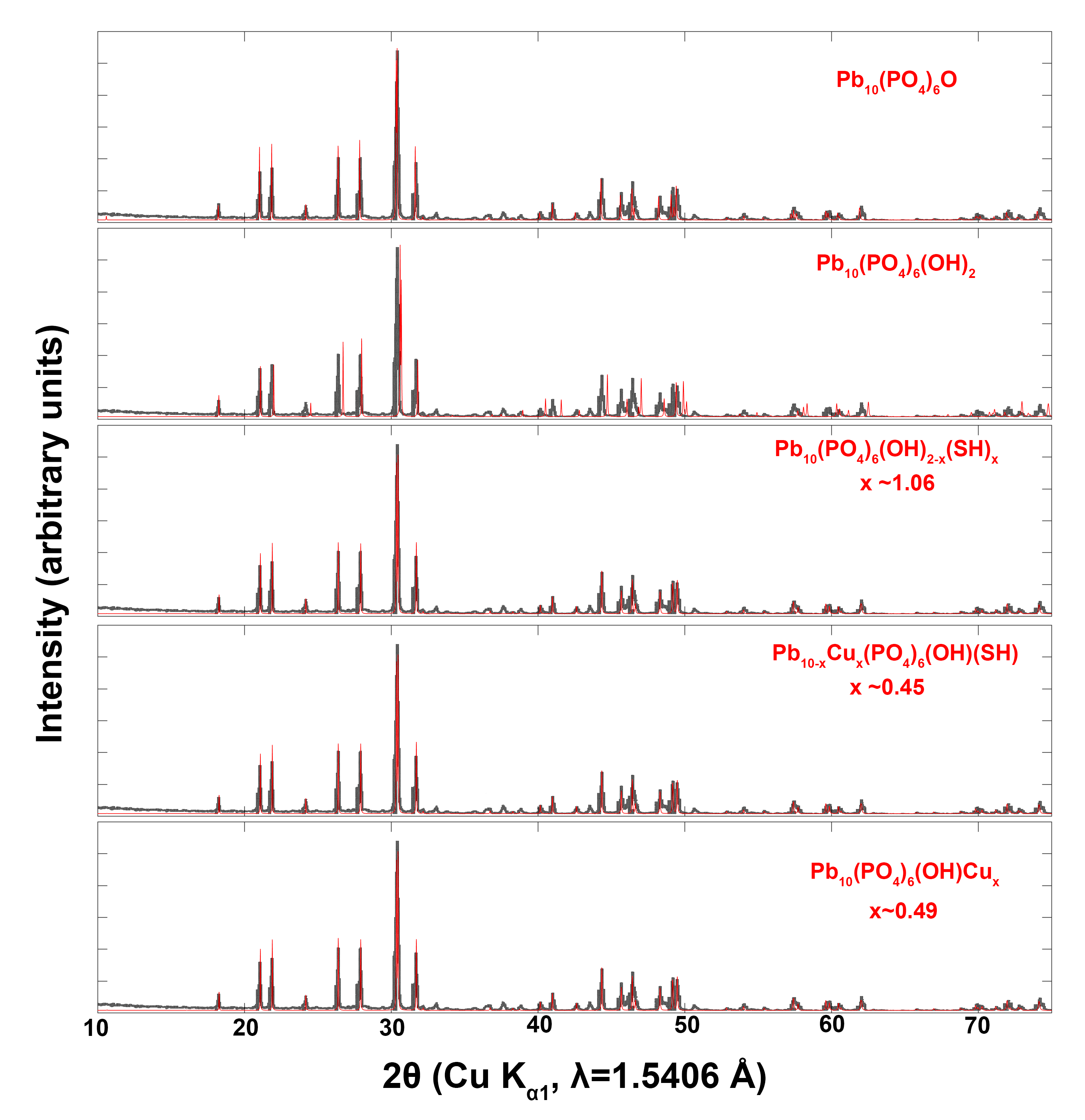}
    \caption{Comparisons of experimental (black) and calculated powder X-ray diffraction patters of structures in this work (red). The calculated patterns are all based on refinements of diffraction experiments.}
    \label{fig:ExpRefinements}
\end{figure*}

\begin{figure*}[t]
    \centering
    \includegraphics[width=0.8\columnwidth]{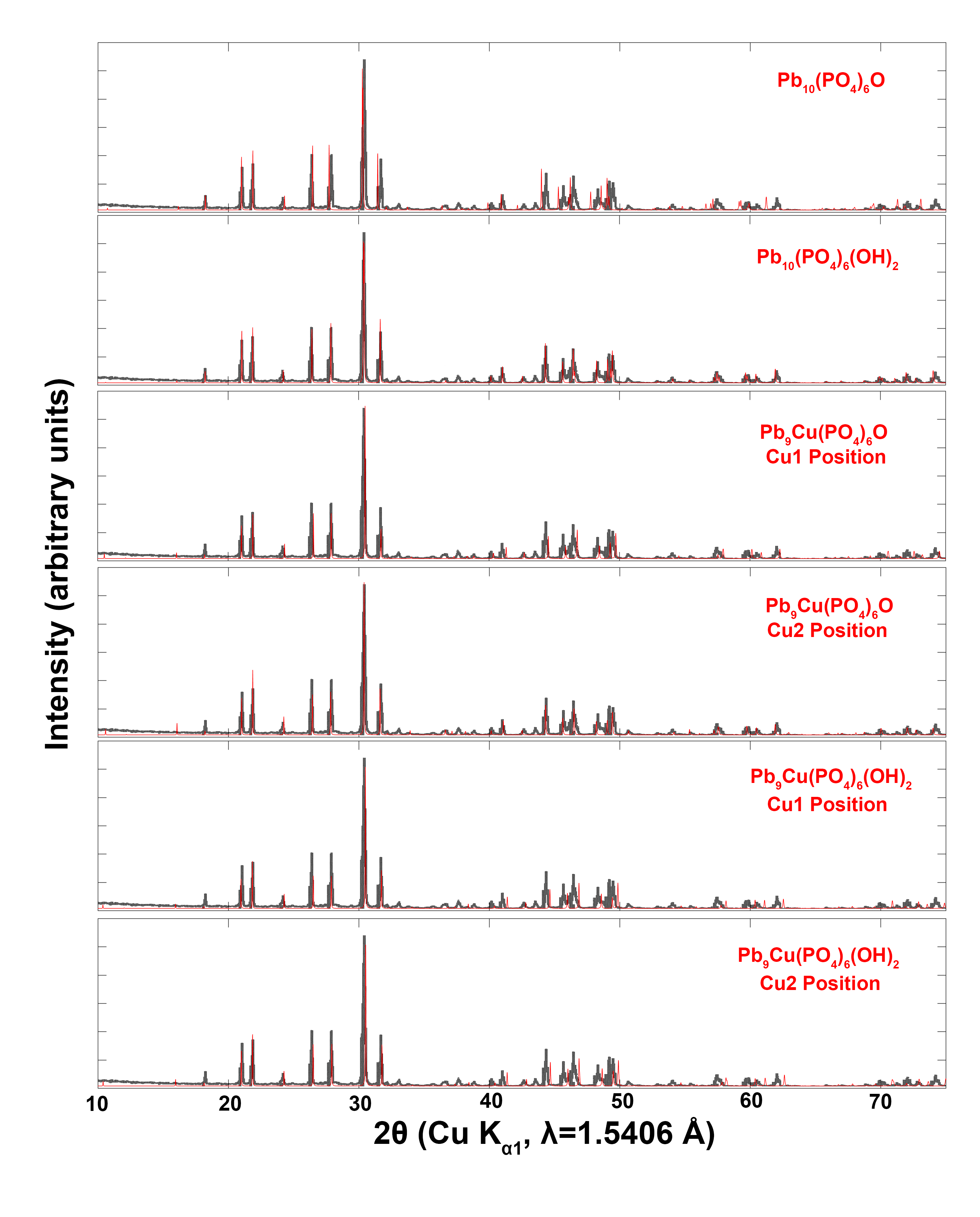}
    \caption{Comparisons of previous \cite{LEE23a,LEE23b} experimental (black) and simulated powder X-ray diffraction patters of relaxed structures in this work (red).}
    \label{fig:CompRefinements}
\end{figure*}

In this section, we compare the experimental data from \cite{LEE23a, LEE23b} with simulated patterns of experimentally determined \Fig{fig:ExpRefinements} and computational structures \Fig{fig:CompRefinements} used in this body of work. The first two comparisons show patterns from crystallographic information files not determined in this paper. Upon first glance, it is understandable why the originals authors came to the conclusion that their structure resembles Pb$_{10}$(PO$_4$)$_6$O. The direct overlap of peaks resembles a good match of the experimental pattern, however, as previously stated this structure is unstable at room temperature.

In contrast, the simulated pattern for Pb$_{10}$(PO$_4$)$_6$(OH)$_2$ seems chemically reasonable and utilizes neutron diffraction to refine the positions of the H atoms, yet falls short in matching well with the data. The slow migration of simulated peaks to higher 2$\theta$ values is indicative of the simulated crystal structure needs to expand slightly to overlap with the experimental data. 

Comparing to the simulated patterns of 3 of our structure solutions reported in this paper, we see negligible changes in the way our data fits the experimental work. This is despite the fact that one of our structures, Pb$_{10}$(PO$_4$)$_6$(OH)Cu$_{0.49}$ is chemically infeasible because it cannot be charged balanced. This leads us to the conclusion that even though the patterns match well, the match of experimental data is still not enough to accurately verify composition or structure.

In \Fig{fig:CompRefinements}, we compare the simulated patterns of computationally relaxed structures with the experimental data. Intriguingly, as we relax the structures of Pb$_{10}$(PO$_4$)$_6$O and Pb$_{10}$(PO$_4$)$_6$(OH)$_2$, both compounds expand, leading to Pb$_{10}$(PO$_4$)$_6$(OH)$_2$ looking like a proper fit. This structure is then likely another chemically reasonable interpretation of the data. 

\begin{figure*}[t]
    \centering
    \includegraphics[width=0.8\columnwidth]{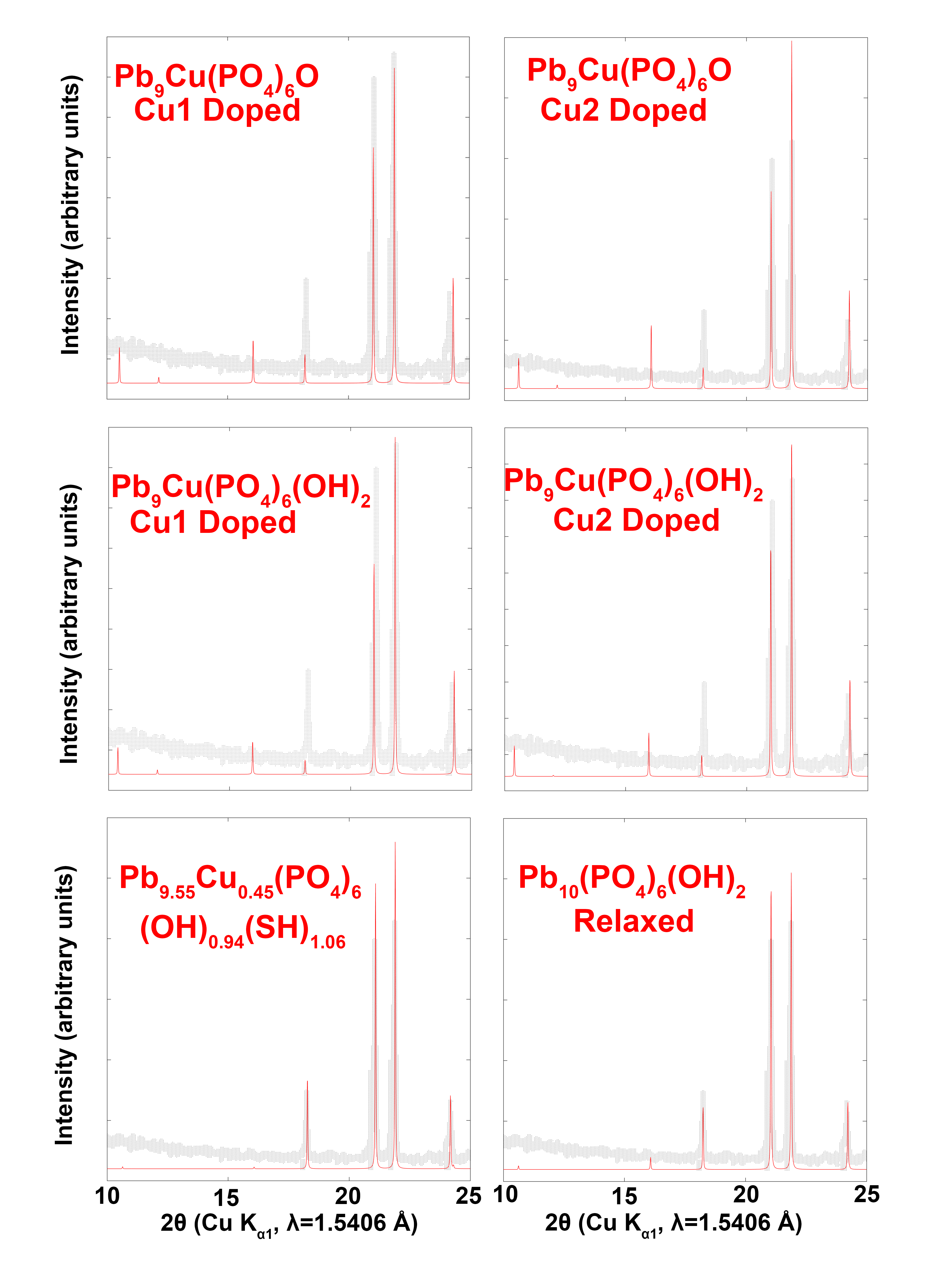}
    \caption{Zoomed in low angle experimental (black) and simulated powder X-ray diffraction patters of selected structures in this work (red).}
    \label{fig:ZoomPXRD}
\end{figure*}

As we probed the Cu doped structures, we noticed that all of these computationally modified structures grow another peak around 15 degrees which varies in intensity based on the anion identity and Cu doping position shown zoomed in with \Fig{fig:ZoomPXRD}. The highest expected intensity at this angle is found in the the Cu2-substituted Pb$_{9}$Cu(PO$_4$)$_6$O pattern, which, except for ignoring this peak is the best-matched pattern to the experimental data of the four. Once again, we see crystallographic uncertainties with ordered Cu dopants. From these simulations, we are skeptical of any ordering of Cu dopants inside the lead apatite structure. 

Another interesting possibility, however, is the prospect of the Cu dopants not having a preferred site selectivity. We note that our model for Pb$_{9.55}$Cu$_{0.45}$(PO$_4$)$_6$(OH)$_{0.94}$ and Pb$_{10}$(PO$_4$)$_6$(OH) $_{1.11}$ Cu$_{0.49}$ have somewhat significant Cu doping, yet they do not have any predicted peaks below 18 degrees. 

In conclusion, laboratory X-ray diffraction techniques such as SCXD and PXRD are not yet enough to come to a conclusion on the exact structure of lead apatite synthesized via the Lanarkite plus Cu$_3$P route, but hope that symbiotic techniques such as neutron scattering and X-ray photoelectron spectroscopy will be used in the near future.

\bibliographystyle{apsrev4-2}
\bibliography{LK99}

\end{document}